\documentclass[preprint,tightenlines,eqsecnum,superscriptaddress,floats,nofootinbib,showpacs,11pt]{revtex4}

\usepackage{amsmath,amssymb,amsfonts}
\usepackage{graphicx}

\newcommand{\cyl}{\mathrm{Cyl}} 
\newcommand{\hilbert}{\mathcal{H}} 
\newcommand{\dual}[1]{{#1}^\star} 
\newcommand{\abs}[1]{{\left|{#1}\right|}} 
\newcommand{\inner}[2]{{\langle {#1}\vert {#2} \rangle}} 
\newcommand{\oinner}[2]{({#1}\vert{#2})} 
\newcommand{\ket}[1]{\vert{#1}\rangle} 
\newcommand{\round}[1]{({#1}\vert} 

\newcommand{\mubar}{{\bar \mu}} 

\newcommand{\Pl}{\ell_{\rm Pl}} 
\newcommand{\Plm}{m_{\rm Pl}} 
\newcommand{\sgn}{{\rm sgn}} 

\newcommand{\secref}[1]{Sec.~\ref{#1}}

\newcommand{\eqnref}[1]{(\ref{#1})}
\newcommand{\figref}[1]{Fig.~\ref{#1}}
\newcommand{\tabref}[1]{Table~\ref{#1}}
\newcommand{\appref}[1]{Appendix~\ref{#1}}
\newcommand{\footref}[1]{Footnote~\ref{#1}}

\begin{document}


\title{Loop quantization of spherically symmetric midisuperspaces and loop quantum geometry of the maximally extended Schwarzschild spacetime}
\author{Dah-Wei Chiou}
\email{chiou@gravity.psu.edu}
\affiliation{Center for Advanced Study in Theoretical Sciences and Center for Condensed Matter Sciences, National Taiwan University, Taipei, Taiwan}
\author{Wei-Tou Ni}
\email{weitou@gmail.com}
\affiliation{Center for Gravitation and Cosmology, Department of Physics, National Tsing Hua University, Hsinchu, Taiwan}
\author{Alf Tang}
\email{alf.tang@nspo.narl.org.tw}
\affiliation{National Space Organization, Hsinchu, Taiwan}

\begin{abstract}
We elaborate on the Ashtekar's formalism for spherically symmetric midisuperspaces and, for loop quantization, propound a new quantization scheme which yields a graph-preserving Hamiltonian constraint operator and by which one can impose the fundamental discreteness of loop quantum gravity \textit{\`{a} la} the strategy of ``improved'' dynamics in loop quantum cosmology (LQC). Remarkable consequences are inferred at the heuristic level of effective dynamics: first, consistency of the constraint algebra regarding the Hamiltonian and diffeomorphism constraints fixes the improved quantization scheme to be of the form reminiscent of the improved scheme in LQC which preserves scaling invariance; second, consistency regarding two Hamiltonian constraints further demands the inclusion of higher order holonomy corrections and fixes a ratio factor of 2 for the improved scheme. It is suggested that the classical singularity is resolved and replaced by a quantum bounce which bridges a classical solution to another classical phase. However, the constraints violate briefly during the bouncing period, indicating that one cannot make sense of symmetry reduction by separating the degrees of freedom of the full theory into spherical and non-spherical ones in the vicinity of the quantum bounce, although the heuristic effective dynamics can still give a reliable semiclassical description of large-scale physics. Particularly, for the Schwarzschild solution in accordance with the Kruskal coordinates, revealing insights lead us to conjecture the complete quantum extension of the Schwarzschild spacetime: the black hole is evaporated via the Hawking radiation and meanwhile the quantum spacetime is largely extended from the classical one via the quantum bounce, suggesting that the information paradox might be resolved.
\end{abstract}

\pacs{04.60.Pp, 04.70.Dy, 03.65.Sq}

\maketitle

\tableofcontents

\newpage

\section{Introduction}
Loop quantum gravity (LQG) \cite{Ashtekar:2004eh,Rovelli:2004tv,Thiemann:2007zz} is considered to be a promising candidate for quantum theory of gravity. It provides mathematically rigorous foundation for quantum gravity and has led to several significant results about the quantum structure of space and time. While its kinematics is well understood, many open issues remain unsettled, mainly in regard to the dynamics and the semiclassical limit. In spite of difficulties in the full theory of LQG, the loop approach has been successfully applied for symmetry-reduced minisuperspaces in the context of loop quantum cosmology (LQC). LQC has yielded valuable insight into the full theory as many issues are better understood in LQC thanks to its formal simplicity. In recent years, LQC has undergone lively progress and become an active area of research \cite{Bojowald:2008zzb,Ashtekar:2011ni,Bojowald:book}. Markedly, it has been shown that, for a variety of models of LQC, the cosmological singularity (big bang, big crunch, big rip \cite{Chiou:2010ne}, etc.) is resolved and replaced by the big bounce, therefore affirming the long-held conviction that singularities in general relativity signal a breakdown of the classical theory and should be resolved by the quantum effects of gravity.

It is natural to ask whether the black hole singularity is resolved as well. To study loop quantum geometry of black holes, the simplest step  is to consider the interior of a Schwarzschild black hole, in which the metric components are homogeneous with Kantowski-Sachs symmetry. By virtue of homogeneity, the loop quantization of the Schwarzschild interior can be formulated as a minisuperspace model in a similar fashion to LQC. This has been developed in \cite{Modesto:2004wm,Ashtekar:2005qt,Modesto:2005zm} and its effective solution has been investigated at the level of heuristic dynamics in \cite{Modesto:2006mx} based on the original quantization strategy (also referred to as ``$\mu_0$-scheme'') and in \cite{Bohmer:2007wi,Chiou:2008nm} based on the modified quantization strategies (``$\mubar$-scheme'' and ``$\mubar'$-scheme''). In the $\mu_0$- and $\mubar$-schemes, the black hole singularity is resolved by the quantum bounce, which bridges the black hole interior with a white hole interior. In the $\mubar'$-scheme, the black hole singularity is resolved and the event horizon is also diffused by the quantum bounce, through which the black hole is connected to a baby black hole with a much smaller mass.

Both $\mu_0$- and $\mubar$-schemes suffer from the problem that the resulting dynamics depends on the arbitrary choice of the finite sized comoving cell to which the spatial integration is restricted to make the Hamiltonian finite. On the other hand, the $\mubar'$-scheme is independent of the choice of the finite sized cell and thus considered to be the correct quantization strategy in accordance with the idea of ``improved dynamics'' first suggested in \cite{Ashtekar:2006wn} and later generalized for Bianchi I models in \cite{Chiou:2007mg,Ashtekar:2009vc}. As opposed to the results of the $\mubar'$-scheme heuristic dynamics, however, the spacetime curvature in the immediate vicinity of the event horizon can be fairly flat and does not necessarily incur any quantum corrections. The reason that the event horizon wrongly receives quantum corrections in the $\mubar'$-scheme is essentially because the finite sized cell collapses as its side surface shrinks when approaching the horizon. The collapse near the horizon is merely an artifact resulting from coordinate singularity; in actual fact, a given finite sized cell will simply pass through the horizon without being collapsed.
This glitch calls into question the minisuperspace treatment for the Schwarzschild interior, even though it is legitimate for the Kantowski-Sachs cosmology as studied in \cite{Chiou:2008eg}.\footnote{The Schwarzschild interior cannot be considered as a self-contained university, as it is extensible beyond its boundary (event horizon). Put differently, classical general relativity does not admit a \emph{vacuum} cosmological solution with Kantowski-Sachs symmetry. With inclusion of matter content, on the other hand, the Kantowski-Sachs cosmology exists and the LQC treatment based on it is sensible (see \cite{Chiou:2008eg} for the case with a scalar field).} Therefore, one is obliged to consider the black hole interior and exterior as a whole in the framework of spherically symmetric midisuperspaces.

Loop theories of spherically symmetric midisuperspaces are of theoretical importance in its own right, as they deal with the simplest field-theory framework which is symmetry-reduced. They provide an arena for testing important issues which are too difficult in the full theory and trivialized in minisuperspace (LQC) models. Particularly, the $SU(2)$ internal gauge is reduced to $U(1)$ and the 3-dimensional diffeomorphism is reduced to 1-dimensional, making the constraint algebra much simpler yet nontrivial. Ashtekar's formalism for spherically symmetric midisuperspaces and its loop quantization have been studied and developed with different degrees of rigor \cite{Bojowald:2004af,Bojowald:2004ag,Bojowald:2005cb,Campiglia:2007pr}. In this paper, we recapitulate results of previous works and elaborate on the geometrical meaning of the Ashtekar variables and constraint algebra. For loop quantization, we suggest that some details have to be modified and propound a new quantization scheme which yields a graph-preserving Hamiltonian constraint operator and by which one can impose the fundamental discreteness of LQG by hand \textit{\`{a} la} the strategy of improved dynamics in LQC.

Rigorous and complete construction for the loop quantum theory of spherically symmetric midisuperspaces is faced with complications and still challenging. Nevertheless, ramifications of loop quantization and other open issues can still be inferred at the heuristic level of semiclassical dynamics. Particularly, an effective solution corresponding to the semiclassical theory has been obtained for the complete Schwarzschild spacetime which covers both the interior and exterior in \cite{Gambini:2008dy}, yielding a singularity-free global structure akin to that in the $\mu_0$- and $\mubar$-schemes of the minisuperspace (interior) treatment in \cite{Modesto:2006mx,Bohmer:2007wi,Chiou:2008nm}.

The formulation of \cite{Gambini:2008dy} follows the strategy of \cite{Campiglia:2007pr} to partially fix the 1-dimensional diffeomorphism gauge in such a way that one is left with a single abelian constraint and a true Hamiltonian. The partial gauge fixing avoids the hard problem of having structure functions in the constraint algebra and thus makes the consistent loop quantum theory possible, but it also obscures significance of the interplay between the diffeomorphism and Hamiltonian constraints.\footnote{Analogously, on top of the strategy of \cite{Gambini:2008dy}, if one further fixes the diffeomorphism gauge completely before quantization, one is led to the quantization of Kucha\v{r} \cite{Kuchar:1994zk}, which does not offer insights about the structure of the quantum spacetime and the question of singularities.} In this paper, we do not fix the diffeomorphism gauge but instead propose a new quantization scheme. The new quantization scheme does not resolve the complications of constructing a consistent loop quantum theory but nevertheless it yields profound insights at the heuristic level of semiclassical dynamics. Remarkably, consistency of constraint algebra in heuristic effective dynamics leads to fascinating consequences. First, requirement that the Poisson bracket between the Hamiltonian and diffeomorphism constraints has to weakly vanish fixes the improved quantization scheme to be of the same form as the $\mubar'$-scheme in minisuperspace (LQC) models \cite{Chiou:2008nm,Chiou:2007mg,Chiou:2008eg}. Second, requirement that the Poisson bracket between any two Hamiltonian constraints (with different lapse functions) also has to weakly vanish further demands the inclusion of higher order holonomy corrections, which was first suggested in \cite{Chiou:2009hk,Chiou:2009yx} in the context of LQC but motivated differently. The second requirement also fixes a ratio factor of 2 for the improved quantization scheme, which remains ambiguous (and is usually set to be 1) in homogenous models.

Analysis of the heuristic effective dynamics suggests that, with loop quantum corrections, the black hole singularity is resolved and replaced by a quantum bounce. The quantum bounce is expected to bridge a classical solution to another classical one, as the constraint algebra is exactly satisfied before and after the bounce, provided that the improved quantization scheme is adopted and higher order holonomy corrections are included to the order of infinity. The constraint algebra, however, breaks down briefly during the transition period of the bounce, implying that the spacetime is no longer invariant under spatial diffeomorphism and change of spacetime foliation after the bounce. Nevertheless, the violation is minuscule at large scale and thus the heuristic effective dynamics remains reliable after the bounce insofar as large-scale physics is concerned.
During the transition of the quantum bounce, the intimate matching between diffeomorphism and Hamiltonian constraints is spoiled as a consequence of imposing fundamental discreteness.\footnote{By contrast, in the context of LQC, imposing discreteness by hand yields no inconsistency, because diffeomorphism is removed by homogeneity and one is left with scaling invariance (with respect to the choice of the finite sized cell). However, as a consequence of imposing discreteness, scaling invariance is softly broken in the quantum theory (the fundamental discreteness enters as a constant step size in the difference equation of Hamiltonian constraint, while triad variables scale up with the finite sized cell).} This indicates that, in the full theory of LQG, a given \emph{weave} state (coherent state of spin networks) which represents a smooth, spherically symmetric space at large scale will eventually manifest granularity of spin networks when approaching the quantum bounce. In other words, close to the quantum bounce, the weave state becomes very granular to the extent that one can no longer make sense of symmetry reduction by separating out spherical degrees from non-spherical fluctuations. After the bounce, however, the weave state evolves to become smooth and spherically symmetric again.

To study the interior and exterior of the Schwarzschild black hole at the same time, we choose the Kruskal coordinates, which cover the entire manifold of the maximally extended Schwarzschild spacetime (i.e. Kruskal extension of the Schwarzschild spacetime) \cite{Kruskal:1959vx,Szekeres:1960}. Although the numerical method remains extremely challenging, inspection of the Hamilton's equation at the level of heuristic effective dynamics still offers revealing insight about the characteristic natures of the loop quantum geometry of the Schwarzschild spacetime. Firstly, the classical singularity is resolved and replaced by a quantum bounce which bridges the black/white hole interior to a different classical phase. Secondly, inclusion of the Hawking radiation is mandatory and thus the Hawking evaporation is expected. These lead us to conjecture that the complete quantum Schwarzschild spacetime manifests the similar structure of the 2-dimensional quantum black holes studied in \cite{Ashtekar:2008jd,Ashtekar:2010hx,Ashtekar:2010qz}: the black hole is evaporated via the Hawking radiation, and the quantum spacetime is largely extended from the classical one as the black hole interior is extended beyond the putative singularity.\footnote{\label{foot:different results}The global structure of the quantum Schwarzschild spacetime conjectured in this paper is very different from that obtained in \cite{Gambini:2008dy}. The discrepancy will be discussed in the end of \secref{sec:preliminary analysis}.} The information paradox due to the Hawking evaporation is believed to be resolved by the quantum extension of the classical spacetime as for the 2-dimensional black holes in \cite{Ashtekar:2008jd,Ashtekar:2010hx,Ashtekar:2010qz}.

This paper is organized as follows. In \secref{sec:Ashtekar's formalism}, we set up Ashtekar's formalism for spherically symmetric spacetimes. In \secref{sec:Kruskal solution}, the classical solution in terms Ashtekar variables for the maximally extended Schwarzschild spacetime is given in accordance with the Kruskal coordinates. In \secref{sec:loop quantization}, loop quantization of spherically symmetric midisuperspaces is discussed. In \secref{sec:heuristic dynamics}, we study the heuristic effective dynamics and conjecture the quantum geometry of the Schwarzschild spacetime. We end with a summary and discussion in \secref{sec:summary}. Supplementary materials are included in the appendices.

\section{Ashtekar's formalism for spherically symmetric spacetimes}\label{sec:Ashtekar's formalism}
In this section, we formulate the canonical theory of gravity in terms of Ashtekar variables for the spacetimes which admit spherical symmetry. We mainly follow the previous works in \cite{Bojowald:2004af,Bojowald:2005cb,Campiglia:2007pr} but give more details and expound on the physical meaning of the canonical variables and constraints.

\subsection{Ashtekar variables and constraints}
The ADM formalism supposes that the 4-dimensional spacetime $\mathcal{M}$ is foliated into a family of spacelike surfaces $\Sigma_t$, labeled by the parameter time coordinate $t$.
Further, we consider the cases that the topology of each of the ``leaves'' $\Sigma_t$ is of the form $\Sigma=I\times S^2$, where $I$ is a 1-dimensional manifold coordinatized by $x$ (called ``radial coordinate'') and $S^2$ is a 2-sphere coordinatized by $\theta$ and $\phi$.\footnote{This covers a wide range of important cases. For the exterior of the Schwarzschild solution, we have $x\in[0,\infty)$ and $x=0$ corresponds to the horizon \cite{Campiglia:2007pr}. For the interior of the Schwarzschild black hole \cite{Chiou:2008nm} or for the Kantowski-Sachs cosmology \cite{Chiou:2008eg}, we have $x\in(-\infty,\infty)$ and one can further reduce the phase space to that of finitely many degrees of freedom by imposing homogeneity. In this paper, we take into account both the interior and exterior of the Schwarzschild solution as a whole, and choose $x\in(-\infty,\infty)$ to be the horizontal coordinates in the Kruskal diagram.}

If $\Sigma$ admits spherical symmetry (i.e., there are three Killing vectors of $S^2$ that form the $so(3)\cong su(2)$ algebra), the Ashtekar connection in the full theory is symmetry-reduced to the form of
\begin{eqnarray}\label{A}
A = A_a^i \tau_i dx^a &=& A_x(x)\tau_3 dx + \left(A_1(x)\tau_1+A_2(x)\tau_2\right)d\theta \nonumber\\ &&\quad+\big[\left(A_1(x)\tau_2-A_2(x)\tau_1\right)\sin\theta + \tau_3\cos\theta\big]d\phi,
\end{eqnarray}
where $A_x$, $A_1$ and $A_2$ are functions of $x$ and $\tau_i$ and $\tau_i=-i\sigma_i/2$ with $\sigma_i$ being the Pauli matrices are orthonormal generators of $SU(2)$ (note $\tau_i\wedge\tau_j\equiv[\tau_i,\tau_j]={\epsilon_{ij}}^k\tau_k$). Correspondingly, the densitized triad takes the form
\begin{eqnarray}\label{E}
\tilde{E} = \tilde{E}^a_i \tau^i \partial_a &=& E^x(x)\tau_3\sin\theta\frac{\partial}{\partial x} + \left(E^1(x)\tau_1+E^2(x)\tau_2\right)\sin\theta\frac{\partial}{\partial\theta}\nonumber\\ &&\quad+\left(E^1(x)\tau_2-E^2(x)\tau_1\right)\frac{\partial}{\partial\phi},
\end{eqnarray}
where again $E^x$, $E^1$ and $E^2$ are functions of $x$.
Note that the full (3+1 dimensional) theory has been reduced by the spherical symmetry to a 1+1 dimensional theory. In the full theory, the connection is a one-form and the densitized triad is a vector density of weight 1. But in the one dimension, under the coordinate transformation $x\rightarrow\bar{x}(x)$, $A_x$ transforms as a scalar density of weight 1, i.e. $\bar{A}_x(\bar{x})=(\partial x/\partial\bar{x})A_x(x)$, while $A_1$ and $A_2$ are scalars; and $E^x$ transforms as a scalar, while $E^1$ and $E^2$ are scalar densities of weight 1.\footnote{\label{foot:density character}It is helpful to keep track of the density character of the canonical variables. For example, it can be used to check that $U(x)$ defined in \eqnref{Gauss constraint}, $D(x)$ in \eqnref{diff constraint} and $C(x)$ in \eqnref{Hamiltonian constraint} are scaler densities of weight 1, weight 2, and weight 1, respectively. (Correspondingly, $\lambda(x)$, $N^x(x)$ and $N(x)$ are a scalar, a scalar density of weight -1, and a scalar, respectively). Furthermore, The minus sign in the integrand in \eqnref{diff constraint} is due to the fact that $A_x$ is a scalar density and $E^x$ is a scalar, while the pairs $(A_1,E^1)$ and $(A_2,E^2)$ are the other way around. Also see \footref{foot:diff} and \appref{app:dimension and weight}.}

In terms of the symmetry-reduced variables, the symplectic structure given by the symplectic two-form reads as:
\begin{eqnarray}\label{Omega}
\Omega &=& \frac{1}{8\pi G\gamma}\int_\Sigma d^3x \delta \tilde{E}^a_i\wedge\delta A_a^i\\
   &=& \frac{1}{8\pi G\gamma}\int_I dx \int d\theta d\phi\sin\theta
   \left(\delta E^x\wedge\delta A_x
   +2\left(\delta E^1\wedge\delta A_1+\delta E^2\wedge\delta A_2\right)\right)\nonumber\\
   &=& \frac{1}{2G\gamma}\int_I dx \left(\delta E^x\wedge\delta A_x
   +2\left(\delta E^1\wedge\delta A_1+\delta E^2\wedge\delta A_2\right)\right),
\end{eqnarray}
which implies
\begin{subequations}
\begin{eqnarray}
  \label{bracket Ax Ex}
  \{A_x(x),E^x(x')\} &=& 2G\gamma\delta(x-x'), \\
  \{A_1(x),E^1(x')\} &=& G\gamma\delta(x-x'), \\
  \{A_2(x),E^2(x')\} &=& G\gamma\delta(x-x'),
\end{eqnarray}
\end{subequations}
and all the other brackets vanish. $G$ is Newton's constant and $\gamma$ is the Barbero-Immirzi parameter.

The connection in \eqnref{A} gives the field strength\footnote{Throughout this paper, we will use ${}'$ and $\dot{}$ to denote derivatives with respect to the radial coordinate $x$ and the parameter time $t$, respectively. However, in some occasions, ${}'$ is merely used to denote a related but different variable, such as $\delta(x-x')$ as $x'$ is in contrast to $x$ or $E\rightarrow E'$ as $E'$ is transformed from $E$. The meaning of ${}'$ should have no confusion when put in the context.}
\begin{eqnarray}\label{Fab}
  F &=& \frac{1}{2}F_{ab}dx^a\wedge dx^b = dA+A\wedge A \nonumber \\
    &=& \left(A_1^2+A_2^2-1\right)\tau_3\sin\theta\, d\theta\wedge d\phi
        -\left[\right(A'_1\tau_2-A'_2\tau_1\left)-A_x\left(A_1\tau_1+A_2\tau_2\right)\right]\sin\theta \,d\phi\wedge dx \nonumber\\
    & & + \left[\right(A'_1\tau_1+A'_2\tau_2\left)+A_x\left(A_1\tau_2-A_2\tau_1\right)\right]dx\wedge d\theta.
\end{eqnarray}

In the full theory \cite{Ashtekar:2004eh}, the dynamics is described by three constraints. First, the Gauss constraint function is given by
\begin{equation}
    \mathcal{C}_G[\Lambda^i(\vec{x})]:=\frac{1}{8\pi G\gamma}\int_\Sigma d^3x \Lambda^i G_i,
\end{equation}
which generates the internal $SU(2)$ gauge transformation with
\begin{equation}
    G_i=\mathcal{D}_a \tilde{E}^a_i := \partial_a\tilde{E}^a_i+{\epsilon_{ij}}^k A_a^jE^a_k
\end{equation}
for any smooth field $\Lambda(\vec{x})$ on $\Sigma$.
With the symmetry-reduced variables in \eqnref{A} and \eqnref{E}, it is easy to show that the two components $G_{i=1}$ and $G_{i=2}$ vanish identically and the only non-vanishing component is
\begin{equation}
    G_{i=3}=\sin\theta\left(E^{x\prime}+2A_1E^2-2A_2E^1\right).
\end{equation}
Thus, the $SU(2)$ gauge is reduced to $U(1)$ and generated by
\begin{equation}\label{Gauss constraint}
    \mathcal{C}_G[\lambda(x)]=\frac{1}{2G\gamma}\int_I dx \lambda\left(E^{x\prime}+2A_1E^2-2A_2E^1\right) =:\int_I dx \lambda(x)U(x),
\end{equation}
where $\lambda(x):=\frac{1}{4\pi}\int d\theta d\phi\sin\theta\Lambda^{i=3}(x,\theta,\phi)$ and note that the factor $(4\pi)^{-1}$ is introduced to counterbalance the angular integral.\footnote{\label{foot:Gauss transform}Under the $U(1)$ gauge transformation, the term $2A_1E^2-2A_2E^1$ in \eqnref{Gauss constraint} corresponds to a rotation on the $(\tau_1,\tau_2)$ plane, while $E^{x\prime}$ gives rise to the inhomogeneous term for the transformation of $A$. That is
\begin{eqnarray*}
\delta E^1(x) &=& \left\{E^1(x), \frac{1}{2G\gamma}\int_I dx' \lambda(x')\left(2A_1(x')E^2(x')-2A_2(x')E^1(x')\right)\right\}\\
&=& -\int dx'\lambda(x')\delta(x'-x)E^2(x')=-\lambda(x)E^2(x),
\end{eqnarray*}
and similarly
\begin{equation*}
\delta E^2(x) = \lambda(x)E^1(x), \qquad \delta A_1(x)=-\lambda(x)A_2(x), \qquad \delta A_2(x)=\lambda(x)A_1(x),
\end{equation*}
while
\begin{equation*}
\delta A_x(x) = \left\{A_x(x), \frac{1}{2G\gamma}\int_I dx' \lambda(x')\partial_{x'}E^x(x') \right\} = \int dx'\lambda(x')\partial_{x'}\delta(x-x') =-\partial_x\lambda(x)
\end{equation*}
and obviously
\begin{equation*}
\delta E^x(x)=0.
\end{equation*}
To summarize, the $SU(2)$ gauge transformation $e^{\vec{\tau}\cdot\vec{\Lambda}(x)}$ is reduced to the $U(1)$ transformation $e^{\tau_3\lambda(x)}$, which rotates $(\tau_1,\tau_2)$ by the angle $\lambda(x)$, and the fields transform accordingly:
\begin{equation*}
E\rightarrow E'=e^{\tau_3\lambda(x)}E\,e^{-\tau_3\lambda(x)}
\end{equation*}
and
\begin{equation*}
A\rightarrow A'=e^{\tau_3\lambda(x)}A\,e^{-\tau_3\lambda(x)}+e^{\tau_3\lambda(x)}de^{-\tau_3\lambda(x)} =e^{\tau_3\lambda(x)}A\,e^{-\tau_3\lambda(x)}-\partial_x\lambda(x)\tau_3dx.
\end{equation*}
}

Second, the diffeomorphism constraint function is given by
\begin{equation}
    \mathcal{C}_\mathrm{Diff}[N^a(\vec{x})]:=\frac{1}{8\pi G\gamma}\int_\Sigma d^3x \left(N^a\tilde{E}^a_iF^i_{ab}-N^aA_a^iG_i\right),
\end{equation}
which generates diffeomorphisms in $\Sigma$ for any smooth vector field $N^a(\vec{x})$.
In terms of the symmetry-reduced variables, again, two components of the integrand coupled with $N^\theta$ and $N^\phi$ vanish identically and the only non-vanishing contribution is given by
\begin{equation}\label{diff constraint}
    \mathcal{C}_\mathrm{Diff}[N^x(x)] =\frac{1}{2G\gamma}\int_I dx N^x \left(2A'_1E^1+2A'_2E^2-A_xE^{x\prime}\right) =: \int_I dx N^x(x)D(x),
\end{equation}
where $N^x(x):=\frac{1}{4\pi}\int d\theta d\phi \sin\theta N^{a=x}(x,\theta,\phi)$. This generates the remnant diffeomorphism along $x$.\footnote{\label{foot:diff}Under an infinitesimal diffeomorphism transformation $x\rightarrow\bar{x}=x+\epsilon(x)$, $A_1$ transforms as a scalar: $A_1(x)\rightarrow\bar{A}_1(\bar{x})=A_1(x)$. Thus,
\begin{eqnarray*}
\delta A_1(x)&=&\bar{A}_1(x)-A(x)=\bar{A}_1(\bar{x}-\epsilon(x))-A_1(x) =\bar{A}_1(\bar{x})-\epsilon(x)\partial_{\bar{x}}\bar{A}_1(\bar{x})-A_1(x)+\mathcal{O}(\epsilon^2) \\ &=&-\epsilon(x)\partial_xA_1(x)+\mathcal{O}(\epsilon^2).
\end{eqnarray*}
And $E^1$ transforms as a scalar density:
$E^1(x)\rightarrow \bar{E}^1(\bar{x})=({\partial x}/{\partial\bar{x}})E^1(x) =\left(1-\partial_x\epsilon(x)\right)E^1(x)+\mathcal{O}(\epsilon^2)$. Thus,
\begin{eqnarray*}
\delta E^1(x)&=&\bar{E}^1(x)-E^1(x) =\bar{E}^1(\bar{x})-\epsilon(x)\partial_{\bar{x}}\bar{E}^1(\bar{x})-E^1(x)+\mathcal{O}(\epsilon^2)\\ &=&-\partial_x\epsilon(x) E^1(x)-\epsilon(x)\partial_xE^1(x)+\mathcal{O}(\epsilon^2).
\end{eqnarray*}
This is in agreement with the $2A_1'E^1$ term in \eqnref{diff constraint} provided that $\bar{x}=x-N^x(x)dt$, as
\begin{eqnarray*}
\delta A_1(x)&=&\left\{A_1(x),\frac{1}{2G\gamma}\int_I dx'2N^x(x')A_1'(x')E^1(x')\right\} =\int dx'N^x(x')A_1'(x')\delta(x-x') =N^x(x)\partial_x A_1(x),\\
\delta E^1(x)&=&\left\{E^1(x),\frac{1}{2G\gamma}\int_I dx'2N^x(x')A_1'(x')E^1(x')\right\} =-\int dx'N^x(x')\partial_{x'}\delta(x-x')E^1(x')\\
&=&\int dx'\delta(x-x')\partial_{x'}\left(N^x(x')E^1(x')\right)
=\partial_xN^x(x) E^1(x)+N^x(x)\partial_xE^1(x)
\end{eqnarray*}
Similarly, $2A_2'E^2$ term corresponds to the diffeomorphism transformation for the pair $(A_2,E^2)$. On the other hand, because $(A_x,E^x)$ has opposite density character, it corresponds to $-A_xE^{x\prime}$ with a minus sign (recall \footref{foot:density character}). Generically, for any scalar density $f(x)$ of density weight $w$, i.e. $f(x)\rightarrow\bar{f}(\bar{x}) = (\partial x/\partial\bar{x})^wf(x)$, we have
\begin{equation*}
\delta f(x)=-w\partial_x\epsilon(x) f(x)-\epsilon(x)\partial_xf(x)+\mathcal{O}(\epsilon^2),
\end{equation*}
which is in agreement with
\begin{equation*}
\delta f(x) =\left\{f(x),\mathcal{C}_\mathrm{Diff}[N^x]\right\} = w\,\partial_xN^x(x)f(x)+N^x(x)\partial_xf(x) =:\mathcal{L}_{N^x}f(x),
\end{equation*}
if $f(x)$ is a composite function of the canonical variables.}

Finally, the Hamiltonian constraint (also called scalar constraint) is given by
\begin{equation}\label{full Hamiltonian constraint}
    \mathcal{C}[N(\vec{x})]:=\frac{1}{16\pi G}\int_\Sigma d^3x N(\vec{x}) e^{-1}
    \left\{{\epsilon_i}^{jk}F^i_{ab}\tilde{E}^a_j\tilde{E}^b_k-2(1+\gamma^2)K_{[a}^iK_{b]}^j\tilde{E}^a_i\tilde{E}^b_j\right\},
\end{equation}
where $f_{[a}f_{b]}:=(f_af_b-f_bf_a)/2$ and $e:=\big|\det \tilde{E}^a_i\big|^{1/2}$ (note that $\det \tilde{E}^a_i=\det q_{ab}=:q$ by the relation $qq^{ab}=\tilde{E}^a_i\tilde{E}^b_j\delta^{ij}$). The triad in \eqnref{E} gives $\det \tilde{E}^a_i=E^x[(E^1)^2+(E^2)^2]\sin^2\theta$.
The extrinsic curvature takes the form
\begin{equation}\label{K}
    K=K_a^i\tau_i dx^a=K_x(x)\tau_3 dx + \left(K_1(x)\tau_1+K_2(x)\tau_2\right)d\theta + \left(K_1(x)\tau_2-K_2(x)\tau_1\right)\sin\theta d\phi,
\end{equation}
where $K_x$, $K_1$ and $K_2$ can be written in terms of $A$'s and $E$'s.
By \eqnref{E} and \eqnref{Fab}, the reduced variables then yield
\begin{eqnarray}\label{Hamiltonian constraint}
  \mathcal{C}[N(x)] &=& \frac{1}{2G} \int_I dx \frac{N(x)}{\sqrt{\abs{E^x}[(E^1)^2+(E^2)^2]}}\nonumber\\
       && \quad \times
           \Big\{ 2E^x\left(E^1A'_2-E^2A'_1\right) + 2A_xE^x\left(A_1E^1+A_2E^2\right)
           +(A_1^2+A_2^2-1)[(E^1)^2+(E^2)^2]\nonumber\\
       &&  \qquad \quad -(1+\gamma^2)\left(2K_xE^x(K_1E^1+K_2E^2) + [(K_1)^2+(K_2)^2][(E^1)^2+(E^2)^2]\right)\Big\}\nonumber\\
       &=:&\int_I dx N(x) C(x),
\end{eqnarray}
where $N(x)=\frac{1}{4\pi}\int d\theta d\phi \sin\theta\, N(x,\theta,\phi)$.

\subsection{Polar type variables}
In what follows, to be better adapted to the $U(1)$ gauge transformation, we follow the ideas in \cite{Bojowald:2004af,Bojowald:2005cb,Campiglia:2007pr} to introduce the ``polar'' type variables.\footnote{In this paper, we adopt slightly different notations for polar type variables, which we believe are better than those used in \cite{Bojowald:2004af,Bojowald:2004ag,Bojowald:2005cb,Campiglia:2007pr,Gambini:2008dy}. See \appref{app:polar variables}.} First, we define $A_\rho$ and $E^\rho$ via
\begin{subequations}\label{polar A E}
\begin{eqnarray}
  A_1 &=& A_\rho\cos\beta, \qquad A_2=A_\rho\sin\beta, \\
  \label{Erho}
  E^1 &=& E^\rho\cos(\alpha+\beta), \qquad E^2=E^\rho\sin(\alpha+\beta).
\end{eqnarray}
\end{subequations}
Note that $A_\rho$, $E^\rho$ and the inner product $(A_1,A_2)\cdot(E^1,E^2)=A_\rho E^\rho\cos\alpha$ are invariant under the $U(1)$ rotation; i.e. the angle $\alpha(x)$ between $(A_1,A_2)$ and $(E^1,E^2)$ is gauge independent, while the angle $\beta(x)$ is pure gauge. With the new variables, \eqnref{Omega} becomes
\begin{subequations}
\begin{eqnarray}
\Omega
   &=& \frac{1}{2G\gamma}\int_I dx \Big\{\delta E^x\wedge\delta A_x
   +2\big(\cos\alpha\,\delta E^\rho\wedge\delta A_\rho +\sin\alpha\,A_\rho\delta E^\rho\wedge\delta\beta\nonumber\\
   \label{Omega 2a}
   &&\qquad\qquad\qquad+\sin\alpha\,E^\rho\delta A_\rho\wedge\delta\beta +\sin\alpha\,E^\rho\delta A_\rho\wedge\delta\alpha +\cos\alpha\,A_\rho E^\rho\delta\alpha\wedge\delta\beta \big)\Big\}\nonumber\\
   &=& \frac{1}{2G\gamma}\int_I dx \Big\{\delta E^x\wedge\delta A_x
   +2\big[\delta(E^\rho\cos\alpha)\wedge\delta A_\rho +\delta(A_\rho E^\rho\sin\alpha)\wedge\delta\beta \big]\Big\}\\
   \label{Omega 2b}
   &=& \frac{1}{2G\gamma}\int_I dx \Big\{\delta E^x\wedge\delta A_x
   +2\big[\delta E^\rho\wedge\delta(A_\rho\cos\alpha) +\delta(A_\rho E^\rho\sin\alpha)\wedge\delta(\alpha+\beta) \big]\Big\}.
\end{eqnarray}
\end{subequations}
Equation \eqnref{Omega 2b} tells that one can transform the set of canonical variables $(A_x,A_1,A_2;E^x,E^1,E^2)$ to the new one $(A_x,\bar{A}_\rho,\eta;E^x,E^\rho,P^\eta)$ by defining
\begin{subequations}\label{Abar eta Peta}
\begin{eqnarray}
  \bar{A}_\rho &:=& 2A_\rho\cos\alpha,\\
  \eta &:=& \alpha+\beta,\\
  P^\eta &:=& 2A_\rho E^\rho\sin\alpha
  = 2A_1E^2-2A_2E^1.
\end{eqnarray}
\end{subequations}
Equivalently, we have
\begin{subequations}
\begin{eqnarray}
  \{A_x(x),E^x(x')\} &=& 2G\gamma\delta(x-x'), \\
  \{\bar{A}_\rho(x),E^\rho(x')\} &=& 2G\gamma\delta(x-x'), \\
  \{\eta(x),P^\eta(x')\} &=& 2G\gamma\delta(x-x'),
\end{eqnarray}
\end{subequations}
and all other brackets for the new canonical variables vanish.\footnote{\label{foot:irrelevant variables}Alternatively, by \eqnref{Omega 2a}, one can transform the original canonical variables to a different canonical set $(A_x,A_\rho,\beta;E^x,P^\rho,P^\beta)$ with $P^\rho:=2E^\rho\cos\alpha$ and $P^\beta:=2A_\rho E^\rho\sin\alpha$ (see \cite{Bojowald:2004af,Bojowald:2005cb}). This canonical transformation is less relevant for our purpose.}

In terms of the polar type variables, the Gauss constraint \eqnref{Gauss constraint} reads as
\begin{equation}\label{U}
U=\frac{1}{2G\gamma}\left(E^{x\prime}+2A_1E^2-2A_2E^1\right)=\frac{1}{2G\gamma}\left(E^{x\prime}+2A_\rho E^\rho\sin\alpha\right) = \frac{1}{2G\gamma}\left(E^{x\prime}+P^\eta\right),
\end{equation}
and the diffeomorphism constraint \eqnref{diff constraint} becomes
\begin{eqnarray}\label{D}
D&=& \frac{1}{2G\gamma}\left(2A'_1E^1+2A'_2E^2-A_xE^{x\prime}\right) = \frac{1}{2G\gamma}\left(2A_\rho'E^\rho\cos\alpha +2A_\rho E^\rho\sin\alpha\,\beta'-A_xE^{x\prime}\right)\nonumber\\
&=& \frac{1}{2G\gamma}\left[\left((2A_\rho\cos\alpha)'+2A_\rho\sin\alpha\,\alpha'\right)E^\rho+2A_\rho E^\rho\sin\alpha\,\beta'-A_xE^{x\prime}\right]\nonumber\\
&=&\frac{1}{2G\gamma}\left(\bar{A}'_\rho E^\rho + \eta'P^\eta-A_xE^{x\prime}\right).
\end{eqnarray}

It is more involved to write the Hamiltonian constraint in terms of canonical variables as we first have to solve $(K_x,K_1,K_2)$ in terms of $A$'s and $E$'s. To make the calculation tractable, we perform change of variables in accordance with \eqnref{Erho} by rotating the orthonormal basis $(\tau_1,\tau_2,\tau_3)$ of $su(2)$ to the new orthonormal basis $(\tau_{1'},\tau_{2'},\tau_3)$ (orientation unchanged):
\begin{subequations}\label{new tau}
\begin{eqnarray}
\tau_{1'}(x) &:=& \tau_1\cos(\alpha+\beta)+\tau_2\sin(\alpha+\beta),\\
\tau_{2'}(x) &:=& -\tau_1\sin(\alpha+\beta)+\tau_2\cos(\alpha+\beta).
\end{eqnarray}
\end{subequations}
It should be noted that the rotation angle is different from point to point in $x$. Both $\tau_{1'}$ and $\tau_{2'}$ are functions of $x$ and we have
\begin{equation}\label{tau'}
\tau_{1'}'=\tau_{2'}(\alpha+\beta)',
\qquad
\tau_{2'}'=-\tau_{1'}(\alpha+\beta)'.
\end{equation}
Equations \eqnref{A} and \eqnref{E} then read as
\begin{eqnarray}\label{A 2}
A = A_a^i \tau_i dx^a &=& A_x(x)\tau_3 dx
+ A_\rho(x)\left(\cos\alpha\,\tau_{1'}(x)-\sin\alpha\,\tau_{2'}(x)\right)d\theta\nonumber\\
&&\quad
+A_\rho(x)\left(\sin\alpha\,\tau_{1'}(x)+\cos\alpha\,\tau_{2'}(x)\right)\sin\theta\,d\phi
+\tau_3\cos\theta\,d\phi,
\end{eqnarray}
\begin{equation}\label{E polar}
E = \tilde{E}^a_i \tau^i \partial_a = E^x(x)\tau_3\sin\theta\frac{\partial}{\partial x} + E^\rho(x)\tau_{1'}(x)\sin\theta\frac{\partial}{\partial\theta} + E^\rho(x)\tau_{2'}(x)\frac{\partial}{\partial\phi},
\end{equation}
as $\tilde{E}^a_i$ is diagonalized in the new basis.
We have $q=\det \tilde{E}^a_i=E^x(E^\rho)^2\sin^2\theta$ and the relation $e_a^i=\sqrt{\abs{q}}\,(\tilde{E}^a_i)^{-1}$ yields the (undensitized) cotriad:
\begin{equation}\label{cotriad}
e= e_a^i \tau_i dx^a = \frac{E^\rho}{\sqrt{\abs{E^x}}}\,\tau_3dx +\sqrt{\abs{E^x}}\,\tau_{1'}(x) d\theta +\sqrt{\abs{E^x}}\,\sin\theta\,\tau_{2'}(x) d\phi.
\end{equation}
Correspondingly, the 3-metric components given by $q_{ab}=e_a^ie_b^j\delta_{ij}$ read as
\begin{equation}
q_{xx}=\frac{(E^\rho)^2}{\abs{E_x}},
\qquad
q_{\theta\theta}=\abs{E^x},
\qquad
q_{\phi\phi}=\abs{E^x}\sin^2\theta.
\end{equation}
The cotriad in \eqnref{cotriad} gives
\begin{eqnarray}\label{de}
de &=& \frac{E^{x\prime}}{2\sqrt{\abs{E^x}}}\,\tau_{1'} dx\wedge d\theta +\sqrt{\abs{E^x}}\,(\alpha+\beta)'\tau_{2'} dx\wedge d\theta \nonumber\\
&&
+\frac{E^{x\prime}}{2\sqrt{\abs{E^x}}}\,\tau_{2'}\sin\theta\, dx\wedge d\phi -\sqrt{\abs{E^x}}\,(\alpha+\beta)'\tau_{1'}\sin\theta\, dx\wedge d\phi\nonumber\\
&&\quad
+\sqrt{\abs{E^x}}\,\tau_{2'}\cos\theta\, d\theta\wedge d\phi.
\end{eqnarray}
where \eqnref{tau'} has been used to compute $d\tau_{1'}$ and $d\tau_{2'}$.
The spin connection $\Gamma$ compatible with the cotriad via $de+\Gamma\wedge e=0$ is then given by the solution:
\begin{equation}\label{Gamma}
\Gamma=\Gamma_a^i\tau_i da
=-(\alpha+\beta)'\tau_3 dx +\frac{E^{x\prime}}{2E^\rho}\,\tau_{2'} d\theta
-\frac{E^{x\prime}}{2E^\rho}\,\tau_{1'}\sin\theta\,d\phi+\tau_3\cos\theta\,d\phi.
\end{equation}
Equations \eqnref{A 2} and \eqnref{Gamma} then give the extrinsic curvature $\gamma K=A-\Gamma$:
\begin{eqnarray}\label{K 2}
&&\gamma K=\gamma K_a^i\tau_i dx^a\nonumber\\
&=& \left[A_x+(\alpha+\beta)'\right]\tau_3 dx\\
&&+\left[A_\rho\cos\alpha\,\tau_{1'} -\left(A_\rho\sin\alpha+\frac{E^{x\prime}}{2E^\rho}\right)\tau_{2'}\right]d\theta
+\left[A_\rho\cos\alpha\,\tau_{2'} +\left(A_\rho\sin\alpha+\frac{E^{x\prime}}{2E^\rho}\right)\tau_{1'}\right]\sin\theta\,d\phi, \nonumber
\end{eqnarray}
which takes the form of \eqnref{K} with
\begin{subequations}\label{K components}
\begin{eqnarray}
\gamma K_x &=& A_x+(\alpha+\beta)',\\
\gamma K_1 &=& A_\rho\cos\beta + \frac{E^{x\prime}}{2E^\rho}\,\sin(\alpha+\beta),\\
\gamma K_2 &=& A_\rho\sin\beta - \frac{E^{x\prime}}{2E^\rho}\,\cos(\alpha+\beta).
\end{eqnarray}
\end{subequations}
If we immediately impose the Gauss constraint \eqnref{U}, we have
\begin{equation}\label{K 3}
\gamma K=\gamma K_a^i\tau_i dx^a
= \left[A_x+(\alpha+\beta)'\right]\tau_3 dx
+A_\rho\cos\alpha\,\tau_{1'} d\theta+A_\rho\cos\alpha\,\tau_{2'}\sin\theta d\phi,
\end{equation}
and correspondingly $\gamma K_1=A_\rho\cos\alpha\,\cos(\alpha+\beta)$ and $\gamma K_2=A_\rho\cos\alpha\,\sin(\alpha+\beta)$.\footnote{Equations \eqnref{K 2} and \eqnref{K 3} give rise to the equivalent Hamiltonian constraint modulo the Gauss constraint. In this paper, \eqnref{K 2} is used to derive \eqnref{C 2}.}

\textit{Remark.} In the above calculation, the reader might be puzzled why the basis $(\tau_{1'},\tau_{2'},\tau_3)$ is treated as $x$-dependent while the original basis $(\tau_1,\tau_2,\tau_3)$ is considered fixed with respect to exterior derivative. Is not the difference between these two merely a gauge transformation? Well, yes and no! What we just did is \emph{not} a gauge transformation. In fact we have stuck with the $(\tau_1,\tau_2,\tau_3)$ basis all the way through; introducing $(\tau_{1'},\tau_{2'},\tau_3)$ is only a procedure for \emph{change of variables} which largely simplifies the calculation (mainly, by diagonalizing $\tilde{E}^a_i$). Alternatively, if one wish, one can indeed regard the change from $(\tau_1,\tau_2,\tau_3)$ to $(\tau_{1'},\tau_{2'},\tau_3)$ as a gauge transformation rather than a change of variables. This way, $(\tau_{1'},\tau_{2'},\tau_3)$ is a fixed basis just like $(\tau_1,\tau_2,\tau_3)$ and we should not take $d\tau_{1'}$ and $d\tau_{2'}$ as we did in \eqnref{de}; consequently, we will not have $(\alpha+\beta)'$ in \eqnref{de} and \eqnref{Gamma}. However, under the gauge transformation \eqnref{new tau}, all the fields have to transform accordingly: i.e., $E\rightarrow E'=e^{-\tau_3(\alpha+\beta)}E\,e^{\tau_3(\alpha+\beta)}$ and $A\rightarrow A'=e^{-\tau_3(\alpha+\beta)}A\,e^{\tau_3(\alpha+\beta)}+e^{-\tau_3(\alpha+\beta)}de^{\tau_3(\alpha+\beta)}$ (see \footref{foot:Gauss transform}). Note that the inhomogeneous term for the above gauge transformation of $A$ yields $(\alpha+\beta)'\tau_3dx$, which is missing in \eqnref{A 2}. Therefore, both methods (change of variables and gauge transformation) yield the same result in the end; particularly, the term $(\alpha+\beta)'\tau_3dx$ for $\gamma K=A-\Gamma$ is from $\Gamma$ in the former approach and from $A$ in the latter. (It should be noted that both the Ashtekar connection $A$ and the spin connection $\Gamma$ do not transform covariantly under the gauge transformation, but their difference $\gamma K$ does!)

From \eqnref{polar A E} and \eqnref{K components}, we can compute the part inside the curly parenthesis of \eqnref{Hamiltonian constraint}:
\begin{eqnarray}
2G\sqrt{q}\,C &:=&  2E^x\left(E^1A'_2-E^2A'_1\right) + 2A_xE^x\left(A_1E^1+A_2E^2\right)
           +(A_1^2+A_2^2-1)[(E^1)^2+(E^2)^2] \nonumber\\
&&  \qquad -(1+\gamma^2)\left(2K_xE^x(K_1E^1+K_2E^2) + [(K_1)^2+(K_2)^2][(E^1)^2+(E^2)^2]\right)\nonumber\\
&=&-\frac{1}{4\gamma^2}\Big\{4\left(\gamma^2+A_\rho^2\right)(E^\rho)^2 +(1+\gamma^2)(E^{x\prime})^2
+8\gamma^2E^x E^\rho A_\rho'\sin\alpha +8E^x E^\rho A_x A_\rho\cos\alpha\nonumber\\
&&\qquad \qquad
+8E^x E^\rho A_\rho\cos\alpha\left((1+\gamma^2)\alpha'+\beta'\right)
+4(1+\gamma^2)E^\rho A_\rho E^{x\prime}\sin\alpha \Big\}.
\end{eqnarray}
Although $\sqrt{q}\,C$ as a whole is $U(1)$ invariant, the above expression still involves $U(1)$-dependent variables $A_x$ and $\beta$. To make the $U(1)$ invariance explicit, it is convenient to add a suitable multiple of the Gauss constraint:
\begin{eqnarray}
2G\gamma E^x\partial_x U &=& 2E^x E^\rho A_\rho'\sin\alpha +2E^x E^\rho A_\rho\cos\alpha\,\alpha' +2E^xA_\rho E^{\rho\prime}\sin\alpha +E^xE^{x\prime\prime},
\end{eqnarray}
so that the new $\sqrt{q}\,C$ reads as
\begin{eqnarray}
2G\sqrt{q}\,C &\rightarrow& 2G\sqrt{q}\,C+2G\gamma E^x\partial_x U\nonumber\\
&=&
-\frac{1}{4\gamma^2}\Big\{4\left(\gamma^2+A_\rho^2\right)(E^\rho)^2 +(1+\gamma^2)(E^{x\prime})^2\nonumber\\
&&\qquad\quad
+8E^x E^\rho A_\rho(A_x+\alpha'+\beta')\cos\alpha\
+4(1+\gamma^2)E^\rho A_\rho E^{x\prime}\sin\alpha \Big\}\nonumber\\
&&
+2E^xA_\rho E^{\rho\prime}\sin\alpha +E^xE^{x\prime\prime},
\end{eqnarray}
in which $A_x$ and $\beta$ appear only through the $U(1)$-invariant quantity $\gamma K_x=A_x+(\alpha+\beta)'$.
Replacing $A_\rho\cos\alpha$ and $A_\rho\sin\alpha$ with \eqnref{Abar eta Peta} and $A_\rho^2$ with \begin{equation}
A_\rho^2 = \frac{\bar{A}_\rho^2}{4}+A_\rho^2\sin^2\alpha =\frac{\bar{A}_\rho^2}{4}+\left(\frac{P^\eta}{2E^\rho}\right)^2,
\end{equation}
we then have
\begin{subequations}
\begin{eqnarray}
2G\sqrt{q}\,C &=&
-(E^\rho)^2 -\frac{\bar{A}_\rho^2(E^\rho)^2}{4\gamma^2} -\frac{E^xE^\rho\bar{A}_\rho}{\gamma^2}\left(A_x+\alpha'+\beta'\right) +\frac{E^xE^{\rho\prime}P^\eta}{E^\rho}\nonumber\\
&& -\frac{(P^\eta)^2}{4\gamma^2} -\frac{1+\gamma^2}{2\gamma^2}\,E^{x\prime}P^\eta -\frac{1+\gamma^2}{4\gamma^2}\,(E^{x\prime})^2 +E^xE^{x\prime\prime},\\
&=&
-(E^\rho)^2 -\frac{\bar{A}_\rho^2(E^\rho)^2}{4\gamma^2} -\frac{E^xE^\rho\bar{A}_\rho}{\gamma^2}\left(A_x+\alpha'+\beta'\right) -\frac{E^xE^{x\prime}E^{\rho\prime}}{E^\rho}\nonumber\\
&& +\frac{(E^{x\prime})^2}{4} +E^xE^{x\prime\prime},
\end{eqnarray}
\end{subequations}
where in the second step we have applied the Gauss constraint $G=0$ by the form of \eqnref{U}. Finally, by \eqnref{Hamiltonian constraint} and noting that $\sqrt{q}=\sqrt{\abs{E^x}[(E^1)^2+(E^2)^2]}=\sqrt{\abs{E^x}}\,E^\rho$, we have
\begin{eqnarray}\label{C}
C
&=& \frac{1}{2G} \left\{
-\frac{E^\rho}{\sqrt{\abs{E^x}}} -\frac{\bar{A}_\rho^2E^\rho}{4\gamma^2\sqrt{\abs{E^x}}} -\sgn(E^x)\frac{\sqrt{\abs{E^x}}\,\bar{A}_\rho}{\gamma^2} \left(A_x+\alpha'+\beta'\right) \right.\nonumber\\
&& \left. \qquad \quad -\sgn(E^x)\frac{\sqrt{\abs{E^x}}\,E^{x\prime}E^{\rho\prime}}{(E^\rho)^2} +\frac{(E^{x\prime})^2}{4\sqrt{\abs{E^x}}\,E^\rho} +\sgn(E^x)\frac{\sqrt{\abs{E^x}}\,E^{x\prime\prime}}{E^\rho} \right\},
\end{eqnarray}
which is now written all in terms of $U(1)$ gauge independent variables: $E^\rho, \bar{A}_\rho$, $E^x$, and $A_x+\alpha'+\beta'$.

\subsection{Further change of variables}\label{sec:further change of variables}
Previously, we have transformed the set of canonical variables $(A_x,A_1,A_2;E^x,E^1,E^2)$ to the new set $(A_x,\bar{A}_\rho,\eta;E^x,E^\rho,P_\eta)$. To better grasp the physical meaning of Ashtekar variables, we can perform change of variables once more to separate $U(1)$-dependent variables from $U(1)$-independent ones. The symplectic form \eqnref{Omega 2b} can be recast as
\begin{eqnarray}
\Omega &=& \frac{1}{2G\gamma}\int_I dx \Big\{\delta E^x\wedge\delta (A_x+\alpha'+\beta')
   +\delta E^\rho\wedge\delta(2A_\rho\cos\alpha)\nonumber\\
  && \qquad\qquad +\delta(2A_\rho E^\rho\sin\alpha+E^{x\prime})\wedge\delta(\alpha+\beta)\Big\}
\end{eqnarray}
by integration by parts. This allows one to canonically transform $(A_x,\bar{A}_\rho,\eta;E^x,E^\rho,P_\eta)$ to the new set of canonical variables $(\bar{A}_x,\bar{A}_\rho,\eta;E^x,E^\rho,\bar{P}_\eta)$ by defining
\begin{subequations}
\begin{eqnarray}
\bar{A}_x &=& A_x+\eta'\equiv A_x + (\alpha+\beta)',\\
\bar{P}^\eta &=& P^\eta+E^{x\prime}.
\end{eqnarray}
\end{subequations}
In terms of the new canonical variables $(\bar{A}_x,\bar{A}_\rho,\eta;E^x,E^\rho,\bar{P}_\eta)$, we have
\begin{subequations}\label{canonical rel}
\begin{eqnarray}
  \{\bar{A}_x(x),E^x(x')\} &=& 2G\gamma\delta(x-x'), \\
  \{\bar{A}_\rho(x),E^\rho(x')\} &=& 2G\gamma\delta(x-x'), \\
  \{\eta(x),\bar{P}^\eta(x')\} &=& 2G\gamma\delta(x-x'),
\end{eqnarray}
\end{subequations}
and all other brackets vanish. Meanwhile, the three constraints given by \eqnref{U}, \eqnref{D} and \eqnref{C} are recast as
\begin{subequations}\label{3 constraints}
\begin{eqnarray}
  \label{U 2}
  U &=& \frac{1}{2G\gamma}\bar{P}^\eta, \\
  \label{D 2}
  D &=& \frac{1}{2G\gamma}\left(\bar{A}'_\rho E^\rho + \eta'\bar{P}^\eta-\bar{A}_xE^{x\prime}\right), \\
  \label{C 2}
  C &=& \frac{1}{2G} \left\{
-\frac{E^\rho}{\sqrt{\abs{E^x}}} -\frac{\bar{A}_\rho^2E^\rho}{4\gamma^2\sqrt{\abs{E^x}}} -\sgn(E^x)\frac{\sqrt{\abs{E^x}}\,\bar{A}_\rho\bar{A}_x}{\gamma^2}\right.\nonumber\\
&& \left. \qquad \quad -\sgn(E^x)\frac{\sqrt{\abs{E^x}}\,E^{x\prime}E^{\rho\prime}}{(E^\rho)^2} +\frac{(E^{x\prime})^2}{4\sqrt{\abs{E^x}}\,E^\rho} +\sgn(E^x)\frac{\sqrt{\abs{E^x}}\,E^{x\prime\prime}}{E^\rho} \right\}.
\end{eqnarray}
\end{subequations}

The Poisson brackets between the constraints are given by
\begin{subequations}\label{constraint algebra}
\begin{eqnarray}
\label{pb a}
\left\{\mathcal{C}_G[\lambda],\mathcal{C}_G[\lambda']\right\}&=&0,\\
\label{pb b}
\left\{\mathcal{C}_G[\lambda],\mathcal{C}[N]\right\}&=&0,\\
\label{pb c}
\left\{\mathcal{C}_G[\lambda],\mathcal{C}_\mathrm{Diff}[N^x]\right\} &=& -\mathcal{C}_G[N^x\lambda']=-\mathcal{C}_G[\mathcal{L}_{N^x}\lambda],\\
\label{pb d}
\left\{\mathcal{C}_\mathrm{Diff}[M^x],\mathcal{C}_\mathrm{Diff}[N^x]\right\}&=& \mathcal{C}_\mathrm{Diff}[M^xN^{x'}-N^xM^{x'}] \equiv \mathcal{C}_\mathrm{Diff}[[M^x,N^x]]\nonumber\\
&=&\mathcal{C}_\mathrm{Diff}[\mathcal{L}_{M^x}N^x]=-\mathcal{C}_\mathrm{Diff}[\mathcal{L}_{N^x}M^x]\\
\label{pb e}
\left\{\mathcal{C}[N],\mathcal{C}_\mathrm{Diff}[N^x]\right\}
&=& -\mathcal{C}[N^xN']=-\mathcal{C}[\mathcal{L}_{N^x}N],\\
\label{pb f}
\left\{\mathcal{C}[N],\mathcal{C}[M]\right\}
&=&\mathcal{C}_\mathrm{Diff}\left[(NM'-MN')\frac{E_x^2}{\abs{q}}\right] -\mathcal{C}_G\left[(NM'-MN')\frac{E_x^2}{\abs{q}}\,\eta'\right].
\end{eqnarray}
\end{subequations}
The right hand sides of \eqnref{constraint algebra} are \emph{weakly zero} (abbreviated as $\approx0$) in the sense that they vanish in the constrained phase space.\footnote{Instead of routine calculation, \eqnref{pb c}, \eqnref{pb d} and \eqnref{pb e} can be obtained immediately by knowing the density characters of $U$, $D$ and $C$ (see \appref{app:dimension and weight}). For example, as $D$ is of density weight 2, according to \footref{foot:diff}, we have
\begin{equation*}
\left\{D(x),\mathcal{C}_\mathrm{Diff}[N^x]\right\} = 2\,\partial_xN^xD+N^x\partial_xD,
\end{equation*}
which follows
\begin{eqnarray*}
\left\{\mathcal{C}_\mathrm{Diff}[M^x],\mathcal{C}_\mathrm{Diff}[N^x]\right\}
&=& \int dx M^x(x)\left[2N^{x'}(x)D(x)+N^x(x)D'(x)\right]
=\int dx\, (M^xN^{x'}-N^xM^{x'})D(x)\\
&\equiv& \mathcal{C}_\mathrm{Diff}[M^xN^{x'}-N^xM^{x'}]
\end{eqnarray*}
by integration by parts. For \eqnref{pb f}, see \appref{app:Poisson bracket} for the detailed calculation.
}
The three constraints are of \emph{first class} in Dirac's terminology, but the constraint algebra is open in the sense that we have structure function instead of structure constants, as the smearing fields in the right hand side of \eqnref{pb f} depend on dynamical fields. (See \cite{Ashtekar:2004eh} for the constraint algebra in the full theory and related comments.)

The Hamiltonian in the full theory is given by $\mathcal{H}=\mathcal{C}[N(\vec{x})] +\mathcal{C}_\mathrm{Diff}[N^a(\vec{x})] +\mathcal{C}_G[\omega^i\cdot t(\vec{x})]$ (see \cite{Ashtekar:2004eh}). With the imposition of spherical symmetry, $\mathcal{H}$ reads as
\begin{eqnarray}
\mathcal{H}&=&\mathcal{C}[N] +\mathcal{C}_\mathrm{Diff}[N^x)] +\mathcal{C}_G[\omega^3\cdot t]\nonumber\\
&\equiv&\int_I dx \left[N(x)C(x)+N^x(x)D(x)+\left(\omega^3(x)\cdot t\right)U(x)\right],
\end{eqnarray}
where $N$, $N^x$ and $\omega^3\cdot t$ are Lagrange multipliers.
The canonical variables evolve with respect to the parameter time $t$ via the Hamilton's equations:
\begin{subequations}\label{Hamilton eqs}
\begin{eqnarray}
\label{Hamilton a}
\dot{\bar{A}}_x &=& \left\{\bar{A}_x,\mathcal{H}\right\} = 2G\gamma\frac{\delta \mathcal{H}}{\delta E^x} \nonumber\\
&=& 2G\gamma\left\{N\frac{\partial C}{\partial E^x} -\left(N\frac{\partial C}{\partial E^{x\prime}}\right)' +\left(N\frac{\partial C}{\partial E^{x\prime\prime}}\right)'' -\left(N^x\frac{\partial D}{\partial E^{x\prime}}\right)' \right\},\\
\label{Hamilton b}
\dot{\bar{A}}_\rho &=& \left\{\bar{A}_\rho,\mathcal{H}\right\} = 2G\gamma\frac{\delta \mathcal{H}}{\delta E^\rho}
= 2G\gamma\left\{N\frac{\partial C}{\partial E^\rho} -\left(N\frac{\partial C}{\partial E^{\rho\prime}}\right)' +N^x\frac{\partial D}{\partial E^\rho}\right\},\\
\label{Hamilton c}
\dot{E}^x &=& \left\{E^x,\mathcal{H}\right\} = -2G\gamma\frac{\delta \mathcal{H}}{\delta \bar{A}_x}
=-2G\gamma\left\{N\frac{\partial C}{\partial \bar{A}_x} + N^x\frac{\partial D}{\partial \bar{A}_x}\right\}\\ 
\label{Hamilton d}
\dot{E}^\rho &=& \left\{E^\rho,\mathcal{H}\right\} = -2G\gamma\frac{\delta \mathcal{H}}{\delta \bar{A}_\rho}
= -2G\gamma\left\{N\frac{\partial C}{\partial \bar{A}_\rho} - \left(N^x\frac{\partial D}{\partial \bar{A}_\rho'}\right)'\right\},\\
\label{Hamilton e}
\dot{\eta} &=& \left\{\eta,\mathcal{H}\right\} = 2G\gamma\frac{\delta \mathcal{H}}{\delta \bar{P}^\eta}
=2G\gamma\left\{N^x\frac{\partial D}{\partial\bar{P}^\eta}+(\omega^3\cdot t)\frac{\partial U}{\partial\bar{P}^\eta}\right\}
=N^x\eta'+(\omega^3\cdot t),\\
\label{Hamilton f}
\dot{\bar{P}}^\eta &=& \left\{\eta,\mathcal{H}\right\}=-2G\gamma\frac{\delta \mathcal{H}}{\delta \eta}
= -2G\gamma\left\{\left(-N^x\frac{\partial D}{\partial \eta'}\right)'\right\}
= \left(N^x\bar{P}^\eta\right)'.
\end{eqnarray}
\end{subequations}
The three constraints $U=0$, $D=0$ and $C=0$ together with the above Hamilton's equations are completely equivalent to Einstein's equations (with the spherical symmetry imposed).

It should be noted that the Gauss constraint $U=0$ simply yields $\bar{P}^\eta=0$, which is the solution to \eqnref{Hamilton f}.
In terms of the canonical variables $(\bar{A}_x,\bar{A}_\rho,\eta;E^x,E^\rho,\bar{P}^\eta)$, the Gauss constraint \eqnref{U 2} involves only the pair $(\eta;\bar{P}^\eta)$, the evolution of which is completely decoupled from that of $(\bar{A}_x,\bar{A}_\rho;E^x,E^\rho)$, as the variables $(\eta;\bar{P}^\eta)$ appear only in \eqnref{Hamilton e} and \eqnref{Hamilton f}, while $(\bar{A}_x,\bar{A}_\rho;E^x,E^\rho)$ appear only in \eqnref{Hamilton a}--\eqnref{Hamilton d}. In \secref{sec:relation}, we will show that the geometry of spacetime is completely specified by $(\bar{A}_x,\bar{A}_\rho;E^x,E^\rho)$, while $(\eta;\bar{P}^\eta)$ correspond solely to the internal degrees of $U(1)$ gauge. The canonical structure in terms of $(\bar{A}_x,\bar{A}_\rho,\eta;E^x,E^\rho,\bar{P}^\eta)$ thus decouples the $U(1)$ degrees of freedom from the geometric/metric ones.

\subsection{Relation between spacetime metric and Ashtekar variables}\label{sec:relation}
In order to read out the spacetime geometry from the fundamental canonical variables, we need a dictionary which translates between the Ashtekar variables and the spacetime metric.

Take the spherically symmetric 3-dimensional Riemannian space $(\Sigma,q)$ and adapting the coordinates $x^a$ to the spherical symmetry: $x^a=(x,\theta,\phi)$, the line element on $\Sigma$ is completely characterized by two functions $\Lambda(t,x)$ and $R(t,x)$:
\begin{equation}\label{dsigma}
d\sigma^2(t)=q_{ab}dx^adx^b=\Lambda^2(t,x)dx^2+R^2(t,x)d\Omega^2,
\end{equation}
where $d\Omega^2=d\theta^2+\sin^2\theta d\phi^2$ and $R$ is the \emph{curvature radius}. By the relation $qq^{ab}=\tilde{E}^a_i\tilde{E}^b_j\delta^{ij}$ and \eqnref{E polar}, we then have
\begin{equation}\label{dict 1}
\abs{E^x}=R^2,
\qquad
E^\rho=R\Lambda.
\end{equation}

On the other hand, to characterize the line element on the 4-dimensional spacetime $(\mathcal{M},g)$, we also need the lapse function $N$ and the shift vector $N^a$:
\begin{equation}\label{ds}
ds^2=g_{\mu\nu}dx^\mu dx^\nu=-N^2dt^2+q_{ab}(dx^a+N^adt)(dx^b+N^bdt).
\end{equation}
Because of the spherical symmetry, only the radial component $N^x$ of the shift vector survives, and both $N(t,x)$ and $N^x(t,x)$ depends only on $t$ and $x$. The extrinsic curvature expressed in terms of the lapse and shift functions is given by
\begin{equation}
K_{ab}=\frac{1}{2N} \left(\frac{\partial N_a}{\partial x^b} +\frac{\partial N_b}{\partial x^a} -\frac{\partial g_{ab}}{\partial t}-2\,\Gamma_{cab}N^c\right)
\equiv\frac{1}{2N}\left(N_{a|b}+N_{b|a}-\dot{g}_{ab}\right).
\end{equation}
For the line element given by \eqnref{dsigma}, $K_{ab}$ reads as (also see \cite{Kuchar:1994zk})
\begin{subequations}
\begin{eqnarray}
K_{xx} &=& -N^{-1}\Lambda\left(\dot{\Lambda}-(\Lambda N^x)'\right),\\
K_{\theta\theta} &=& -N^{-1}R\left(\dot{R}-R'N^x\right),
\qquad
K_{\phi\phi}=\sin^2\theta\, K_{\theta\theta},
\end{eqnarray}
\end{subequations}
and all off-diagonal components vanish.
By \eqnref{cotriad}, \eqnref{K 3} and the relation $K_a^i=-{K_a}^be_b^i=-K_{ab}\tilde{E}^b_j\delta^{ij}$ (where $e^b_j$ is the inverse of $e_b^j$),\footnote{It is usually given as $K_a^i={K_a}^be_b^i$ in the literature without the extra negative sign. The sign is merely a convention, which flips when the orientation of the parameter time $t$ is reversed. In this paper, we add the extra minus sign to be in accord with the Kruskal coordinates.} we then have
\begin{subequations}\label{dict 2}
\begin{eqnarray}
\bar{A}_x &\equiv& A_x+(\alpha+\beta)' = \gamma K_x^3 =-\gamma K_{xx}e^x_3 =-\gamma K_{xx}\frac{\sqrt{\abs{E^x}}}{E^\rho}\nonumber\\
&=& -\gamma K_{xx}\Lambda^{-1} =\gamma N^{-1}\left(\dot{\Lambda}-(\Lambda N^x)'\right),\\
\bar{A}_\rho &\equiv& 2A_\rho\cos\alpha = 2\gamma K_\theta^{1'} =-2\gamma K_{\theta\theta}e^\theta_{1'} =-2\gamma K_{\theta\theta}\frac{1}{\sqrt{\abs{E^x}}}\nonumber\\
&=&-2\gamma K_{\theta\theta}R^{-1} = 2\gamma N^{-1}\left(\dot{R}-R'N^x\right).
\end{eqnarray}
\end{subequations}

Equations \eqnref{dict 1} and \eqnref{dict 2} make up a dictionary between the metric variables $(\Lambda,R;N,N^x)$ and the Ashtekar variables $(\bar{A}_x,\bar{A}_\rho;E^x,E^\rho)$. More precisely, while $(E^x,E^\rho)$ gives the \emph{intrinsic} geometry of the leaf $\Sigma_t$, $(\bar{A}_x,\bar{A}_\rho)$ tells its \emph{extrinsic} geometry (i.e. how $\Sigma_t$ is imbedded in $\mathcal{M}$). On the other hand, the remaining canonical variables $\eta$ and $\bar{P}^\eta$ correspond solely to the internal $U(1)$ degrees of freedom and are completely decoupled from the metric degrees.

\subsection{Remarks on falloff conditions and boundary terms}
In the full theory, for simplicity, it is assumed that the spatial surface $\Sigma$ is a compact 3-manifold without boundary. Modifications are required to incorporate boundary terms when $\Sigma$ is non-compact or has boundary (see \cite{Ashtekar:2004eh} for comments and references). In spherically symmetric cases, the 1-dimensional manifold $I$ is usually taken to be non-compact or with boundary and therefore we have to pay attention to the behavior of the canonical variables at boundary (or at infinity if $I$ is non-compact). For the maximally extended Schwarzschild spacetime, which has two (left and right) infinities, the falloff conditions of the canonical variables and the corresponding boundary terms have been studied in \cite{Kuchar:1994zk}.

Let $(x,y,z)$ be a system of coordinates on $\Sigma$ which is asymptotically Cartesian. Such a system is related to a spherical system of coordinates $(r,\theta,\phi)$ through the standard flat space formulae, i.e. $(x,y,z) =(r\sin\theta\cos\phi,r\sin\theta\sin\phi,r\cos\theta)$. Consistency then demands that, at $r\rightarrow\pm\infty$, the canonical variables have to satisfy the falloff conditions:
\begin{subequations}
\begin{eqnarray}
\label{falloff a}
\Lambda(t,r)&=&1+2GM_\pm(t)\abs{r}^{-1}+\mathcal{O}^\infty(\abs{r}^{-(1+\epsilon)}),\\
\label{falloff b}
R(t,r)&=&\abs{r}+\mathcal{O}^\infty(\abs{r}^{-\epsilon}),\\
P_\Lambda(t,r)&=&\mathcal{O}^\infty(\abs{r}^{-\epsilon}),\\
P_R(t,r)&=&\mathcal{O}^\infty(\abs{r}^{-(1+\epsilon)}),
\end{eqnarray}
\end{subequations}
where $P_\Lambda$ and $P_R$ are conjugate momenta of $\Lambda$ and $R$, respectively, and $f(x^a)=\mathcal{O}^\infty(r^{-n})$ means that $f$ falls off like $r^{-n}$, $f_{,a}$ like $r^{-(n+1)}$, and so on for higher spatial derivatives. Meanwhile, the Lagrange multipliers have to satisfy
\begin{subequations}\label{falloff e and f}
\begin{eqnarray}
\label{falloff e}
N(t,r)&=&N_\pm(t)+\mathcal{O}^\infty(\abs{r}^{-\epsilon}),\\
\label{falloff f}
N^x(t,r)&=&\mathcal{O}^\infty(\abs{r}^{-\epsilon}).
\end{eqnarray}
\end{subequations}
It turns out $M_\pm$ are the ADM energy measured at right and left infinities (i.e. the Schwarzschild mass) as the boundary terms, and $N_\pm$ correspond to the Lagrange multipliers on the boundary.

Our main purpose is to study the Kruskal extension of the Schwarzschild black hole (and the loop quantum corrections on it). In \secref{sec:Kruskal solution}, we will see that the Schwarzschild solution in terms of Kruskal coordinates leads to $N_\pm=0$. In this particular situation, the boundary terms vanish and no furthermore modifications have to be taken into account.

\section{Maximally extended Schwarzschild spacetime}\label{sec:Kruskal solution}
In order to study the interior and exterior of the Schwarzschild black hole at the same time in Ashtekar's formalism, we need a foliation and a system of coordinates for $\Sigma_t$, such that $\Sigma_t$ includes both interior and exterior and all dynamical variables are well-behaved everywhere on $\Sigma_t$ (except the singularity). The standard Schwarzschild solution in terms of Schwarzschild coordinates is unsuitable for this purpose, as some of the metric components are ill-behaved across the horizon. The appropriate coordinates are the Kruskal coordinates, which have the advantage that they cover the entire spacetime manifold of the maximally extended Schwarzschild solution and are well-behaved everywhere except the singularity \cite{Kruskal:1959vx,Szekeres:1960}. The maximally extended Schwarzschild spacetime is also called the Kruskal extension of the Schwarzschild solution.

\subsection{Kruskal extension of the Schwarzschild black hole}
The Schwarzschild metric is given by
\begin{equation}
ds^2=-\left(1-\frac{2GM}{R}\right)dT^2 + \left(1-\frac{2GM}{R}\right)^{-1}dR^2 + R^2d\Omega^2,
\end{equation}
where $T$ is the Killing time and $R$ is the curvature radius.
Kruskal coordinates are defined, from the Schwarzschild coordinates $(T,R,\theta,\phi)$, by transforming $(T,R)$ to the new coordinates $(t,r)$ via:
\begin{subequations}\label{Kruskal transform I}
\begin{eqnarray}
t &=& \left(\frac{R}{2GM}-1\right)^{1/2}e^{R/4GM}\sinh\left(\frac{T}{4GM}\right),\\
x &=& \left(\frac{R}{2GM}-1\right)^{1/2}e^{R/4GM}\cosh\left(\frac{T}{4GM}\right),
\end{eqnarray}
\end{subequations}
for the exterior region $R>2GM$, and
\begin{subequations}\label{Kruskal transform II}
\begin{eqnarray}
t &=& \left(1-\frac{R}{2GM}\right)^{1/2}e^{R/4GM}\cosh\left(\frac{T}{4GM}\right),\\
x &=& \left(1-\frac{R}{2GM}\right)^{1/2}e^{R/4GM}\sinh\left(\frac{T}{4GM}\right),
\end{eqnarray}
\end{subequations}
for the interior region $0<R<2GM$.
In the Kruskal coordinates $(t,x,\theta,\phi)$, the metric of the extended Schwarzschild spacetime is given by
\begin{equation}\label{Kruskal sol}
  ds^2 = \frac{32G^3M^3}{R}\,e^{-R/2GM}\left(-dt^2+dx^2\right) + R^2d\Omega^2,
\end{equation}
where the curvature radius $R=R(t,x)$ is a function of $t$ and $x$ via
\begin{equation}\label{t2-x2}
t^2-x^2=\left(1-\frac{R}{2GM}\right)e^{R/2GM},
\end{equation}
or more explicitly
\begin{equation}\label{Lambert function}
R(t,x)=2GM\left[1+W\!\left(\frac{x^2-t^2}{e}\right)\right]
\end{equation}
with $W(x)$ being the Lambert $W$ function.\footnote{\label{foot:Lambert fun}The Lambert $W$ function, also called the Omega function or product logarithm, is the inverse function of $f(W)=We^W$. The derivative of $W$ is given by
\begin{equation*}
W'(x)=\frac{W(x)}{x\left[1+W(x)\right]}.
\end{equation*}}

\begin{figure}
\begin{picture}(450,230)(0,0)

\put(-65,-30)
{
\scalebox{0.9}{\includegraphics{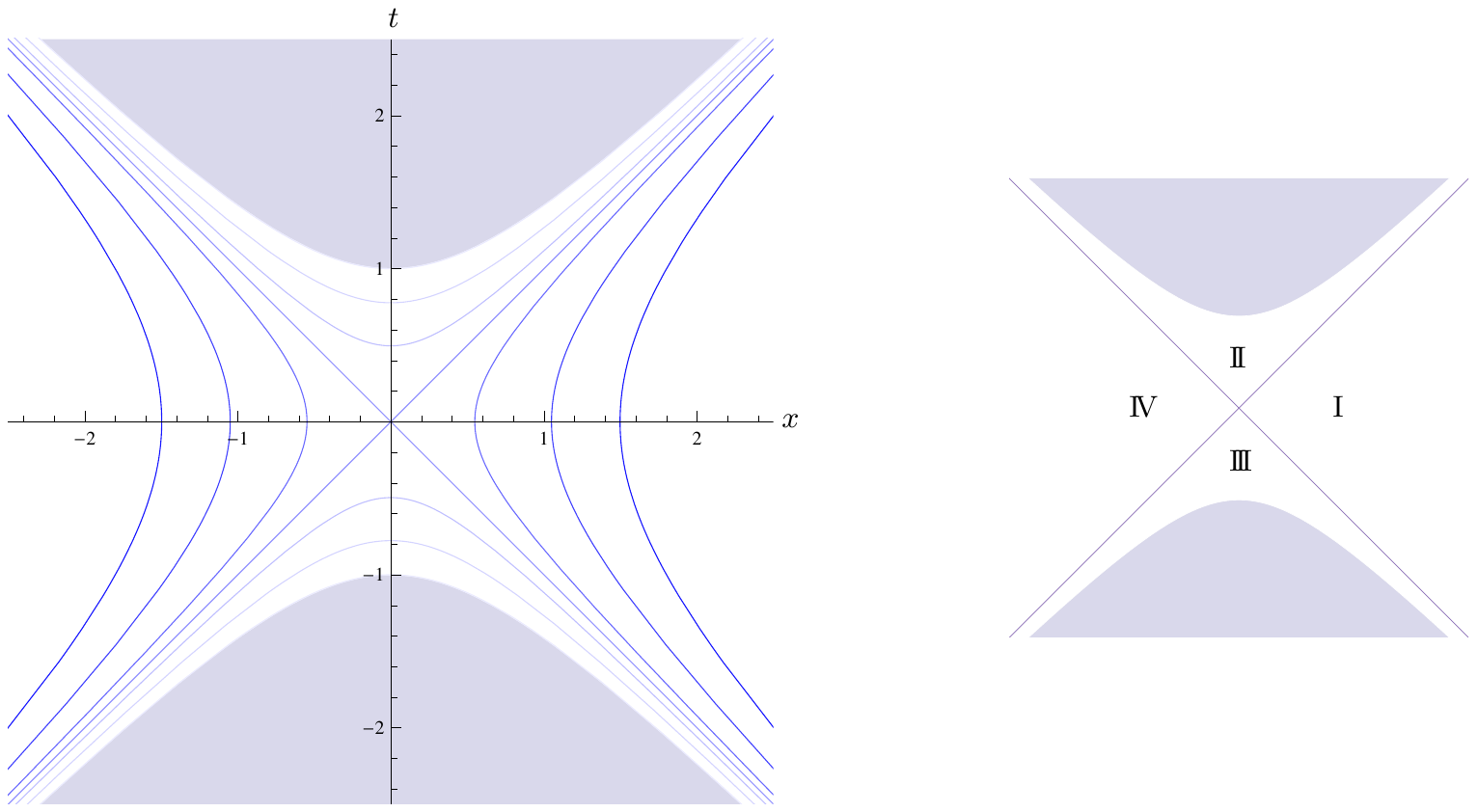}}
}

\end{picture}
\caption{The Kruskal diagram --- the Schwarzschild solution in Kruskal coordinates $x$ and $t$ (with $\theta$ and $\phi$ suppressed). Contours of constant $R$ are hyperbolae as shown (smaller $R$ in lighter color/shade); particularly, $t^2=x^2$ corresponds to the event horizons ($R=2GM$) and $t^2-x^2=1$ corresponds to the black and white hole singularities ($R=0$). Contours of constant $T$ are straight lines passing the origin (not shown except for the event horizons, of which $t=-x$ corresponds to $T=-\infty$ and $t=x$ to $T=+\infty$). Four regions are also indicated. The Kruskal diagram can be conformally deformed into the Penrose diagram as in \figref{fig:Penrose diagrams} (a) and (c).}\label{fig:Kruskal diagram}
\end{figure}

The transformation between Schwarzschild coordinates and Kruskal coordinates is defined for $R>0$, $R\neq2GM$, and $-\infty<T<\infty$, which is the range for which the Schwarzschild coordinates make sense. However, the Kruskal coordinates can be extended beyond the range of the Schwarzschild coordinates and the allowed values are $-\infty<x<\infty$ and $t^2-x^2<1$.
The maximally extended Schwarzschild spacetime can be divided into four regions as depicted in the Kruskal diagram (with $\theta$ and $\phi$ suppressed) in \figref{fig:Kruskal diagram}. The four regions are separated by event horizons at $t^2-x^2=0$ (i.e. $R=2GM$). Region I is the exterior region described by \eqnref{Kruskal transform I}; Region II is the black hole interior described by \eqnref{Kruskal transform II}; Region III is the parallel exterior region described by \eqnref{Kruskal transform I} with $(t,x)$ replaced by $(-t,-x)$; and Region IV is the white hole interior described by \eqnref{Kruskal transform II} with $(t,x)$ replaced by $(-t,-x)$. The extended spacetime has two singularities at $t^2-x^2=1$ (i.e. $R=0$) for the black and white holes respectively.

\subsection{Classical canonical solution}
To have a canonical solution corresponding to the Kruskal extension of the Schwarzschild black hole, we adapt the foliation in accordance with the Kruskal coordinates: that is, let the leaves $\Sigma_t$ be the horizontal lines in the Kruskal diagram and the vertical coordinate $t$ be the parameter time of the foliation.

Comparing \eqnref{Kruskal sol} with \eqnref{ds}, we then have
\begin{subequations}\label{lapse shift}
\begin{eqnarray}
\label{sol N}
N(t,x) &=& \sqrt{\frac{32G^3M^3}{R}}\ e^{-R/4GM},\\
\label{sol Nx}
N^x(t,x) &=& 0;
\end{eqnarray}
\end{subequations}
comparing with \eqnref{dsigma}, we have
\begin{equation}\label{sol Lambda}
\Lambda(t,x)=\sqrt{\frac{32G^3M^3}{R}}\ e^{-R/4GM},
\end{equation}
and $R$ given in \eqnref{Kruskal sol} is identical to the curvature radius $R$ defined in \eqnref{dsigma}.
Equations \eqnref{dict 1} and \eqnref{dict 2} then yield
\begin{subequations}\label{cl sol}
\begin{eqnarray}
E^x(t,x) &=& R^2,\\
E^\rho(t,x) &=& R\Lambda = \sqrt{32G^3M^3}\,R^{1/2}\,e^{-R/4GM},\\
\bar{A}_x(t,x) &=& \gamma N^{-1}\dot{\Lambda} = -\gamma \left(\frac{1}{2R}+\frac{1}{4GM}\right)\dot{R},\\
\bar{A}_\rho(t,x) &=& 2\gamma N^{-1}\dot{R} = \frac{2\gamma}{\sqrt{32G^3M^3}}\,R^{1/2}\dot{R}\,e^{R/4GM}.
\end{eqnarray}
\end{subequations}

It is tedious but routine to show that \eqnref{cl sol} with \eqnref{Lambert function} satisfies $D=0$ and $C=0$ given by \eqnref{D 2} and \eqnref{C 2} and the Hamilton's equations \eqnref{Hamilton eqs} with the lapse and shift functions given by \eqnref{lapse shift}.\footnote{This has been explicitly verified by the authors with the help of \textit{Mathematica}.} Thus, \eqnref{cl sol} is the solution in terms of Ashtekar variables for the Kruskal extension of the Schwarzschild spacetime.

It should be noted that \eqnref{sol Nx} trivially satisfies \eqnref{falloff f}. Meanwhile, \eqnref{sol N} together with \eqnref{falloff b} leads to $N(t,r)=\mathcal{O}^\infty(\abs{r}^{-1/2}e^{-\abs{r}})$, which implies that $N_\pm(t)=0$ in \eqnref{falloff e} and thus the boundary terms vanish. Furthermore, by $\Lambda(t,r)=\Lambda(t,x)\abs{dx/dr}$, we can compute $\Lambda(t,r)=1+2GM_\pm\abs{r}^{-1}+\mathcal{O}^\infty(\abs{r}^{1+\epsilon})$, which satisfies \eqnref{falloff a} with $M_\pm(t)=M$.

\section{Loop quantization of spherically symmetric midisuperspaces}\label{sec:loop quantization}
As the Ashtekar's formalism for the spherically symmetric spacetime has been formulated for classical theory and the classical canonical solution has been obtained for the maximally extended Schwarzschild spacetime in the previous sections, we now turn our attention to loop quantum theory of the spherically symmetric midisuperspace. The framework has been developed in \cite{Bojowald:2004af,Bojowald:2004ag,Bojowald:2005cb}. Kinematics of the quantum theory is well understood, but much remains to be done for dynamics. Our intention here is not to construct a rigorous loop quantum theory but rather to review and bring out conceptual and technical issues. In \secref{sec:loop representation}, we recapitulate the ideas of previous works for kinematics but meanwhile suggest modifying some details. In \secref{sec:Hamiltonian constraint}, we discuss the dynamics and propose a new strategy to construct a graph-preserving Hamiltonian constraint operator. In \secref{sec:improved dynamics}, we propound a new quantization scheme inspired by the ``improved dynamics'' in LQC and show that it is a more sophisticated and sensible way to impose the fundamental discreteness of LQG when the remnant diffeomorphism is concerned.

\subsection{Loop representation}\label{sec:loop representation}
Quantization in the loop representation is based on \emph{cylindrical functions} of connections through holonomies. Let $g$ be a graph on the spatial manifold $I$ composed of a set of edges $\{e\}$ and points $\{p\}$. To begin with, we keep the graph $g$ generic: edges of $g$ may or may not overlap with one another; points are not necessarily the endpoints of edges and may or may not intersect edges. When a point intersects an edge, the intersection point is called a vertex.\footnote{From the point of view of the full theory, the edges $e$ are edges along the radial ($\partial_x$) direction, while the points $p$ are (collapsed) edges along the homogeneous ($\partial_\theta, \partial_\phi$) directions. The vertex in the full theory corresponds to the intersection of a point with edges.} The vector space of cylindrical functions with support on a given graph $g$ is denoted as $\cyl_g$, an element of which, in terms of the connection variables $A_x$, $\bar{A}_\rho$ and $\eta$, is given by
\begin{eqnarray}\label{Psig}
\Psi_g(A)&\equiv&\inner{A}{\Psi_g}\nonumber\\
&=&\sum_{k_e,\mu_p,n_p}\, \prod_{e\in e(g)}\exp\left(\frac{i k_e}{2}\int_e A_x(x)dx\right)
\prod_{p\in p(g)}\exp\left(\frac{i\mu_p}{2}\,\bar{A}_\rho(p)\right)\exp\left(in_p\eta(p)\right),
\end{eqnarray}
which is an \emph{almost periodic function} of $\frac{1}{2}\int_e A_x(x)dx$ and $\frac{1}{2}A_\rho(p)$ as well as a periodic function of $\eta$. In summation, edge labels $k_e\in\mathbb{R}$ and point labels $\mu_v\in\mathbb{R}$ and $n_v\in\mathbb{Z}$ all run over a \emph{finite} number of values. Note that $A_x$ is a scalar density of weight 1 and is integrated along an edge to yield (radial) holonomy, while $\bar{A}_\rho$ and $\eta$ are scalars of weight 0 and the cylindrical function is via their ``point holonomies''.
As far as cylindrical functions are concerned, an arbitrary graph can always be regarded as a set of non-overlapping edges and the union of all endpoints of the edges (thus, all points are now vertices and henceforth we rename $\mu_p$ and $n_p$ as $\mu_v$ and $n_v$).\footnote{The overlapped segment is regarded as a new edge whose $k_e$ is the sum of edge labels of overlapped edges. If an endpoint is not a given point in the first place, it is then considered as a point with $\mu_p=n_p=0$; on the other hand, a dangling point is considered as an endpoint (a conjunction point) of two adjacent  ``zero edges'' with $k_e=0$.} Allowing edge and point labels to be zero, any cylindrical function can be viewed with support on an (irregular and finite) 1-dimensional lattice in $I$.

\textit{Remark.} The vertex labels are integers as $\eta$ takes values in a circle $S^1$ (i.e. $\eta\equiv\eta+2\pi$). The other two labels $k_e$ and $\mu_v$, on the other hand, are real numbers. Contrast to the full theory, in which the edge labels of spin networks are quantized to half integers, the edge labels $k_e$ here are not quantized. This is because the $SU(2)$ internal gauge group is reduced to $U(1)$ in the spherical symmetry-reduced theory and the weight of irreducible representations of $U(1)$ is continuous. In previous works \cite{Bojowald:2004af,Bojowald:2004ag,Bojowald:2005cb}, the holonomy $h^{(e)}:=\exp\frac{1}{2}i\int_eA_x(x)dx$ is regarded as an $SU(2)$-valued $\tau_3$-holonomy and accordingly $k_e$ are quantized to integers. In this paper, we adopt a different approach by directly considering cylindrical functions of the symmetry-reduced connection variables from the outset without adhering to any $SU(2)$ structure inherited from the full theory. Our approach is essentially in the same spirit of that in LQC and its merit will be clear later as it makes possible the graph-preserving Hamiltonian constraint operator in \secref{sec:Hamiltonian constraint}, which in turn facilitates the improved quantization scheme in \secref{sec:improved dynamics}.

The vector space of all cylindrical functions is denoted as $\cyl$, which is the \emph{projective limit} of $\cyl_g$ for all $g$. The Cauchy completion of $\cyl$ is the kinematical Hilbert space $\hilbert$, which is spanned by an orthonormal basis of (symmetry-reduced) spin network states $\ket{g;k_e,\mu_v,n_v}$:
\begin{eqnarray}\label{spin network}
\inner{A}{g;k_e,\mu_v,n_v}
&=&\prod_{e\in e(g)}\exp\left(\frac{i k_e}{2}\int_e A_x(x)dx\right)
\prod_{v\in v(g)}\exp\left(\frac{i\mu_v}{2}\,\bar{A}_\rho(v)\right)\exp\left(in_v\eta(v)\right),
\end{eqnarray}
where $v\in v(g)$ are vertices of $g$ and labels $\{k_e,\mu_v,n_v\}$ are called the \emph{coloring} of the spin network. A spin network defined on $g$ can be regarded as a spin network with support on a larger graph $\bar{g}\supset g$ by assigning trivial labels to the edges and vertices which are not in $g$. For any two graphs $g$ and $g'$, let $\bar{g}=g\cup g'$ and we have
\begin{eqnarray}\label{inner product}
\inner{g;k_e,\mu_v,n_v}{g';k'_e,\mu'_v,n'_v}
&\equiv& \inner{\bar{g};k_e,\mu_v,n_v}{\bar{g};k'_e,\mu'_v,n'_v}\nonumber\\
&=&\prod_{e\in e(\bar{g})} \delta_{k_e,k'_e}\prod_{v\in v(\bar{g})}\delta_{\mu_v,\mu'_v}\delta_{n_v,n'_v}.
\end{eqnarray}
It should be noted that even though $k_e$ and $\mu_v$ take continuous values, $\delta_{k_e,k'_e}$ and $\delta_{\mu_v,\mu'_v}$ are Kronecker deltas rather than Dirac delta functions, as \eqnref{Psig} takes the form of summation instead of integral. The kinematical Hilbert space $\hilbert$ is spanned by $\hilbert_g$, the subspace of $\hilbert$ with support on $g$, and each $\hilbert_g$ is given by $L^2(\mathbb{R}_\mathrm{Bohr}^\abs{e},d^\abs{e}{k_e}_\mathrm{Bohr}) \otimes L^2(\mathbb{R}_\mathrm{Bohr}^\abs{v},d^\abs{v}{\mu_v}_\mathrm{Bohr}) \otimes \mathbb{Z}^\abs{v}$, where $\mathbb{R}_\mathrm{Bohr}$ is the Bohr compactification of $\mathbb{R}$ and $\abs{e}$ and $\abs{v}$ are numbers of edges and vertices of $g$.

The action of the flux operators that quantize the momenta $E^x$, $E^\rho$ and $P^\eta$ are given by (with the Planck length $\Pl:=\sqrt{G\hbar}$\ )
\begin{subequations}\label{flux ops}
\begin{eqnarray}
\hat{E}^x(x)\ket{g;k_e,\mu_v,n_v} &=& \gamma\Pl^2\frac{k_{e^+(x)}+k_{e^-(x)}}{2}\,\ket{g;k_e,\mu_v,n_v},\\
\int_\mathcal{I}dx \hat{E}^\rho(x)\ket{g;k_e,\mu_v,n_v} &=& \gamma\Pl^2 \sum_{v\in v(g)\cap\mathcal{I}}\mu_v\,\ket{g;k_e,\mu_v,n_v},\\
\int_\mathcal{I}dx \hat{P}^\eta(x)\ket{g;k_e,\mu_v,n_v} &=& 2\gamma\Pl^2 \sum_{v\in v(g)\cap\mathcal{I}}n_v\,\ket{g;k_e,\mu_v,n_v},
\end{eqnarray}
\end{subequations}
where $e^\pm(x)$ are two edges or two parts of a single edge meeting at $x$ and $\mathcal{I}\subset I$ is an arbitrary region of $I$.\footnote{If $x$ does not intersect any edges, we have $k_{e^+(x)}=k_{e^-(x)}=0$. If $x$ is the right (resp. left) endpoint of $e$ but does not touch another edge, we have $k_{e^+(x)}=0$ (resp. $k_{e^-(x)}=0$).} Note that $E^\rho$ and $P^\eta$ are of density weight $1$ and thus their corresponding operators are to be smeared over a region of $I$, whereas $E^x$ is of density weight 0 and its corresponding operator is not smeared.

Consequently, after regularization procedure as in \cite{Bojowald:2004ag}, the volume operator corresponding to the volume $V(\mathcal{I})=4\pi\int_\mathcal{I}dx\sqrt{\abs{E^x}}\,E^\rho$ of a region $\mathcal{I}\subset I$ is given by\footnote{Note that both $A_\rho$ (as well as $\bar{A}_\rho$) and $E^\rho$ are defined to be non-negative as in \eqnref{polar A E}, and thus only $\mu_v\geq0$ is allowed. However, it is technically easier to allow all values $\mu_v\in\mathbb{R}$ and in the end require physical states to be symmetric under the large gauge transformation $\mu_v\mapsto-\mu_v$ (see \cite{Ashtekar:2005qt}). Classically, flipping the signs of $\bar{A}_\rho$ and $E^\rho$ simultaneously yields the equivalent solution, as the constraints \eqnref{3 constraints} admits the large gauge $(\bar{A}_\rho,E^\rho)\rightarrow(-\bar{A}_\rho,-E^\rho)$.}
\begin{equation}
\hat{V}(\mathcal{I})=4\pi\int_\mathcal{I}dx|\hat{E}^\rho|\sqrt{|\hat{E}^x|}\,,
\end{equation}
where $\hat{E}^\rho(x)$ is the distribution-valued operator:
\begin{equation}
\hat{E}^\rho(x)\ket{g;k_e,\mu_v,n_v} = \gamma\Pl^2 \sum_{v\in v(g)}\delta(x-v)\,\mu_v\ket{g;k_e,\mu_v,n_v}.
\end{equation}
The volume operator then has $\ket{g;k_e,\mu_v,n_v}$ as its eigenstates:
\begin{equation}
\hat{V}(\mathcal{I})\ket{g;k_e,\mu_v,n_v} = 4\pi\gamma^{3/2}\Pl^3 \sum_{v\in v(g)\cap\mathcal{I}} \abs{\mu_v}\sqrt{\frac{\abs{k_{e^+(v)}+k_{e^-(v)}}}{2}}\ \ket{g;k_e,\mu_v,n_v}.
\end{equation}

The Gauss constraint with \eqnref{U} is classically given by
\begin{equation}
\mathcal{C}_G[\lambda]=\frac{1}{2G\gamma}\int_I dx\, \lambda (E^{x'}+P^\eta)\approx0,
\end{equation}
and quantized to
\begin{equation}
\hat{\mathcal{C}}_G[\lambda]\ket{g;k_e,\mu_v,n_v}=\frac{\Pl^2}{2G} \sum_{v\in v(g)} \lambda(p)\left(k_{e^+(v)}-k_{e^-(v)}+2n_v\right)\ket{g;k_e,\mu_v,n_v}=0,
\end{equation}
which is solved exactly by
\begin{equation}\label{sol of np}
n_v=-\frac{1}{2}\left(k_{e^+(v)}-k_{e^-(v)}\right).
\end{equation}
Substituting \eqnref{sol of np} back to \eqnref{spin network}, we obtain the general form of the $U(1)$ gauge invariant spin networks $\ket{g;k_e,\mu_v}\in\hilbert_\mathrm{inv}^G$:
\begin{subequations}\label{spin network}
\begin{eqnarray}
\inner{A}{g;k_e,\mu_v}
&=&\prod_{e\in e(g)}\exp\left(\frac{i k_e}{2}\int_e (A_x+\eta')dx\right)
\prod_{v\in v(g)}\exp\left(\frac{i\mu_v}{2}\,\bar{A}_\rho(v)\right)\\
&=&\prod_{e\in e(g)}\exp\left(\frac{i k_e}{2}\int_e \bar{A}_x dx\right)
\prod_{v\in v(g)}\exp\left(\frac{i\mu_v}{2}\,\bar{A}_\rho(v)\right),
\end{eqnarray}
\end{subequations}
which are cylindrical functions of the $U(1)$ invariant connections $\bar{A}_x$ and $\bar{A}_\rho$. In accordance with the canonical relations \eqnref{canonical rel}, the flux operators in \eqnref{flux ops} acting on the $U(1)$-invariant spin networks are given as
\begin{subequations}\label{flux ops 2}
\begin{eqnarray}
\label{flux op Ex}
\hat{E}^x(x)\ket{g;k_e,\mu_v} &=& \gamma\Pl^2\frac{k_{e^+(x)}+k_{e^-(x)}}{2}\,\ket{g;k_e,\mu_v},\\
\int_\mathcal{I} dx\hat{E}^\rho(x)\ket{g;k_e,\mu_v} &=& \gamma\Pl^2 \sum_{v\in v(g)\cap\mathcal{I}}\mu_v\,\ket{g;k_e,\mu_v},\\
\int_\mathcal{I}dx \hat{\bar{P}}^\eta(x)\ket{g;k_e,\mu_v} &=& 0,
\end{eqnarray}
\end{subequations}
where the last equation is equivalent to the Gauss constraint. Similarly, the volume operator acting on the $U(1)$-invariant spin networks is given by
\begin{equation}\label{V op 2}
\hat{V}(\mathcal{I})\ket{g;k_e,\mu_v} = 4\pi\gamma^{3/2}\Pl^3 \sum_{v\in v(g)\cap\mathcal{I}} \abs{\mu_v}\sqrt{\frac{\abs{k_{e^+(v)}+k_{e^-(v)}}}{2}}\ \ket{g;k_e,\mu_v}.
\end{equation}

The next is to impose the diffeomorphism constraint $\mathcal{C}_\mathrm{Diff}[N^x]\approx0$. Following the strategy used in the full theory (see Section 6.2 of \cite{Ashtekar:2004eh} for the details), the procedure of \emph{group averaging} can be applied in the similar fashion to obtain diffeomorphism invariant states. Although one begins with a state in $\cyl$, after group averaging, the resulting diffeomorphism invariant state is distributional and belongs to $\dual{\cyl}$, the algebraic dual of $\cyl$. The subspace of all elements of $\dual{\cyl}$ which are invariant under diffeomorphism is denoted as $\dual{\cyl}_\mathrm{Diff}$. In the end, every element $\round{\Psi}\in\dual{\cyl}_\mathrm{Diff}$ can be uniquely decomposed as
\begin{equation}
\round{\Psi}=\sum_{[g]}\round{\Psi_{[g]}},
\end{equation}
where $[g]$ runs through the diffeomorphism classes of graphs. If two graphs $g$ and $g'$ are not in the same diffeomorphism class, $\round{\Psi_{[g]}}$ and $\round{\Psi_{[g']}}$ are orthogonal to each other, i.e. $\oinner{\Psi_{[g]}}{\Psi_{[g']}}=0$. The basis states of $\hilbert_\mathrm{Diff}$, the Cauchy completion of $\dual{\cyl}_\mathrm{Diff}$, are first labeled by the diffeomorphism class of graphs and then distinguished only by colorings of edges and vertices. The colorings labeled for spin networks in $\hilbert$ are not necessarily orthonormal in $\hilbert_\mathrm{Diff}$, due to the nontrivial action of the discrete graph symmetry group.\footnote{The graph symmetry group in 1-dimensional manifolds is much simpler than that in 3-dimensional manifolds, yet non-triviality still remains.} The states of an orthonormal basis of $\hilbert_\mathrm{Diff}$ are called (symmetry-reduced) spin-knot states or $s$-knot states (by adopting the same name used in the full theory), each of which is specified by a graph diffeomorphism class $[g]$ and a discrete coloring of edges and vertices, which is different from the coloring of spin networks.

Disregarding the technicalities due to the graph symmetry group, one can view a spin-knot as a spin network state by dismissing any reference to its localization in $I$. Similar to the interpretation in the full theory (see Figure 6.8 of \cite{Rovelli:2004tv}), a spin-knot can be abstractly interpreted as an ensemble of chunks of volumes, given in \eqnref{V op 2}, arranged in a radial order with adjacent surfaces of areas, given in \eqnref{flux op Ex}.\footnote{More precisely, each of the chunks of volumes takes the shape of a spherical shell, as the adjacent surface takes the shape of a sphere.} A spin-knot state possesses the information about the volumes and the areas that separate these volumes, but any information of localization of the chunks and adjacent surfaces is irrelevant. However, contrary to the full theory, where the eigenvalues of volumes and areas are quantized, the volume and area eigenvalues are only ``partially'' quantized: that is, they take continuous values but the values are Bohr compactified. This is a consequence that we reduce the internal $SU(2)$ gauge to $U(1)$ from the outset.

\subsection{Hamiltonian constraint}\label{sec:Hamiltonian constraint}
The most complicated and difficult part is to quantize the Hamiltonian constraint. In the full theory, a standard procedure is established to quantize \eqnref{full Hamiltonian constraint} by taking care of three complications \cite{Ashtekar:2004eh}: first, curvature components $F_{ab}^i$ are expressed in terms of holonomies which can be directly promoted to operators; second, the operator corresponding to inverse volume element $e^{-1}$ is obtained from a commutator between holonomies and the volume operator \textit{\`{a} la} Thiemann's trick; third, the Lorentzian part of $\mathcal{C}[N]$ (which involves extrinsic curvature $K_a^i$) is expressed as a commutator between the total volume and the Euclidian part of $\mathcal{C}[N]$. This procedure can also be adapted to symmetry-reduced models (see \cite{Bojowald:2008zzb,Ashtekar:2011ni,Bojowald:book} for the case of LQC). For the spherically symmetric models, the resultant Hamiltonian constraint operator has been derived and given in Equation (41) of \cite{Bojowald:2005cb}, which essentially is constructed in terms of the volume operator $\hat{V}(v)$ and the holonomy operators acting on the vertices of spin networks; that is
\begin{equation}
\hat{\mathcal{C}}[N]=\sum_v N(v)\, \hat{C}_v,
\end{equation}
and $\hat{C}_v$ is expressed in terms of $\hat{V}(v)$ and the holonomy operators
\begin{subequations}\label{holonomies}
\begin{eqnarray}
\hat{h}_{x,\pm}(v,\epsilon)&=&\exp\left(\frac{1}{2}\int_{x(v)}^{x(v)\pm\epsilon}\!A_x(x)dx\,\tau_3\right),\\
\hat{h}_\theta(v,\delta)&=&\exp\left(\frac{\delta}{2}\,\bar{A}_\rho(v)\,\tau_{1'}\right),\\
\hat{h}_\phi(v,\delta)&=&\exp\left(\frac{\delta}{2}\,\bar{A}_\rho(v)\,\tau_{2'}\right),
\end{eqnarray}
\end{subequations}
where $\epsilon$ is the coordinate length of the radial edge starting at the vertex $v$ and ending at the coordinate $x(v)\pm\epsilon$ and the sign $\pm$ denotes the orientation of the radial  edge running to the right ($+$) or left ($-$) of the vertex $v$. The parameters $\epsilon$ and $\delta$ are used for regularization.

The $SU(2)$ values arising from the exponentials of $\tau_3,\tau_{1'},\tau_{2'}$ in \eqnref{holonomies} will be traced out in the end and thus yield only combinatorial numbers. Apart from the combinatorial numbers, the operator $\hat{C}_v$ essentially involves only the volume operator $\hat{V}(v)$ and the two holonomy operators for the radial and point holonomies:
\begin{subequations}
\begin{eqnarray}
\label{hx}
\hat{h}_{x,\pm}^\epsilon(v)&=&\exp\left(\frac{i}{2}\int_{x(v)}^{x(v)\pm\epsilon}\bar{A}_x(x)dx\right),\\
\label{hrho}
\hat{h}_\rho^\delta(v)&=&\exp\left(\frac{i\delta}{2}\,\bar{A}_\rho(v)\right),
\end{eqnarray}
\end{subequations}
where we have replace $A_x$ with $\bar{A}_x$ for $\hat{h}_{x,\pm}^\epsilon(v)$ as from now on we will restrict ourselves to the $U(1)$-invariant spin networks.
When acting on a vertex $v$ of a $U(1)$-invariant spin network $\ket{g;k_e,\mu_v}$, the aforementioned operators only interfere with the vertex $v$, its two neighboring vertices and its two connecting edges. We therefore drop all other labels and denote
\begin{equation}
\ket{\mu_-,k_-,\mu,k_+,\mu_+}=\quad
\begin{picture}(160,18)(0,0)
\put(0,3){\line(1,0){160}}
\multiput(40,3)(40,0){3}{\circle*{3}}
\put(15,8){\dots}
\put(56,7){$k_-$}
\put(96,7){$k_+$}
\put(135,8){\dots}
\put(37,-6){$\mu_-$}
\put(77,-6){$\mu$}
\put(117,-6){$\mu_+$}
\put(78,7){$v$}
\end{picture}
\;.
\end{equation}
The actions of the operators are then given explicitly by
\begin{eqnarray}
\hat{V}(v)\ket{\mu_-,k_-,\mu,k_+,\mu_+} &=& 4\pi\gamma^{3/2}\Pl^3\, \abs{\mu}\sqrt{\frac{\abs{k_++k_-}}{2}} \ \ket{\mu_-,k_-,\mu,k_+,\mu_+},\\
\label{action of hrho}
\hat{h}_\rho^\delta(v)\ket{\mu_-,k_-,\mu,k_+,\mu_+} &=& \ket{\mu_-,k_-,\mu+\delta,k_+,\mu_+},
\end{eqnarray}
and
\begin{subequations}\label{action of hx}
\begin{eqnarray}
\hat{h}_{x,+}^\epsilon(v)\ket{\mu_-,k_-,\mu,k_+,\mu_+} &=&
\left\{
  \begin{array}{lr}
  \cdots, & \text{if } \epsilon>\abs{\Delta x_+},\\
  \ket{\mu_-,k_-,\mu,k_+\!+\!1,\mu_+}, & \text{if } \epsilon=\abs{\Delta x_+},\\
  \ket{\mu_-,k_-,\mu,k_+\!+\!1,0,k_+,\mu_+}, & \text{if } \epsilon<\abs{\Delta x_+},
  \end{array}
\right.\\
\hat{h}_{x,-}^\epsilon(v)\ket{\mu_-,k_-,\mu,k_+,\mu_+} &=&
\left\{
  \begin{array}{lr}
  \cdots, & \text{if } \epsilon>\abs{\Delta x_+},\\
  \ket{\mu_-,k_-\!-\!1,\mu,k_+,\mu_+}, & \text{if } \epsilon=\abs{\Delta x_-},\\
  \ket{\mu_-,k_-,0,k_-\!-\!1,\mu,k_+,\mu_+}, & \text{if } \epsilon<\abs{\Delta x_-},
  \end{array}
\right.
\end{eqnarray}
\end{subequations}
where $\Delta x_\pm$ is the coordinate distance from the vertex $v$ labeled by $\mu$ to the neighboring vertex labeled by $\mu_\pm$.
When $\epsilon<\abs{\Delta x_\pm}$, the endpoint of the radial holonomy does not fall on vertices of the original state and thus creates a new vertex in between labeled by $0$. Note that $\hat{V}(v)$ is hermitian, while $\hat{h}_\rho^\delta(v)$ and $\hat{h}_{x,\pm}^\epsilon(v)$ are unitary.

The Hamiltonian constraint operator implemented with \eqnref{action of hx} changes the graph of a given spin network state by creating new vertices and thus makes the dynamics rather complicated. Alternatively, we can construct a graph-preserving Hamiltonian constraint operator, which changes only the coloring of spin network states, by devising a different regularization scheme modified from \eqnref{hx}:
\begin{equation}\label{new hx}
\hat{h}_{x,\pm}^\epsilon(v)=\exp\left(\frac{i\epsilon}{2\,\abs{\Delta x_\pm}}\int_{e^\pm(v)}\bar{A}_x(x)dx\right),
\end{equation}
which, when acting on spin network states, gives
\begin{subequations}\label{new action of hx}
\begin{eqnarray}
\hat{h}_{x,+}^\epsilon(v)\ket{\mu_-,k_-,\mu,k_+,\mu_+} &=& \ket{\mu_-,k_-,\mu,k_+\!+\!\epsilon\,\abs{\Delta x_+}^{-1},\mu_+},\\
\hat{h}_{x,-}^\epsilon(v)\ket{\mu_-,k_-,\mu,k_+,\mu_+} &=& \ket{\mu_-,k_-\!-\!\epsilon\,\abs{\Delta x_-}^{-1},\mu,k_+,\mu_+}.
\end{eqnarray}
\end{subequations}
The regularization variable $\epsilon$ is now introduced to change the edge label of the connecting edge, instead of attaching a new edge starting from the vertex. The action of \eqnref{new action of hx} can be viewed as evenly stretching the new edge in \eqnref{action of hx} to the whole extent of the connecting edge $e^\pm(v)$. The quantization scheme \eqnref{new action of hx} is possible only if we allow edge labels $k_e$ to be real numbers (see the remark in \secref{sec:loop representation}).\footnote{One might argue that the quantization scheme \eqnref{new action of hx} may not be legitimate, as the connecting edge of a given spin network could be quite long and the holonomy along the long edge no longer well approximates the curvature strength at the vertex, thus, leading to breakdown of approaching classical dynamics in the classical regime. This, however, is not a problem. A spin network state which represents a smooth classical geometry is called a \emph{weave} state \cite{Rovelli:2004tv}, in which the physical lengths of edges are supposed to be so small that the granular structure of space is completely negligible as far as attainable technology is concerned. The quantization scheme \eqnref{new action of hx} indeed approximates the classical dynamics very well when acting on the weave state.}

In the full theory, the Hamiltonian constraint operator, albeit defined upon spin networks with reference to coordinates in the first place, turns out to be well-defined and independent of the regularization at the level for diffeomorphism invariant states, i.e. spin-knots, as a consequence of the intimate interplay between diffeomorphism invariance and short-scale discreteness \cite{Ashtekar:2004eh,Rovelli:2004tv}. The remarkable feature of LQG that the dependence of the regulating parameter disappears at the level of spin-knots is no longer the case in symmetry-reduced theories, because the diffeomorphism invariance as well as the $SU(2)$ gauge invariance of the full theory are partially (in the case of spherically symmetric models) or completely (in the case of LQC) broken, and as a result the short-scale geometry is only ``partially'' discrete (in the sense of Bohr compactification). That said, \eqnref{action of hrho} and either of \eqnref{action of hx} and \eqnref{new action of hx} are well-defined only upon spin networks, and cannot be carried over for spin-knots. On the other hand, the Hamiltonian constraint operator is ill-defined in the limit $\epsilon,\delta\rightarrow0$, because of the very nature of loop representation that connection operators do not exist.\footnote{If one \emph{formally} take the limit $\epsilon,\delta\rightarrow0$ on the Hamiltonian constraint operator, it will give rise to the Wheeler-DeWitt quantization, which is essentially different from the loop quantization (see \cite{Ashtekar:2007em} for the case of LQC).} (However, see Section 5.3 of \cite{Bojowald:2005cb} for more comments on and alternative approaches for regularization issues and anomalies.)

In order to faithfully manifest the discreteness of quantum geometry of LQG in symmetry-reduced models, we follow the strategy devised in LQC: keep the regulating parameters $\epsilon$ and $\delta$ \emph{finite} by hand to imprint the fundamental discreteness of LQG in a sophisticated manner such that the Hamiltonian constraint is diffeomorphism-invariant. The next subsection is devoted to issues of this strategy.

\subsection{Improved dynamics}\label{sec:improved dynamics}
In LQC, to impose the fundamental discreteness, if the regulating parameters are simply prescribed to be small constant, it has been demonstrated that this prescription (i.e. $\mu_0$-scheme) leads to wrong semiclassical behavior, due to the underlying fact that the prescription is not independent of the choice of the comoving finite sized cell to which the spatial integration is restricted to make the Hamiltonian finite. To fix this problem, a more sophisticated quantization scheme of ``improved'' dynamics (i.e. $\mubar$-scheme or, more precisely, $\mubar'$-scheme for anisotropic models) was formulated, in which the regulating parameters are taken to be \emph{adaptive} discreteness variables. The quantization scheme of improved dynamics was first developed in \cite{Ashtekar:2006wn} and later generalized and elaborated on for many other models (see \cite{Chiou:2007mg,Ashtekar:2009vc} for Bianchi I models and \cite{Chiou:2008nm,Chiou:2008eg} for Kantowski-Sachs models).

In our case of spherically symmetric midisuperspaces, the same problem of wrong semiclassical behavior is anticipated if we simply take $\epsilon$ and $\delta$ to be constant, because the 1-dimensional diffeomorphism invariance will be broken as both \eqnref{action of hx} and \eqnref{new action of hx} depend on the radial coordinate (either implicitly or explicitly). Therefore, we have to formulate a new quantization scheme which resolves coordinate dependence.

Ashtekar's formalism for spherically symmetric models has a direct correspondence to that for the Kantowski-Sachs spacetime if the homogeneity is formally imposed. It is not surprising that the improved quantization scheme for spherically symmetric models bears close resemblance to that for the Kantowski-Sachs model as formulated in \cite{Chiou:2008eg}. Following the same ideas (of the $\mubar'$-scheme) in Appendix B of \cite{Chiou:2008eg}, areas of rectilinear loops of holonomy have to be shrunk to the \emph{area gap} obtained in the full theory of LQG. Let $\mubar_x$ and $\mubar_\rho$ be the coordinate lengths of the holonomy paths in the radial and homogeneous directions, it is required (by the so-called $\mubar'$-scheme) that
\begin{subequations}
\begin{eqnarray}
\label{mubar idea a}
\left(\mubar_x\sqrt{g_{xx}}\,\right)\left(\mubar_\rho\sqrt{g_{\Omega\Omega}}\,\right)
&=&\mubar_x\mubar_\rho\Lambda R =2\Delta,\\
\label{mubar idea b}
\left(\mubar_\rho\sqrt{g_{\Omega\Omega}}\,\right)^2
&=&\mubar_\rho^2 R^2 =\Delta,
\end{eqnarray}
\end{subequations}
where $\Delta=\xi\gamma\Pl^2$ is the area gap of LQG and $\xi=2\sqrt{3}\,\pi$ for the standard choice (but other choices are also possible). By \eqnref{dict 1}, the discreteness variables are solved as
\begin{equation}\label{mu bar}
\mubar_x=2\sqrt{\frac{\abs{E^x}\Delta}{\left(E^\rho\right)^2}}\equiv2\frac{\sqrt{\Delta}}{\Lambda},
\qquad
\mubar_\rho=\sqrt{\frac{\Delta}{\abs{E^x}}}\equiv\frac{\sqrt{\Delta}}{R}.
\end{equation}
Note that we have shrunk the area in \eqnref{mubar idea a} to $2\Delta$ while the area in \eqnref{mubar idea b} to $\Delta$. The difference of the factor 2 gives rise to the extra factor 2 for $\mubar_x$ in \eqnref{mu bar}. This factor is demanded by the consistency of diffeomorphism and Hamiltonian constraints, as we will see in \secref{sec:higher order}. This extra factor was not introduced in \cite{Chiou:2008eg} for the Kantowski-Sachs models, since the diffeomorphism invariance is broken (and reduced to scaling invariance) by homogeneity and thus the ratio between the areas of \eqnref{mubar idea a} and \eqnref{mubar idea b} is not fixed.\footnote{\label{foot:factor 2}Heuristically, the factor 2 in \eqnref{mubar idea a} can be understood as a tradeoff that \eqnref{mubar idea a} actually represents \emph{two} holonomy loops in $(\partial_x,\partial_\theta)$ and $(\partial_x,\partial_\phi)$ directions, but the two degrees of $\theta$ and $\phi$ are collapsed to one by the spherical symmetry.}

Replacing $\delta$ in \eqnref{action of hrho} and $\epsilon$ in \eqnref{new action of hx} with $\pm\mubar_\rho$ and $\pm\mubar_x$ prescribed in \eqnref{mu bar}, respectively, we can accordingly construct the new holonomies $\hat{h}^{\pm\mubar_\rho}_\rho(v)$ and $\hat{h}^{\pm\mubar_x}_{x,\pm}(v)$, which, when acting on spin network states, are given by
\begin{subequations}\label{improved h}
\begin{eqnarray}
\hat{h}^{\pm\mubar_\rho}_\rho(v)\ket{\mu_-,k_-,\mu,k_+,\mu_+} &=& \ket{\mu_-,k_-,\mu\pm\sqrt{\frac{2\,\xi}{\abs{k_++k_-}}}\,,k_+,\mu_+},\\
\hat{h}_{x,+}^{\pm\mubar_x}(v)\ket{\mu_-,k_-,\mu,k_+,\mu_+} &=& \ket{\mu_-,k_-,\mu,k_+\pm2\sqrt{\abs{\frac{\xi k_+}{\mu\mu_+}}}\,,\mu_+},\\
\hat{h}_{x,-}^{\pm\mubar_x}(v)\ket{\mu_-,k_-,\mu,k_+,\mu_+} &=& \ket{\mu_-,k_-\mp2\sqrt{\abs{\frac{\xi k_-}{\mu\mu_-}}}\,,\mu,k_+,\mu_+},
\end{eqnarray}
\end{subequations}
in accordance with the actions of flux operators given by \eqnref{flux ops 2}.\footnote{The holonomy operators implemented in \eqnref{improved h} are reminiscent of Equation (3.15) in \cite{Ashtekar:2009vc} for LQC of Bianchi I models. The techniques of algebraic simplification in \cite{Ashtekar:2009vc} by introducing new variables could be carried over to handle the complicated operators in a manageable fashion for further rigorous developments of the quantum theory. By the way, the same reasoning following Equation (3.19) of \cite{Ashtekar:2009vc} can be used to argue that the operators in \eqnref{improved h} are well-defined even though the denominators of the radicands could be zero.} Note that the new holonomies $\hat{h}^{\mubar_\rho}_\rho$ and $\hat{h}^{\mubar_x}_{x,\pm}$ remain unitary, but $(\hat{h}^{\mubar_x}_{x,\pm})^{-1}\neq\hat{h}^{-\mubar_x}_{x,\pm}$ while $(\hat{h}^{\mubar_\rho}_\rho)^{-1}=\hat{h}^{-\mubar_\rho}_\rho$. The improved quantization scheme with \eqnref{improved h} is now independent of coordinates. However, a minor complication arises from the improved quantization scheme: the regulating factor $\delta^{-2}$ appearing in the leading factor of Equation (41) in \cite{Bojowald:2005cb} is now replaced by $\mubar_\rho^{-2}$, which is given by $\abs{E^x}/\Delta$. As a result, the flux operator $\hat{E}^x(v)$ acting at the vertex $v$ is also needed to express this factor.
Once this minor issue is taken care of, the resultant Hamiltonian constraint operator is cast in terms of $\hat{h}^{\pm\mubar_\rho}_\rho(v)$, $\hat{h}_{x,\pm}^{\pm\mubar_x}(v)$, $\hat{V}(v)$ and $\hat{E}^x(v)$. Therefore, information of localization of spin-network states is completely irrelevant and the dynamics dictated by the Hamiltonian constraint operator is diffeomorphism invariant.

Although the improved quantization scheme respects the diffeomorphism invariance, it gives rise to the problematic feature: $[\hat{C}_v,\hat{C}_{v'}]\neq0$ if $v$ and $v'$ are adjacent to each other. Consequently, given different smearing functions, the Hamiltonian constraint operators do not commute, i.e. $[\hat{C}[N],\hat{C}[M]]\neq0$, even restricted to the sector of diffeomorphism invariant graphs. The physical implication is that the resulting dynamics depends on foliation of the spacetime. This problem results from the imposition of the fundamental discreteness, which breaks down the intimate matching between the diffeomorphism and Hamiltonian constraints in the full theory. However, inspired by the ideas for in \cite{Chiou:2009hk,Chiou:2009yx} for LQC, one might speculate that a linear superposition of the Hamiltonian constraint operators in generic $j$ representations might conspire to give the new Hamiltonian constraint operator by which $[\hat{C}_v,\hat{C}_{v'}]=0$. That is, if we consider all generic $j$ representations rather than a fixed one (particularly, $j=1/2$), the notion of \emph{permissibility} of the classical regulator could be restored (at least at large scale). As studied in \cite{Chiou:2009yx}, inclusion of higher $j$ representations corresponds to higher order holonomy corrections. While rigorous construction of the quantum theory is rather difficult, without further ado, in \secref{sec:heuristic dynamics}, we will investigate the ramifications of the improved dynamics with higher order holonomy corrections at the level of heuristic effective dynamics.

\section{Heuristic effective dynamics}\label{sec:heuristic dynamics}
In order to see the ramifications of the loop quantum corrections without going into the detailed construction of the quantum theory, one can study the effective dynamics at the heuristic level by viewing the dynamics as classical but governed by the new ``holonomized'' Hamiltonian constraint, which, to capture the loop quantum corrections, is modified from \eqnref{C 2} by replacing the connection variables with the holonomized ones:
\begin{subequations}\label{holonomization}
\begin{eqnarray}
\bar{A}_x &\rightarrow& \frac{\sin(\mubar_x\bar{A}_x)}{\mubar_x} \equiv \frac{e^{i\mubar_xA_x}-e^{-i\mubar_xA_x}}{2i\mubar_x},\\
\bar{A}_\rho &\rightarrow& \frac{\sin(\mubar_\rho\bar{A}_\rho)}{\mubar_\rho} \equiv \frac{e^{i\mubar_\rho A_\rho}-e^{-i\mubar_\rho A_\rho}}{2i\mubar_\rho}.
\end{eqnarray}
\end{subequations}
This prescription incorporates ``holonomy corrections'' but ignores corrections due to Thiemann's trick (particularly, the ``inverse triad corrections''). It was argued in \cite{Ashtekar:2007em} that, for LQC, the corrections due to Thiemann's trick is negligible on physical grounds.
Various studies on LQC have suggested (and verified in some particular models) that the heuristic effective dynamics with holonomized Hamiltonian constraint indeed gives a very good approximation to the quantum evolution for the states which are semiclassical at large scale \cite{Ashtekar:2006wn,Date:2005nn,Singh:2005xg,Bojowald:2006gr}. For the spherically symmetric models, it is also expected that the heuristic effective dynamics can provide us considerable insight about the quantum evolution.

\subsection{Higher order holonomy corrections}\label{sec:higher order}

\begin{figure}
\begin{picture}(450,190)(0,0)

\put(-15,-25)
{
\scalebox{0.75}{\includegraphics{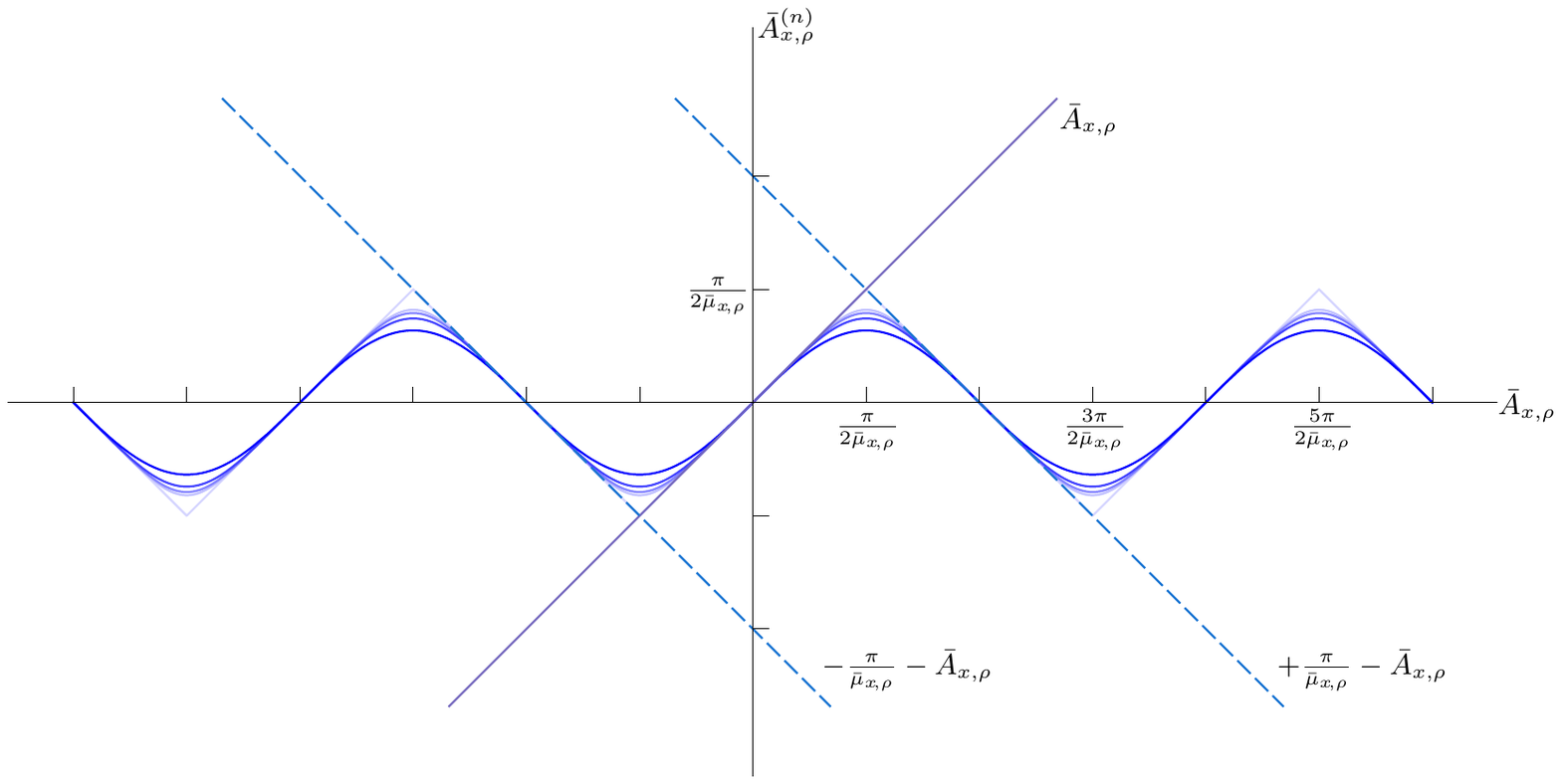}}
}

\end{picture}
\caption{The higher order holonomized connections $\bar{A}_{x,\rho}^{(n)}$ as functions of $\bar{A}_{x,\rho}$ (for given $\mubar_{x,\rho}$). The periodic functions $\bar{A}_{x,\rho}^{(n)}$ (shown with $n=0,1,2,3$) asymptote to the triangle wave $\bar{A}_{x,\rho}^{(\infty)}$. $\bar{A}_{x,\rho}^{(\infty)}$ agrees with $\bar{A}_{x,\rho}$ for $-\pi/2<\bar{\mu}_{x,\rho}\bar{A}_{x,\rho}<\pi/2$.}\label{fig:holonomized connections}
\end{figure}

Following the ideas in \cite{Chiou:2009hk,Chiou:2009yx}, the prescription in \eqnref{holonomization} can be refined in a more elaborate way to include higher order holonomy corrections. This to replace the connection variables with the $n$th order holonomized ones:
\begin{subequations}\label{higher order holonomies}
\begin{eqnarray}
\bar{A}_x &\rightarrow& \bar{A}_x^{(n)}:=\frac{1}{\mubar_x}\sum_{k=0}^n \frac{(2k)!}{2^{2k}(k!)^2(2k+1)} \left(\sin(\mubar_x\bar{A}_x)\right)^{2k+1},\\
\bar{A}_\rho &\rightarrow& \bar{A}_\rho^{(n)}:=\frac{1}{\mubar_\rho}\sum_{k=0}^n \frac{(2k)!}{2^{2k}(k!)^2(2k+1)} \left(\sin(\mubar_\rho\bar{A}_\rho)\right)^{2k+1},
\end{eqnarray}
\end{subequations}
which can be made arbitrarily close to $\bar{A}_x$ an $\bar{A}_\rho$ (as $n\rightarrow\infty$) for $-\pi/2<\mubar_{x,\rho}\bar{A}_{x,\rho}<\pi/2$ but remain functions of the holonomies $\hat{h}_x^{\pm\mubar_x}:=\exp(\pm i\mubar_x\bar{A}_x)$ and $\hat{h}_\rho^{\pm\mubar_\rho}:=\exp(\pm i\mubar_\rho\bar{A}_\rho)$ as well the discreteness variables $\mubar_x$ and $\mubar_\rho$ (see \figref{fig:holonomized connections}).
The higher order corrections correspond to higher powers of $\sin(\mubar_x\bar{A}_x)$ and $\sin(\mubar_\rho\bar{A}_\rho)$, which might be understood as a result of generic $j$ representations for holonomies in the Hamiltonian constraint operator in the quantum theory \cite{Chiou:2009yx}.

It is noteworthy that the density weight of $\bar{A}_x^{(n)}$ is the same as that of $\bar{A}_x$, and $\bar{A}_\rho^{(n)}$ the same as $\bar{A}_\rho$ (see \appref{app:dimension and weight}). Substituting \eqnref{higher order holonomies} into \eqnref{C 2}, we obtain the Hamiltonian constraint with the $n$th order holonomy corrections:
\begin{eqnarray}\label{C nth}
  C^{(n)} &:=& \frac{1}{2G} \left\{
-\frac{E^\rho}{\sqrt{\abs{E^x}}} -\frac{(\bar{A}_\rho^{(n)})^2E^\rho}{4\gamma^2\sqrt{\abs{E^x}}} -\sgn(E^x)\frac{\sqrt{\abs{E^x}}\,\bar{A}_\rho^{(n)}\bar{A}_x^{(n)}}{\gamma^2}\right.\nonumber\\
&& \left. \qquad \quad -\sgn(E^x)\frac{\sqrt{\abs{E^x}}\,E^{x\prime}E^{\rho\prime}}{(E^\rho)^2} +\frac{(E^{x\prime})^2}{4\sqrt{\abs{E^x}}\,E^\rho} +\sgn(E^x)\frac{\sqrt{\abs{E^x}}\,E^{x\prime\prime}}{E^\rho} \right\},
\end{eqnarray}
which has the same density weight as the Hamiltonian constraint $C$. Consequently, the constraint algebra given by \eqnref{pb e} remains unchanged if $\mathcal{C}[N]$ is replaced by $\mathcal{C}^{(n)}[N]:=\int_I dx N(x)C^{(n)}(x)$:
\begin{equation}\label{pb e infty}
\left\{\mathcal{C}^{(n)}[N],\mathcal{C}_\mathrm{Diff}[N^x]\right\}
= -\mathcal{C}^{(n)}[N^xN']=-\mathcal{C}^{(n)}[\mathcal{L}_{N^x}N],
\end{equation}
indicating that the diffeomorphism invariance is still respected regardless of the order $n$ of holonomy corrections.

Unfortunately, as the imposition of fundamental discreteness breaks down the intimate matching between diffeomorphism and Hamiltonian constraints, the constraint algebra given by \eqnref{pb f} does not hold any more if $\mathcal{C}[N]$ is replaced by $\mathcal{C}^{(n)}[N]$. This can be seen as the Poisson brackets between the triad variables and the holonomized connections are no longer the same as before in \eqnref{canonical rel} but instead given by:
\begin{subequations}\label{holonomized pb}
\begin{eqnarray}
  \left\{\Bar{A}_x^{(n)}(x),E^x(y)\right\} &=& 2G\gamma\delta(x-y)\cos(\mubar_x\bar{A}_x) \sum_{k=0}^n\frac{(2k)!}{2^{2k}(k!)^2}\left(\sin(\mubar_x\bar{A}_x)\right)^{2k},\\
  \left\{\Bar{A}_\rho^{(n)}(x),E^\rho(y)\right\} &=& 2G\gamma\delta(x-y)\cos(\mubar_\rho\bar{A}_\rho) \sum_{k=0}^n\frac{(2k)!}{2^{2k}(k!)^2}\left(\sin(\mubar_\rho\bar{A}_\rho)\right)^{2k},\\
  \left\{\Bar{A}_x^{(m)}(x),\Bar{A}_\rho^{(n)}(y)\right\} &=& 2G\gamma\delta(x-y)
  \sum_{k=0}^m\sum_{l=0}^n \frac{(2k)!}{2^{2k}(k!)^2} \frac{(2l)!}{2^{2l}(l!)^2} \left(\sin(\mubar_x\bar{A}_x)\right)^{2k}\left(\sin(\mubar_\rho\bar{A}_\rho)\right)^{2l}\nonumber\\
  && \quad \times \left[\frac{\cos(\mubar_x\bar{A}_x)}{\mubar_\rho}\left(\bar{A}_\rho\cos(\mubar_\rho\bar{A}_\rho) -\frac{\sin(\mubar_\rho\bar{A}_\rho)}{(2l+1)\mubar_\rho}\right) \frac{\partial\mubar_\rho}{\partial E^x}
  \right.\nonumber\\
  && \quad \quad \left.-\frac{\cos(\mubar_\rho\bar{A}_\rho)}{\mubar_x}\left(\bar{A}_x\cos(\mubar_x\bar{A}_x) -\frac{\sin(\mubar_x\bar{A}_x)}{(2k+1)\mubar_x}\right) \frac{\partial\mubar_x}{\partial E^\rho}\right],\\
  \left\{\Bar{A}_x^{(n)}(x),E^\rho(y)\right\} &=& \left\{\Bar{A}_\rho^{(n)}(x),E^x(y)\right\} =0,\\
  \left\{\Bar{A}_x^{(m)}(x),\Bar{A}_x^{(n)}(y)\right\} &=& \left\{\Bar{A}_\rho^{(m)}(x),\Bar{A}_\rho^{(n)}(y)\right\} = \left\{E^x(x),E^\rho(y)\right\} =0,
\end{eqnarray}
\end{subequations}
where we keep the discreteness variables $\mubar_x$ and $\mubar_\rho$ as generic functions of $E^x$ and $E^\rho$.

In the limit $n\rightarrow\infty$, the holonomized connections \eqnref{higher order holonomies} take the form of
\begin{subequations}\label{triangle wave}
\begin{eqnarray}
\bar{A}_x^{(\infty)}&=&\frac{\pi}{2\mubar_x}f_\wedge(\mubar_x\bar{A}_x),\\
\bar{A}_\rho^{(\infty)}&=&\frac{\pi}{2\mubar_\rho}f_\wedge(\mubar_\rho\bar{A}_\rho),
\end{eqnarray}
\end{subequations}
where $f_\wedge(x)$ is a triangle wave function of $x$ with period $2\pi$ and $f_\wedge(0)=0$, $f_\wedge(\pm\pi/2)=\pm1$ (see \figref{fig:holonomized connections}). Therefore, $\bar{A}_x^{(\infty)}$ and $\bar{A}_\rho^{(\infty)}$ agree exactly with $\bar{A}_x$ and $\bar{A}_\rho$ respectively, before the evolution approaches the critical conditions $\mubar_x\bar{A}_x=\pm\pi/2$ and $\mubar_\rho\bar{A}_\rho=\pm\pi/2$, which signal occurrence of the quantum bounce.
Furthermore, by the identity
\begin{equation}\label{identity}
\sum_{k=0}^\infty\frac{(2k)!}{2^{2k}(k!)^2}\left(\sin x\right)^{2k} = \abs{\cos x}^{-1},
\end{equation}
the Poisson brackets in \eqnref{holonomized pb} become
\begin{subequations}\label{holonomized pb infty}
\begin{eqnarray}
  \label{pb infty a}
  \left\{\Bar{A}_x^{(\infty)}(x),E^x(y)\right\} &=& 2G\gamma\delta(x-y)f_\sqcap(\mubar_x\bar{A}_x),\\
  \label{pb infty b}
  \left\{\Bar{A}_\rho^{(\infty)}(x),E^\rho(y)\right\} &=& 2G\gamma\delta(x-y)f_\sqcap(\mubar_\rho\bar{A}_\rho),\\
  \label{pb infty c}
  \left\{\Bar{A}_x^{(\infty)}(x),\Bar{A}_\rho^{(\infty)}(y)\right\} &=& 2G\gamma\delta(x-y)
  \left[ f_\sqcap(\mubar_x\bar{A}_x)\left(f_\sqcap(\mubar_\rho\bar{A}_\rho)\bar{A}_\rho -\bar{A}_\rho^{(\infty)}\right)\frac{1}{\mubar_\rho}\frac{\partial\mubar_\rho}{\partial E^x}
  \right.\nonumber\\
  &&\qquad\qquad\qquad -
  \left. f_\sqcap(\mubar_\rho\bar{A}_\rho)\left(f_\sqcap(\mubar_x\bar{A}_x)\bar{A}_x -\bar{A}_x^{(\infty)}\right)\frac{1}{\mubar_x}\frac{\partial\mubar_x}{\partial E^\rho}
  \right],
\end{eqnarray}
\end{subequations}
and all other brackets vanish, where $f_\sqcap(x):=\sgn(\cos x)=\pm1$ is a square wave function with period $2\pi$.

Before and after the bounce, by \eqnref{triangle wave}, we have
\begin{subequations}\label{triangle wave 2}
\begin{eqnarray}
\bar{A}_x^{(\infty)}&=&
\left\{
  \begin{array}{lr}
  \bar{A}_x, & \text{before bounce},\\
  \pm\frac{\pi}{\mubar_x}-\bar{A}_x, & \text{after bounce},
  \end{array}
\right.\\
\bar{A}_\rho^{(\infty)}&=&
\left\{
  \begin{array}{lr}
  \bar{A}_\rho, & \text{before bounce},\\
  \pm\frac{\pi}{\mubar_\rho}-\bar{A}_\rho, & \text{after bounce},
  \end{array}
\right.
\end{eqnarray}
\end{subequations}
and
\begin{equation}\label{square wave}
f_\sqcap(\mubar_x\bar{A}_x)=f_\sqcap(\mubar_\rho\bar{A}_\rho)=
\left\{
  \begin{array}{lr}
  +1, & \text{before bounce},\\
  -1, & \text{after bounce},
  \end{array}
\right.
\end{equation}
as $f_\sqcap(\mubar_x\bar{A}_x)$ and $f_\sqcap(\mubar_\rho\bar{A}_\rho)$ flip signs when the evolution jumps over the bounce.\footnote{\label{foot:same sign}As the evolution approaches the putative singularity, $\bar{A}_x$ and $\bar{A}_\rho$ increase toward infinity until each of $\mubar_x\bar{A}_x$ and $\mubar_\rho\bar{A}_\rho$ reaches the critical value $\pm\pi/2$. Once the critical value is reached, $\big|\bar{A}_x^{(n)}\big|$ (resp. $\big|\bar{A}_\rho^{(n)}\big|$) starts to decrease and $f_\sqcap(\mubar_x\bar{A}_x)$ (resp. $f_\sqcap(\mubar_\rho\bar{A}_\rho)$) flips its sign, thus causing the bounce. It is possible, however, that when $\mubar_x\bar{A}_x$ approaches $\pm\pi/2$, $\mubar_\rho\bar{A}_\rho$ approaches $\mp\pi/2$ with an opposite sign. We will discuss this case in \secref{sec:further modification} but assume $\mubar_x\bar{A}_x$ and $\mubar_\rho\bar{A}_\rho$ are of the same sign in this subsection.}
Consequently, before the bounce, the Poisson brackets in \eqnref{holonomized pb infty} are the same as the classical ones in \eqnref{canonical rel}. On the other hand, after the bounce, \eqnref{holonomized pb infty} reads as
\begin{subequations}\label{holonomized pb infty after bounce}
\begin{eqnarray}
  \label{pb infty a after bounce}
  \left\{\Bar{A}_x^{(\infty)}(x),E^x(y)\right\} &=& -2G\gamma\delta(x-y),\\
  \label{pb infty b after bounce}
  \left\{\Bar{A}_\rho^{(\infty)}(x),E^\rho(y)\right\} &=& -2G\gamma\delta(x-y),\\
  \label{pb infty c after bounce}
  \left\{\Bar{A}_x^{(\infty)}(x),\Bar{A}_\rho^{(\infty)}(y)\right\} &=& \pm2\pi G\gamma\delta(x-y) \left(\frac{1}{\mubar_\rho^2}\frac{\partial\mubar_\rho}{\partial E_x} - \frac{1}{\mubar_x^2}\frac{\partial\mubar_x}{\partial E_\rho}\right) =0,
\end{eqnarray}
\end{subequations}
where we have used the particular prescription \eqnref{mu bar} to yield zero on the right hand side of \eqnref{pb infty c after bounce} and note that the extra factor $2$ for $\mubar_x$ in \eqnref{mu bar} is essential.
Therefore, after the bounce, the Poisson brackets in \eqnref{holonomized pb infty after bounce} are the same as the classical ones in \eqnref{canonical rel} if $\bar{A}_x$ and $\bar{A}_\rho$ are replaced by $-\bar{A}_x^{(\infty)}$ and $-\bar{A}_\rho^{(\infty)}$.
The extra minus sign indicates that the evolution after the bounce behaves as if time was reversed.
It is remarkable that, before and after the bounce, the diffeomorphism constraint \eqnref{D 2} can be cast in terms of $\bar{A}_x^{(\infty)}$ and $\bar{A}_x^{(\infty)}$ as
\begin{subequations}\label{D across bounce}
\begin{eqnarray}
D &=& \frac{1}{2G\gamma}\left(\bar{A}'_\rho E^\rho + \eta'\bar{P}^\eta-\bar{A}_xE^{x\prime}\right)\nonumber\\
\label{D across bounce a}
&=&
\left\{
  \begin{array}{lr}
  \frac{1}{2G\gamma}\left(\bar{A}_\rho^{(\infty)'} E^\rho + \eta'\bar{P}^\eta-\bar{A}^{(\infty)}_xE^{x\prime}\right), & \text{before bounce},\\
  \frac{1}{2G\gamma}\left[\left(\pm\frac{\pi}{\mubar_\rho}-\bar{A}_\rho^{(\infty)}\right)' E^\rho + \eta'\bar{P}^\eta-\left(\pm\frac{\pi}{\mubar_x}-\bar{A}_x^{(\infty)}\right)E^{x\prime}\right], & \text{after bounce},
  \end{array}
\right.\\
\label{D across bounce b}
&=&
\left\{
  \begin{array}{lr}
  \frac{1}{2G\gamma}\left(\bar{A}_\rho^{(\infty)'} E^\rho + \eta'\bar{P}^\eta-\bar{A}^{(\infty)}_xE^{x\prime}\right), & \text{before bounce},\\
  \frac{1}{2G\gamma}\left(\left(-\bar{A}_\rho^{(\infty)}\right)'E^\rho + \eta'\bar{P}^\eta-\left(-\bar{A}_x^{(\infty)}\right)E^{x'}\right), & \text{after bounce},
  \end{array}
\right.
\end{eqnarray}
\end{subequations}
where the prescription \eqnref{mu bar} is used and again the extra factor 2 is essential.
Once more, after the bounce, $-\bar{A}_x^{(\infty)}$ and $-\bar{A}_\rho^{(\infty)}$ play the roles of $\bar{A}_x$ and $\bar{A}_\rho$ as diffeomorphism is concerned.

Equation \eqnref{D across bounce} is used to derive \eqnref{CN CM infty 3}, which leads to
\begin{equation}\label{pb f infty}
\left\{\mathcal{C}^{(\infty)}[N],\mathcal{C}^{(\infty)}[M]\right\}
\doteq\mathcal{C}_\mathrm{Diff}\left[(NM'-MN')\frac{E_x^2}{\abs{q}}\right] -\mathcal{C}_G\left[(NM'-MN')\frac{E_x^2}{\abs{q}}\,\eta'\right],
\end{equation}
where the symbol $\doteq$ is used to indicate that the identity breaks down briefly during the transition period of the bounce, as $f_\sqcap(\mubar_x\bar{A}_x)$ and $f_\sqcap(\mubar_\rho\bar{A}_\rho)$ flip signs at close but slightly different epochs during the transition. The breakdown leads to the problem that the two constraints $C^{(\infty)}=0$ and $D=0$ are violated after the bounce, implying that the spacetime is no longer invariant under spatial diffeomorphism and change of spacetime foliation after the bounce.
However, if the characteristic parameter used to describe the solution in the classical regime is much larger than the Planck unit (for the Schwarzschild spacetime, it is the case that the Schwarzschild mass $M$ is much bigger than the Planck mass $\Plm:=\sqrt{\hbar/G}$\,), the transition period is so brief compared to the characteristic time of the whole spacetime (i.e. $GM$ for the Schwarzschild spacetime) that violation of $C^{(\infty)}=0$ and $D=0$ is negligible at large scale. Therefore, if we go to the limit $n\rightarrow\infty$, the intimate matching between diffeomorphism and Hamiltonian constraints is restored at large scale.

Finally, it is trivial to show
\begin{equation}\label{pb b infty}
\left\{\mathcal{C}_G[\lambda],\mathcal{C}^{(n)}[N]\right\}=0.
\end{equation}
To summarize, with \eqnref{pb e infty}, \eqnref{pb f infty} and \eqnref{pb b infty}, the constraints $\mathcal{C}_G[\lambda]$, $\mathcal{C}_\mathrm{Diff}[N^x]$ and $\mathcal{C}^{(\infty)}[N]$ yield the same constraint algebra as in the classical case \eqnref{constraint algebra} except for the brief breakdown in the transition of the bounce. The brief breakdown in the transition period causes the two constraints $C^{(\infty)}=0$ and $D=0$ to be slightly violated after the bounce, whereas the Gauss constraints $G=0$ remains exact at any epoch and everywhere. This is a consequence of the fact that imposition of fundamental discreteness spoils the intimate matching between diffeomorphism and Hamiltonian constraints. However, the violation of $C^{(\infty)}=0$ and $D=0$ can be restored at large scale if we go to the limit $n\rightarrow\infty$ and therefore the heuristic effective dynamics dictated by the modified Hamiltonian constraint given by \eqnref{C nth} with $n=\infty$ still provides a reliable description for the large-scale physics.
Details of the resultant dynamics, however, depend on the foliation of spacetime, as by \eqnref{dict 2} the quantities $\mubar_x\bar{A}_x$ and $\mubar_\rho\bar{A}_\rho$, which indicate how strong the quantum correction is, read as
\begin{subequations}\label{meaning of mubar A}
\begin{eqnarray}
\mubar_x\bar{A}_x
&=& -2\gamma\sqrt{\Delta}\, K_{xx} =2\gamma\sqrt{\Delta}\ \frac{\dot{\Lambda}-(\Lambda N^x)'}{N\Lambda},\\
\mubar_\rho\bar{A}_\rho
&=&-2\gamma\sqrt{\Delta}\, K_{\theta\theta} = 2\gamma\sqrt{\Delta}\ \frac{\dot{R}-R'N^x}{NR},
\end{eqnarray}
\end{subequations}
which still involve $N$ and $N^x$.\footnote{\label{foot:Hubble rates}Particularly, if we choose $N^x=0$, we have
\begin{equation*}
\mubar_x\bar{A}_x = 2\gamma\sqrt{\Delta}\ \frac{1}{\Lambda}\frac{d\Lambda}{d\tau},
\qquad
\mubar_\rho\bar{A}_\rho = 2\gamma\sqrt{\Delta}\ \frac{1}{R}\frac{d R}{d\tau},
\end{equation*}
where $d\tau=Ndt$ and $\tau$ is the proper time. For the models which admit a foliation in which the spatial slices are homogeneous (such as \cite{Chiou:2008nm,Chiou:2008eg}), both $\mubar_x\bar{A}_x$ and $\mubar_\rho\bar{A}_\rho$ acquire physical meanings: the former is the Hubble rate in $(\partial_\theta,\partial_\phi)$ direction (times $2\gamma\sqrt{\Delta}$\,) and the latter is the Hubble rate in $x$ direction. This also explains why the extra factor 2 is needed (\textit{cf}. \footref{foot:factor 2}). For generic cases without homogeneity, the dependence on the foliation can be interpreted as reflection of the semiclassical traits (spreading, squeezing, etc.) of the weave states at the level of effective dynamics; i.e. different choices of foliation correspond to different weave states with different semiclassical properties. In \secref{sec:preliminary analysis}, we choose the foliation in accordance with the Kruskal coordinates to investigate the quantum Schwarzschild spacetime, since the Kruskal coordinates give a well-behaved foliation as discussed in \secref{sec:Kruskal solution}. Other well-behaved foliations are also possible and they could yield qualitatively distinct effective solutions (see comments in the end of \secref{sec:preliminary analysis}).}

Until now there is no systematic procedure which leads us to the dynamics of the symmetry-reduced theory from that of the full theory of LQG, but symmetry reduction in general can be understood on the suppositions that in the full theory there is a regime where the nonsymmetric degrees of freedom do not affect too much the dynamics of the symmetric degrees and that the states of concern happen to be within such regime. In other words, the symmetric degrees of freedom can be treated as ``heavy'' degrees of freedom while the nonsymmetric degrees treated as ``light'' ones, in the sense of the Born-Oppenheimer approximation.
The fact that in spherically symmetric theories the constraints violate briefly during the transition period suggests that the degrees of freedom of the full theory cannot be separated into spherical (heavy) and non-spherical (light) ones in the vicinity of the quantum bounce.
That is, in the full theory of LQG, a given \emph{weave} state (coherent state of spin networks) which, after coarse-graining, represents a smooth, spherically symmetric space at large scale will eventually manifest granularity of spin networks when approaching the quantum bounce. After the bounce, the weave state evolves to become smooth and spherically symmetric again if the transition period is very brief. Note that in the heuristic effective dynamics it is possible that $f_\sqcap(\mubar_x\bar{A}_x)$ and $f_\sqcap(\mubar_\rho\bar{A}_\rho)$ do not flip signs at close epochs (or even one of them never flips signs); this situation suggests a different scenario in the full theory: approaching the putative singularity, the smooth weave state descends into granular spin networks instead of being bounced back to a smooth weave.

\subsection{Further phenomenological modification}\label{sec:further modification}
In \secref{sec:higher order}, we assume that $\mubar_x\bar{A}_x$ and $\mubar_\rho\bar{A}_\rho$ approach the critical value $\pm\pi/2$ with the same sign as commented in \footref{foot:same sign}. However, it is possible that when $\mubar_x\bar{A}_x$ approaches $\pm\pi/2$, $\mubar_\rho\bar{A}_\rho$ approaches $\mp\pi/2$ with an opposite sign. The extended Schwarzschild spacetime given by \eqnref{cl sol} gives an example. When this is the case, \eqnref{pb infty c after bounce} is modified to
\begin{equation}\label{pb infty c after bounce modified}
  \left\{\Bar{A}_x^{(\infty)}(x),\Bar{A}_\rho^{(\infty)}(y)\right\} = \mp2\pi G\gamma\delta(x-y) \left(\frac{1}{\mubar_\rho^2}\frac{\partial\mubar_\rho}{\partial E_x} + \frac{1}{\mubar_x^2}\frac{\partial\mubar_x}{\partial E_\rho}\right)
\end{equation}
and \eqnref{D across bounce a} modified to
\begin{equation}\label{D across bounce modified}
D=
\left\{
  \begin{array}{lr}
  \frac{1}{2G\gamma}\left(\bar{A}_\rho^{(\infty)'} E^\rho + \eta'\bar{P}^\eta-\bar{A}^{(\infty)}_xE^{x\prime}\right), & \text{before bounce},\\
  \frac{1}{2G\gamma}\left[\left(\mp\frac{\pi}{\mubar_\rho}-\bar{A}_\rho^{(\infty)}\right)' E^\rho + \eta'\bar{P}^\eta-\left(\pm\frac{\pi}{\mubar_x}-\bar{A}_x^{(\infty)}\right)E^{x\prime}\right], & \text{after bounce}.
  \end{array}
\right.
\end{equation}
We do not have $\left\{\Bar{A}_x^{(\infty)}(x),\Bar{A}_\rho^{(\infty)}(y)\right\}=0$ and \eqnref{D across bounce b} any more, and consequently \eqnref{pb f infty} is no longer valid.

There is a way to fix this problem by modifying \eqnref{mu bar} with inclusion of an extra factor of powers of $E^x$. For the first case that $\mubar_x\bar{A}_x$ and $\mubar_\rho\bar{A}_\rho$ approach the critical value $\pm\pi/2$ with the same sign, \eqnref{mu bar} can be generalized to a more generic prescription:
\begin{subequations}\label{mu bar modified case 1}
\begin{eqnarray}
\mubar_x &=&
\left\{
\begin{array}{lr}
\frac{2}{1-2l}\frac{\sqrt{\abs{E^x}\Delta}}{E^\rho}\,, & \text{for } \abs{E^x}\geq \mathcal{L}^2,\\
\frac{2}{1-2l}\left(\frac{\abs{E^x}}{\mathcal{L}^2}\right)^l\frac{\sqrt{\abs{E^x}\Delta}}{E^\rho}\,, & \text{for } \abs{E^x}\leq \mathcal{L}^2,
\end{array}
\right.\\
\mubar_\rho &=&
\left\{
\begin{array}{lr}
\sqrt{\frac{\Delta}{\abs{E^x}}}\,, & \text{for } \abs{E^x}\geq \mathcal{L}^2,\\
\left(\frac{\abs{E^x}}{\mathcal{L}^2}\right)^l\sqrt{\frac{\Delta}{\abs{E^x}}}\,, & \text{for } \abs{E^x}\leq \mathcal{L}^2,
\end{array}
\right.\\
&&\text{for}\ l<\frac{1}{2}.
\end{eqnarray}
\end{subequations}
Note that \eqnref{mu bar modified case 1} reduces to the original proscription \eqnref{mu bar} for $l=0$.
For the second case that $\mubar_x\bar{A}_x$ and $\mubar_\rho\bar{A}_\rho$ approach the critical value $\pm\pi/2$ with the opposite sign, \eqnref{mu bar} has to be modified as
\begin{subequations}\label{mu bar modified case 2}
\begin{eqnarray}
\mubar_x &=&
\left\{
\begin{array}{lr}
\frac{2}{2l-1}\frac{\sqrt{\abs{E^x}\Delta}}{E^\rho}\,, & \text{for } \abs{E^x}\geq \mathcal{L}^2,\\
\frac{2}{2l-1}\left(\frac{\abs{E^x}}{\mathcal{L}^2}\right)^l\frac{\sqrt{\abs{E^x}\Delta}}{E^\rho}\,, & \text{for } \abs{E^x}\leq \mathcal{L}^2,
\end{array}
\right.\\
\mubar_\rho &=&
\left\{
\begin{array}{lr}
\sqrt{\frac{\Delta}{\abs{E^x}}}\,, & \text{for } \abs{E^x}\geq \mathcal{L}^2,\\
\left(\frac{\abs{E^x}}{\mathcal{L}^2}\right)^l\sqrt{\frac{\Delta}{\abs{E^x}}}\,, & \text{for } \abs{E^x}\leq \mathcal{L}^2,
\end{array}
\right.\\
&&\text{for}\ l>\frac{1}{2}.
\end{eqnarray}
\end{subequations}
Here, we introduce a new constant length scale $\mathcal{L}$ satisfying $\Delta<\mathcal{L}^2\ll G^2M^2$ with $M$ being the mass constant of the spacetime, assume the bounce takes place when $\abs{E^x}<\mathcal{L}^2$, and require $l<1/2$ for \eqnref{mu bar modified case 1} and $l>1/2$ for \eqnref{mu bar modified case 2} to ensure the convention $\mubar_x,\mubar_\rho>0$. Note that, in the new prescription of \eqnref{mu bar modified case 1} and \eqnref{mu bar modified case 2}, the dimensions and density weights of $\mubar_x$ and $\mubar_\rho$ are the same as those of the old ones given by \eqnref{mu bar}. Therefore, the constraint algebra \eqnref{pb e infty} is unchanged with the new prescription. Furthermore, it is not difficult to show that, with \eqnref{mu bar modified case 1}, \eqnref{pb infty c after bounce} and \eqnref{D across bounce b} are still satisfied, and with \eqnref{mu bar modified case 2}, \eqnref{pb infty c after bounce modified} leads to $\left\{\Bar{A}_x^{(\infty)}(x),\Bar{A}_\rho^{(\infty)}(y)\right\}=0$ and \eqnref{D across bounce modified} leads to \eqnref{D across bounce b}. The good features obtained in \secref{sec:higher order}, \eqnref{pb f infty} in particular, are restored.

When the new prescription \eqnref{mu bar modified case 2} is used for the Schwarzschild spacetime in Kruskal coordinates, the classical solution \eqnref{cl sol} yields
\begin{subequations}
\begin{eqnarray}
\mubar_x\bar{A}_x &=&
\left\{
\begin{array}{lr}
-\gamma\sqrt{\Delta}\, \frac{R^{-1/2}\dot{R}\left(2GM+R\right)}{8\sqrt{2}\,(2l-1)(GM)^{5/2}} \,e^{R/4GM}, & \text{for}\ \abs{E^x}\geq\mathcal{L}^2,\\
-\frac{\gamma\sqrt{\Delta}}{\mathcal{L}^{2l}}\, \frac{R^{2l-1/2}\dot{R}\left(2GM+R\right)}{8\sqrt{2}\,(2l-1)(GM)^{5/2}} \,e^{R/4GM}, & \text{for}\ \abs{E^x}\leq\mathcal{L}^2,
\end{array}
\right.\\
\mubar_\rho\bar{A}_\rho &=&
\left\{
\begin{array}{lr}
\gamma\sqrt{\Delta}\, \frac{R^{-1/2}\dot{R}}{2\sqrt{2}\,(GM)^{3/2}} \,e^{R/4GM},  & \text{for}\ \abs{E^x}\geq\mathcal{L}^2,\\
\frac{\gamma\sqrt{\Delta}}{\mathcal{L}^{2l}}\, \frac{R^{2l-1/2}\dot{R}}{2\sqrt{2}\,(GM)^{3/2}} \,e^{R/4GM},  & \text{for}\ \abs{E^x}\leq\mathcal{L}^2,
\end{array}
\right.
\end{eqnarray}
\end{subequations}
which follows
\begin{equation}\label{ratio of mubar A}
\frac{\mubar_x\bar{A}_x}{\mubar_\rho\bar{A}_\rho}=-\frac{1}{4l-2}\left(1+\frac{R}{2GM}\right).
\end{equation}
By \eqnref{t2-x2}, \eqnref{Lambert function} and \footref{foot:Lambert fun}, we have
\begin{equation}
\dot{R}=\frac{8tG^2M^2}{t^2-x^2}\left(\frac{R}{2GM}-1\right)R^{-1}
=\mp\frac{8tG^2M^2}{R}\,e^{-R/2GM},
\end{equation}
which then leads to
\begin{subequations}\label{mubar A}
\begin{eqnarray}
\mubar_x\bar{A}_x &=&
\left\{
\begin{array}{lr}
\pm t\gamma\sqrt{\Delta}\, \frac{R^{-3/2}\left(2GM+R\right)}{\sqrt{2}\,(2l-1)(GM)^{1/2}} \,e^{-R/4GM}, & \text{for}\ \abs{E^x}\geq\mathcal{L}^2,\\
\pm\frac{t\gamma\sqrt{\Delta}}{\mathcal{L}^{2l}}\, \frac{R^{2l-3/2}\left(2GM+R\right)}{\sqrt{2}\,(2l-1)(GM)^{1/2}} \,e^{-R/4GM}, & \text{for}\ \abs{E^x}\leq\mathcal{L}^2,
\end{array}
\right.\\
\mubar_\rho\bar{A}_\rho &=&
\left\{
\begin{array}{lr}
\mp\frac{4t\gamma\sqrt{\Delta}}{\sqrt{2}}\,R^{-3/2} (GM)^{1/2} \,e^{-R/4GM}, & \text{for}\ \abs{E^x}\geq\mathcal{L}^2,\\
\mp\frac{4t\gamma\sqrt{\Delta}}{\mathcal{L}^{2l}\sqrt{2}}\,R^{2l-3/2} (GM)^{1/2} \,e^{-R/4GM}, & \text{for}\ \abs{E^x}\leq\mathcal{L}^2.
\end{array}
\right.
\end{eqnarray}
\end{subequations}
In the limit $t^2-x^2\rightarrow\pm1$ (i.e. $R\rightarrow0$) towards the (black/while hole) singularity,
\begin{subequations}\label{mubar A limit}
\begin{eqnarray}
\mubar_x\bar{A}_x &\mathop{\longrightarrow}\limits_{t^2-x^2\rightarrow\pm1}& \pm\frac{4t\gamma\sqrt{\Delta\,GM}}{\sqrt{2}\,(4l-2)\mathcal{L}^{2l}}\,R^{2l-3/2},\\
\mubar_\rho\bar{A}_\rho &\mathop{\longrightarrow}\limits_{t^2-x^2\rightarrow\pm1}&
\mp\frac{4t\gamma\sqrt{\Delta\,GM}}{\sqrt{2}\,\mathcal{L}^{2l}}\,R^{2l-3/2}.
\end{eqnarray}
\end{subequations}
As the strength of loop quantum corrections are indicated by $\mubar_x\bar{A}_x$ and $\mubar_\rho\bar{A}_\rho$, both in \eqnref{mubar A limit} are supposed to blow up when computed with the classical solution. This restricts the value of $l$ to be
\begin{equation}
1/2<l<3/4.
\end{equation}
Moreover, one can further fix $l=3/4-\epsilon$ with a small positive number $\epsilon$ to have \eqnref{ratio of mubar A} read as
\begin{equation}\label{ratio of mubarA}
\frac{\mubar_x\bar{A}_x}{\mubar_\rho\bar{A}_\rho}= -\left(1+4\epsilon+\mathcal{O}(\epsilon^2)\right)\left(1+\frac{R}{2GM}\right)\approx-(1+4\epsilon),
\quad \text{for }R\ll2GM,
\end{equation}
so that $\mubar_x\bar{A}_x$ gets bounced almost at the same time as $\mubar_\rho\bar{A}_\rho$ and thus the breakdown in \eqnref{pb f infty} during the transition period is optimally mitigated.

It should be emphasized that, until now, the modification in \eqnref{mu bar modified case 1} and \eqnref{mu bar modified case 2} is regarded as a phenomenological prescription devoid of any first-principle motivation and cannot be implemented in the quantum theory as we did in \secref{sec:improved dynamics} as it introduces a new length scale $\mathcal{L}$. This modification is mandatory only for the situation in which, towards the singularity, $\mubar_x\bar{A}_x$ and $\mubar_\rho\bar{A}_\rho$ are of opposite signs, or equivalently $d\Lambda/d\tau$ and $dR/d\tau$ are of opposite signs according to \eqnref{meaning of mubar A}.\footnote{Note that, towards a spacelike singularity, $d\Lambda/d\tau$ and $dR/d\tau$ dominate the terms involving spatial derivatives in \eqnref{meaning of mubar A}, provided that the coordination and $N^x$ are well-behaved. Also see \footref{foot:Hubble rates}.} That is, at classical level, a given comoving cell is collapsing in the homogeneous ($\partial_\theta,\partial_\phi$) directions while stretching in the radial ($\partial_x$) direction (or the other way around), reminiscent of the Kasner (vacuum) solution to the Bianchi I cosmology (see \cite{Chiou:2008eg} for the analogy between the Bianchi I and the Kantowski-Sachs models). In the context of Bianchi I models, it has been shown in \cite{Chiou:2007sp} that inclusion of (perfect fluid) matter with $w<1$ has the effect of ``isotropizing'' the universe and loop quantum corrections take effect in the ``isotropized phase'' rather than in the ``Kasner phase'' if the matter content is abundant enough.\footnote{The analysis in \cite{Chiou:2007sp} is based on the $\mubar$-scheme, instead of the $\mubar'$-scheme, but the conclusion that loop quantum corrections take effect in the isotropized phase when the matter content is abundant should still hold for the $\mubar'$-scheme.} Likewise, it is anticipated that, in spherically symmetric models, with inclusion of abundant matter of $w<1$, the solution is isotropized and loop quantum corrections take place in the isotropized phase. Therefore, for the cases with abundant matter content (collapsing black hole, Reissner-Nordstr\"{o}m black hole, etc.), $\mubar_x\bar{A}_x$ and $\mubar_\rho\bar{A}_\rho$ approach the critical value $\pm\pi/2$ with the same sign and the original prescription in \eqnref{mu bar} can be used and the further modification in \eqnref{mu bar modified case 1} other than $l=0$ may not be necessary.

As commented in the end of \secref{sec:higher order}, the spherical symmetry is expected to break down in the vicinity of the quantum bounce from the perspective of the full theory of LQG. For the case devoid of or deficient in matter content, the quantum bounce takes place in the Kasner phase, in which the spacetime is highly anisotropic; consequently the breakdown could be exaggerated by the Kasner-like anisotropy and become even earlier and severer. Accordingly, the modification in \eqnref{mu bar modified case 2}, albeit an \textit{ad hoc} prescription, can be interpreted as a correction which makes amends for the severe breakdown due to Kasner-like anisotropy, as $\mathcal{L}$ signals the length scale at which the weave state starts to become granular and $l$ characterizes the degree of coherence of the weave state.

\subsection{Preliminary analysis of the quantum Schwarzschild spacetime}\label{sec:preliminary analysis}
At the heuristic level of effective dynamics, we treat the dynamics as classical but governed by the holonomized Hamiltonian constraint $C^{(n)}$ given by \eqnref{C nth} with the $n$th order holonomy corrections. The corresponding Hamilton's equations are given in \eqnref{Hamilton eqs nth order}.
In the case of $n=\infty$, we have
\begin{subequations}\label{derivatives of A infty}
\begin{eqnarray}
\label{dA dE}
\frac{\partial\bar{A}_{x,\rho}^{(\infty)}}{\partial (E^x,E^\rho)} &=&
\left[\frac{\cos(\mubar_{x,\rho}\bar{A}_{x,\rho})}
{\abs{\cos(\mubar_{x,\rho}\bar{A}_{x,\rho})}}\bar{A}_{x,\rho}-\bar{A}_{x,\rho}^{(\infty)}\right] \frac{1}{\mubar_{x,\rho}}\frac{\partial\mubar_{x,\rho}}{\partial(E^x,E^\rho)}\nonumber\\
&\equiv&
\left\{
  \begin{array}{lr}
  0, & \text{before } \mubar_{x,\rho}\bar{A}_{x,\rho} \text{ reaches } \pm\pi/2,\\
  \mp\frac{\pi}{\mubar_{x,\rho}^2}\frac{\partial\mubar_{x,\rho}}{\partial(E^x,E^\rho)}, & \text{after } \mubar_{x,\rho}\bar{A}_{x,\rho} \text{ reaches } \pm\pi/2,
  \end{array}
\right.\\
\label{dAx dAx}
\frac{\partial\bar{A}_x^{(\infty)}}{\partial\bar{A}_x}&=&
\frac{\cos(\mubar_x\bar{A}_x)}{\abs{\cos(\mubar_x\bar{A}_x)}}\equiv f_\sqcap(\mubar_x\bar{A}_x),\\
\label{dArho dArho}
\frac{\partial\bar{A}_\rho^{(\infty)}}{\partial\bar{A}_\rho}&=&
\frac{\cos(\mubar_\rho\bar{A}_\rho)}{\abs{\cos(\mubar_\rho\bar{A}_\rho)}}\equiv f_\sqcap(\mubar_\rho\bar{A}_\rho),
\end{eqnarray}
\end{subequations}
by \eqnref{higher order holonomies}, \eqnref{identity} and \eqnref{triangle wave 2}. Taking \eqnref{derivatives of A infty} into \eqnref{Hamilton eqs nth order}, we obtain the Hamilton's equations when $C$ is replaced by $C^{(\infty)}$.

In particular, \eqnref{Hamilton c} and \eqnref{Hamilton d}, which in the classical case read as
\begin{subequations}
\begin{eqnarray}
\dot{E}^x &=& N\left(\sgn(E_x)\frac{\sqrt{\abs{E^x}}\,\bar{A}_\rho}{\gamma}\right)+N^xE^{x\prime},\\
\dot{E}^\rho &=& N\left(\sgn(E_x)\frac{\sqrt{\abs{E^x}}\,\bar{A}_x}{\gamma} +\frac{\bar{A}_\rho E^\rho}{2\gamma\sqrt{\abs{E^x}}}\right)+\left(N^xE^\rho\right)',
\end{eqnarray}
\end{subequations}
are now replaced by
\begin{subequations}\label{dot Ex Erho}
\begin{eqnarray}
\dot{E}^x &=& Nf_\sqcap(\mubar_x\bar{A}_x)\left(\sgn(E_x)\frac{\sqrt{\abs{E^x}}\,\bar{A}_\rho^{(\infty)}}{\gamma}\right) +N^xE^{x\prime},\\
\dot{E}^\rho &=& Nf_\sqcap(\mubar_\rho\bar{A}_\rho)
\left(\sgn(E_x)\frac{\sqrt{\abs{E^x}}\,\bar{A}_x^{(\infty)}}{\gamma} +\frac{\bar{A}_\rho^{(\infty)} E^\rho}{2\gamma\sqrt{\abs{E^x}}}\right)+\left(N^xE^\rho\right)'.
\end{eqnarray}
\end{subequations}
If we choose $N^x=0$, it is obvious that $E^x$ gets bounced as $f_\sqcap(\mubar_x\bar{A}_x)$ flips signs, and $E^\rho$ gets bounced as $f_\sqcap(\mubar_\rho\bar{A}_\rho)$ flips signs.

With the Hamilton's equations at hand, the numerical solution can be obtained by the method of \emph{finite-difference time-domain} (FDTD) \cite{FiniteDifference}. In the $n=\infty$ limit, however, the numerical method is hindered by the discontinuity appearing on the right hand sides of the Hamilton's equations. One can bypass this problem and still get faithful numerical solutions by keeping $n$ large but finite.\footnote{It is shown in \cite{Chiou:2009yx} that, in the quantum theory of LQC, the evolution of coherent states follows \emph{smooth} trajectories despite the kink of the connection variable with the $n=\infty$ holonomy corrections. In a sense, the discontinuity resulting from the $n=\infty$ limit is smeared by the quantum spreading. To reflect the smearing at the level of heuristic dynamics, one can simply truncate $n$ to a large but finite value (the larger $n$ effectively gives rise to the trajectory of a sharper coherent state).}

Particularly, to obtain the loop quantum geometry of the maximally extended Schwarzschild spacetime, we choose the gauges
\begin{subequations}
\begin{eqnarray}
N(t,x) &=& \frac{\sqrt{32G^3M^3}}{\abs{E^x}^{1/4}}\ e^{-\sqrt{\abs{E^x}}/4GM},\\
N^x(t,x) &=& 0,
\end{eqnarray}
\end{subequations}
in agreement with \eqnref{lapse shift}, and accordingly use the classical solution \eqnref{cl sol} to yield the initial condition in the classical regime (say, for the $t=0$ spatial slice). That is, we assume that the weave state, after coarse-graining, manifests semiclassical traits (spreading, squeezing, etc.) in accordance with the Kruskal coordinates at large scale; this is a sensible prescription since the Kruskal coordinates give a well-behaved foliation which covers both the interior and exterior without introducing boundary terms, i.e. \eqnref{falloff e and f} satisfied with $N_\pm=0$ (also see \footref{foot:Hubble rates}).
With the gauges $N$ and $N^x$ fixed and the initial condition given, the Hamilton's equations \eqnref{Hamilton eqs nth order} in principle can be numerically solved for a given large $n$.
The numerical task is however tremendously demanding on the numerical accuracy and thus it remains challenging and may require more sophisticated algorithms.\footnote{The FDTD method is already very demanding on numerical accuracy even for solving the classical Hamilton's equations \eqnref{Hamilton eqs}. For $M\approx10^2\!-\!10^3\Plm$ ($\Plm$ is the Planck mass), the numerical method requires 256-bit accuracy, which is about 77 digits, in order to approach the singularity close enough (until the quantum corrections take effect). The GNU Multiple Precision Arithmetic Library (GMP) \cite{GMP} and the GNU MPFR Library \cite{MPFR} are needed for arbitrary-precision arithmetic and floating-point computations. To make sense of semiclassicality, $M$ has to be much larger. The large $M$ and higher order quantum corrections together demand even much higher accuracy.}

Even though the numerical solution is extremely difficult, it still offers considerable insight about the natures of the loop quantum geometry of the Schwarzschild spacetime by inspecting the Hamilton's equation of the heuristic effective dynamics. Firstly, \eqnref{dot Ex Erho} implies that $E^x$ and $E^\rho$ get bounced as $f_\sqcap(\mubar_x\bar{A}_x)$ and $f_\sqcap(\mubar_\rho\bar{A}_\rho)$ flip signs respectively. By the prescription \eqnref{mu bar modified case 2}, \eqnref{ratio of mubarA} further tells that $E^x$ and $E^\rho$ get bounced almost around the same epoch, provided that the parameter $l$ is tuned to be close to but smaller than $3/4$. This suggests that, for a well-behaved weave state which gives the extended Schwarzschild spacetime solution in the classical regime, the (black/white hole) singularity is resolved by the loop quantum corrections and replaced by a quantum bounce which bridges the classical solution to another classical phase.

Secondly, the quantum effects are expected to become significant only for the region $\abs{E^x}<\mathcal{L}^2$. However, \eqnref{mubar A} indicates that any spacetime point in the region of $\abs{E^x}\geq(2GM)^2\gg\mathcal{L}^2$ eventually receives quantum corrections in the late times (i.e. when $t$ is large enough). In other words, the slices of $\mubar_x\bar{A}_x\approx\pm\pi/2$ and $\mubar_\rho\bar{A}_\rho\approx\mp\pi/2$ inevitably intersect the event horizons. This seems to signal the breakdown (of the semiclassical treatment) of the quantum theory, as the spacetime curvature around the event horizon is fairly flat and should not incur any quantum corrections. This problem can be regarded as an indication that the inclusion of the Hawking radiation mechanism is mandatory. That is, only when the Hawking radiation is taken into account can the loop quantum theory be consistent. Since the Hawking radiation is not considered, the quantum corrections on the late-time horizons and beyond are not trustable.

\begin{figure}
\begin{picture}(450,360)(0,0)

\put(-30,-50)
{
\scalebox{0.8}{\includegraphics{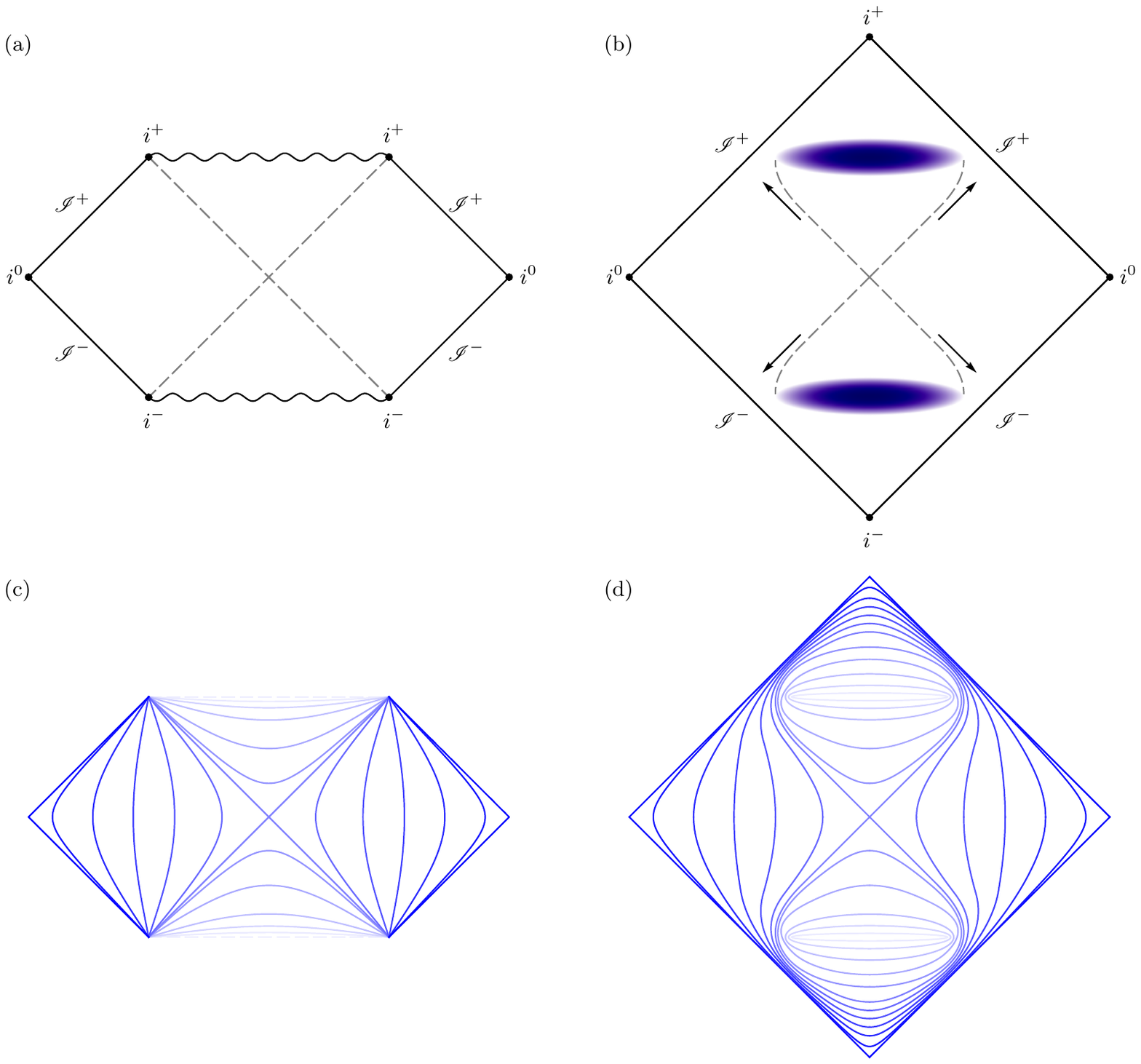}}
}

\end{picture}
\caption{\textbf{(a)}: The Penrose diagram of the maximally extended Schwarzschild spacetime. \textbf{(b)}: The (conjectured) Penrose diagram of the loop quantum extension of (a). The putative singularities are resolved and replaced by the quantum bounces (shaded areas) and the generalized dynamical horizons (dashed lines) deviate from the classical event horizons of the black/white hole in the late/early time. The black and white holes are evaporated via the Hawking radiation (denoted by the arrows). (More precisely, evaporation of the white hole is the time reversal of the black hole evaporation; the arrows pointing backwards in time denote forward radiation of antiparticles.) \textbf{(c)} and \textbf{(d)}: Contours of constant $R=\sqrt{\abs{E^x}}$ of (a) and (b) (smaller $R$ in lighter color/shade).}\label{fig:Penrose diagrams}
\end{figure}

To summarize, assuming that the weave state is carefully chosen such that its semiclassical traits accord with the Kruskal coordinates and semiclassicality is upheld even across the putative singularity, two important consequences are observed: first, the classical singularity is resolved and replaced by a quantum bounce which bridges the black/white hole interior to a different classical phase; second, the Hawking radiation should be taken into account (on the late-time horizons) and thus the Hawking evaporation is expected. As both resolution of the classical singularity and inclusion of the Hawking radiation are expected, the complete quantum Schwarzschild spacetime is conjectured to have a global structure akin to that of the 2-dimensional black holes investigated in \cite{Ashtekar:2008jd,Ashtekar:2010hx,Ashtekar:2010qz}. That is, the black hole is evaporated via the Hawking radiation, and the quantum spacetime is largely extended from the classical one as the classical singularity is resolved and the black hole interior is extended. The conjectured Penrose diagram is depicted in \figref{fig:Penrose diagrams}. Note that the black hole is simply evaporated; it is neither connected to a white hole nor to a baby black hole, as opposed to the results obtained in \cite{Chiou:2008nm} (minisuperspace treatment) and \cite{Gambini:2008dy} (midisuperspace treatment). As in the 2-dimensional quantum black holes in \cite{Ashtekar:2008jd,Ashtekar:2010hx,Ashtekar:2010qz}, the information that is classically lost in the process of Hawking evaporation is recovered, primary because the quantum spacetime is sufficiently larger than the classical counterpart.

It should be noted that the scenario described above is expected only if the weave state manifests semiclassical traits in accordance with the Kruskal coordinates. Other scenarios are possible if one considers a weave state whose semiclassical traits are manifested differently (recall \footref{foot:Hubble rates}). A particular example is the solution obtained in \cite{Gambini:2008dy}, which gives a singularity-free quantum spacetime akin to that in the $\mu_0$- and $\mubar$-schemes of the minisuperspace (interior) treatment \cite{Modesto:2006mx,Bohmer:2007wi,Chiou:2008nm} and does not need to evoke the Hawking radiation. In \cite{Gambini:2008dy}, the 1-dimensional diffeomorphism gauge is partially fixed in such a way that the boundary conditions are given very similar to those in the Kruskal coordinates but one does \emph{not} work exactly in Kruskal coordinates asymptotically. From the viewpoint of this paper, the gauge choice in \cite{Gambini:2008dy} amounts to different semiclassical traits not exactly in accord with Kruskal coordinates and the free parameters in the solution could be interpreted as reflection of semiclassical traits (analogous to $\mathcal{L}$ and $\l$ in our treatment).\footnote{Similarly, in the $\mu_0$- and $\mubar$-schemes of the minisuperspace (interior) treatment \cite{Modesto:2006mx,Bohmer:2007wi,Chiou:2008nm}, there appears to be an additional parameter in the effective solution which gives rise to the difference between the mass of the white hole and that of the black hole (essentially, the solution depends on the gauge choice the finite sized cell). This again can be interpreted as reflection of semiclassical traits of the given weave state (see \cite{Chiou:2008nm} for more comments).} Finally, as commented in the end of \secref{sec:higher order}, if semiclassicality breaks down severely in the quantum regime, the smooth weave state could simply descend into a granular spin network state.

\section{Summary and discussion}\label{sec:summary}
In Ashtekar's formalism for spherically symmetric midisuperspaces, the $SU(2)$ internal gauge in the full theory is reduced to $U(1)$ and the 3-dimensional diffeomorphism is reduced to 1-dimensional. By change of variables, the canonical structure in terms of $(\bar{A}_x,\bar{A}_\rho,\eta;E^x,E^\rho,\bar{P}^\eta)$ decouples the $U(1)$ degrees of freedom from the geometric/metric ones, and accordingly the Gauss constraint is completely decoupled from the diffeomorphism and Hamiltonian constraints. While $(\eta;\bar{P}^\eta)$ corresponds solely to the internal degrees of $U(1)$ gauge, $(E^x,E^\rho)$ gives the intrinsic geometry via \eqnref{dict 1} and $(\bar{A}_x,\bar{A}_\rho)$, together with the lapse $N$ and shift $N^x$, gives the extrinsic geometry via \eqnref{dict 2}. Particularly, the classical solution of the maximally extended Schwarzschild spacetime can be explicitly recast in terms of Ashtekar variables in accordance with the Kruskal coordinates.

For the loop quantum theory of spherically symmetric midisuperspaces, we adopt a new approach, which is different from that in the previous works \cite{Bojowald:2004af,Bojowald:2004ag,Bojowald:2005cb} but directly in the same spirit of LQC, by considering cylindrical functions of the symmetry-reduced connection variables from the outset as in \eqnref{Psig}. Imposition of the Gauss constraint yields the $U(1)$-invariant states as spin networks defined in \eqnref{spin network}. Further imposition of the diffeomorphism constraint via group averaging gives rise to spin-knots, which can be abstractly interpreted as ensembles of chunks of volumes arranged in a radial order with adjacent surfaces of areas without any reference to localization of the chunks and adjacent surfaces. Unlike the full theory of LQG, the volume and area eigenvalues are only partially quantized in the sense of Bohr compactification due to the fact that $SU(2)$ is symmetry-reduced to $U(1)$.

For the dynamics, we propose a new quantization scheme to construct a graph-preserving Hamiltonian constraint operator, which changes only the coloring of spin network states and upon which one can impose the fundamental discreteness of LQG by hand \textit{\`{a} la} the strategy of improved dynamics in LQC. The strategy of improved dynamics is demanded as for respecting the diffeomorphism invariance, which is absent but reduced to scaling invariance in the context of LQC. On the other hand, imposition of fundamental discreteness spoils the intimate matching between diffeomorphism and Hamiltonian constraints in the full theory of LQG, and as a consequence the Hamiltonian constraint operators with different lapse functions do not commute, i.e. $[\hat{C}[N], \hat{C}[M]]\neq0$, and the resulting dynamics depends on foliation of the spacetime. However, this problem seems to be mitigated at large scale, if one includes higher order holonomy corrections, which can be understood as a result of generic $j$ representations for holonomies in the Hamiltonian constraint operator.

While rigorous and complete construction of the quantum theory remains challenging, significant insights can still be obtained at the heuristic level of effective dynamics. Consistency of constraint algebra regarding the Hamiltonian and diffeomorphism constraints fixes the improved quantization scheme to be of the form of \eqnref{mu bar}, reminiscent of the $\mubar'$-scheme in LQC for the Kantowski-Sachs models. Moreover, consistency of constraint algebra regarding two Hamiltonian constraints (with different lapse functions) further demands the inclusion of higher order holonomy corrections as in \eqnref{higher order holonomies} to the order of infinity and meanwhile fixes a ratio factor 2 in \eqnref{mu bar} for the improved quantization scheme.

At the heuristic level of effective dynamics, it is suggested that the black hole singularity is resolved and replaced by a quantum bounce which bridges a classical solution to another classical phase, as the constraint algebra is exactly satisfied before and after the bounce if \eqnref{mu bar} and \eqnref{higher order holonomies} with $n=\infty$ are adopted. As indicated in \eqnref{pb f infty}, however, the constraint algebra breaks down briefly during the transition period of the bounce, as $f_\sqcap(\mubar_x\bar{A}_x)$ and $f_\sqcap(\mubar_\rho\bar{A}_\rho)$ flip signs at close but slightly different epochs. As a result, the two constraints $C^{(\infty)}=0$ and $D=0$ are violated after the bounce, implying that the spacetime is no longer invariant under spatial diffeomorphism and change of spacetime foliation after the bounce. Nevertheless, if the characteristic parameter used to describe the solution in the classical regime is much larger than the Planck unit, the transition period (which is of the scale of Planck time) is so brief (compared to the characteristic time) that the violation is minuscule and therefore the heuristic effective dynamics still gives a reliable description of large-scale physics after the bounce.

The fact that the constraints violate briefly during the transition period indicates that one cannot make sense of symmetry reduction by separating the degrees of freedom of the full theory into spherical (heavy) and non-spherical (light) ones in the vicinity of the quantum bounce.
It suggests that, in the full theory of LQG, a given weave state which represents a smooth, spherically symmetric space at large scale will eventually manifest granularity of spin networks when approaching the quantum bounce. After the bounce, the weave state evolves to become smooth and spherically symmetric again provided that the transition period is very brief.
In the case that $f_\sqcap(\mubar_x\bar{A}_x)$ and $f_\sqcap(\mubar_\rho\bar{A}_\rho)$ do not flip signs at close epochs or even one of them never flips signs, however, a different scenario in the full theory is also possible. In this case, semiclassicality breaks down severely and the smooth weave state could simply descend into a granular spin network state, or other quantum effects (such as the Hawking radiation) should be taken into account to maintain the semiclassicality.

At the heuristic level of semiclassical dynamics, further phenomenological modification can be prescribed as in \eqnref{mu bar modified case 1} and \eqnref{mu bar modified case 2}. This modification is devoid of any first-principle motivation but phenomenologically can be interpreted as quantum corrections due to semiclassical traits (spreading, squeezing, etc.) of the weave state, as $\mathcal{L}$ signals the length scale at which the weave state starts to become granular and $l$ characterizes the degree of coherence of the weave state. For the cases devoid of or deficient in matter content, the quantum bounce takes place in the Kasner phase and the phenomenological modification \eqnref{mu bar modified case 2} is mandatory, if the semiclassicality is to be maintained across the putative singularity. On the other hand, for the cases abundant in matter content, the phenomenological modification \eqnref{mu bar modified case 1} might be optional but the prescription with fine-tuned $\mathcal{L}$ and $l$ could further mitigate the breakdown of semiclassicality. Semiclassicality of the weave state could be upheld even across the putative singularity, thus giving a bouncing scenario, if the weave state is carefully chosen (i.e. $\mathcal{L}$ and $l$ are fine-tuned).

The heuristic effective dynamics can be used to investigate the quantum extension of the Schwarzschild spacetime in accordance with the classical solution in the Kruskal coordinates. Although the full-fledged numerical method is extremely challenging, considerable insight can still be obtained by inspecting the effective Hamilton's equations: first, the classical singularity is resolved and replaced by a quantum bounce which bridges the black/white hole interior to a different classical phase; second, it is mandatory to take into account the Hawking radiation and thus the Hawking evaporation is expected. These lead us to conjecture that the complete quantum spacetime of the Schwarzschild solution resembles the quantum spacetime structure of the 2-dimensional black holes studied in \cite{Ashtekar:2008jd,Ashtekar:2010hx,Ashtekar:2010qz}: the black hole is evaporated via the Hawking radiation, and the quantum spacetime is largely extended from the classical one as depicted in \figref{fig:Penrose diagrams}.
The information that is classically lost through the Hawking evaporation is eventually recovered, because the quantum spacetime is sufficiently larger than the classical counterpart.

It should be kept in mind, however, that other scenarios are still possible if the weave state exhibits different semiclassical traits. Particularly, the solution in \cite{Gambini:2008dy} gives a complete quantum spacetime akin to that in the $\mu_0$- and $\mubar$-schemes of the midisuperspace treatment and does not evoke Hawking radiation. This very different scenario is not contradictory but in fact complementary to that conjectured in this paper, as the quantum evolution indeed depends on details (such as semiclassical traits) of the smooth weave state. The free parameters of the solution in \cite{Gambini:2008dy} could be better understood, if one can rephrase the partial gauge fixing in \cite{Gambini:2008dy} from the perspective of this paper. Furthermore, different scenarios (including those descending into granular spin networks) are in fact connected in the quantum theory in the sense that, at the quantum level, the outcome of a given weave state can branch into different scenarios with various probabilities.

More insight about the natures of loop quantum corrections can be obtained if one applies the same formulation of this paper to other spherically symmetric models. For collapsing black holes, it would be very instructive to know how the formation of the trapped surface is altered as well as how the classical singularity is resolved by the quantum corrections. Moreover, for the Reissner-Nordstr\"{o}m (charged) black hole, one would have the chance to study the quantum corrections on the timelike singularity, which is not possible in the minisuperspace formalism.

\begin{acknowledgments}
The authors would like to thank the attendants of the Mini-Workshop on Canonical Approaches and Numerical Methods in Gravity at NCTS/NTHU in Taiwan for valuable discussion and comments. D.W.C. is supported by the Center for Advanced Study in Theoretical Sciences at NTU and W.T.N. by the National Science Council under the Grant No. NSC101-2112-M-007-007.
\end{acknowledgments}

\appendix

\section{Dimensions and density weights}\label{app:dimension and weight}
From the point of view of the 1-dimensional field theory over $x$, many variables transform as scalars or scalar densities under the 1-dimensional diffeomorphism on $x$. As noted in \footref{foot:density character}, it is very helpful to keep track of the density character and here we list the density weights for various variables in \tabref{table:dimension and density}. The transformation rule under diffeomorphism for scalar densities with different weights is given in \footref{foot:diff}. For convenience, we also list the dimensions for various constants and variables at the same time.
We work in the units with $c=1$ but keep $G$ and $\hbar$ explicit. In this unit system, the Planck length is given by $\sqrt{G\hbar/c^3}=\sqrt{G\hbar}=:\Pl$\, and the Planck mass by $\sqrt{\hbar\,c/G}=\sqrt{\hbar/G}=:\Plm$. We keep the dimensions of the coordinates $t$ and $x$ arbitrary and denote the dimension of length as $L$.
\begin{table}
\begin{tabular}{lc|c|c}
  &  & dimension & density weight\footnote{The Lagrange multipliers as well as the $U(1)$ gauge angle $\lambda$ transform as scalar densities with respect to the coordinate transformation $x\rightarrow\bar{x}(x)$ (together with $t\rightarrow\bar{t}=t$) only in the sense that geometry of ($SO(3,1)$ bundle over) the spacetime is required to remain unchanged. As far as the diffeomorphism of the spatial slice alone is concerned, however, they are regarded as independent variables --- i.e., they commute with the diffeomorphism constraint $D$. For $\lambda$ and variables involving Lagrange multipliers, we denote their density wights in parenthesis to indicate the subtlety.} \\ \hline\hline
constants  &  $\gamma,\xi$  &  $1$  &  --- \\
  &  $\Pl^2:= G\hbar,\Delta,\mathcal{L}^2$  &  $L^2$  &  --- \\
  &  $\Plm:=\sqrt{\hbar/G}$  &  $L/G$  &  --- \\ \hline
mass  &  $M$  &  $L/G$  &  --- \\ \hline
Ashtekar variables  &  $A_x,\bar{A}_x,\bar{A}_x^{(n)}$  &  $1/x$ & 1\\
  &  $A_1,A_2,A_\rho,\bar{A}_\rho,\bar{A}_\rho^{(n)}$  &  $1$  & 0\\
  &  $\alpha,\beta,\eta$  &  $1$  & 0\\ \hline
triad variables  &  $E^x$  &  $L^2$  &  0\\
  &  $E^1,E^2,E^\rho$  &  $L^2/x$  &  1\\
  &  $P^\eta,\bar{P}^\eta$  &  $L^2/x$  &  1\\
  &  $q:=\det E_i^a=E_x(E^\rho)^2$  &  $L^6/x^2$  &  2\\ \hline
spin network labels  &  $k_e,\mu_v,n_v$  &  1  &  0\\ \hline
discreteness variables  &  $\epsilon,\,\mubar_x$  &  $x$  &  -1\\
  &  $\delta,\,\mubar_\rho$  &  1  &  0\\
  &  $\mubar_x\bar{A}_x,\, \mubar_\rho\bar{A}_\rho$  &  1  &  0\\ \hline
3-metric  components &  $R$  &  $L$  &  0\\
  &  $\Lambda$  &  $L/x$  &  1\\ \hline
constraints  &  $U$  &  $\hbar/x$  &  1\\
  &  $D$  &  $\hbar/x^2$  &  2\\
  &  $C, C^{(n)}$  &  $\hbar/xL$  &  1\\ \hline
$U(1)$ gauge angle & $\lambda$ & 1 & (0)\\ \hline
Lagrange multipliers  &  $\omega^3\cdot t$  &  $1/t$  &  (0)\\
  &  $N^x$  &  $x/t$  &  (-1)\\
  &  $N$  &  $L/t$  &  (0)\\ \hline
Hamiltonian density  &  $\left(\omega^3\cdot t\right)U\!+\!N^xD\!+\!NC$  & $\hbar/xt$  &  (1) \\ \hline
Hamiltonian  &  $\mathcal{H}$  &  $\hbar/t$  &  ---
\end{tabular}
\caption{Dimensions and density weights of constants and variables.}\label{table:dimension and density}
\end{table}

\section{Notations for polar type variables}\label{app:polar variables}

In this paper, we adopt slightly different notations for polar type variables, which we believe are more succinct, self-explanatory and less misleading than those used in \cite{Bojowald:2004af,Bojowald:2004ag,Bojowald:2005cb,Campiglia:2007pr,Gambini:2008dy}. To reduce proliferation of notations, we also avoid defining new variables which are not used repeatedly. Furthermore, we also introduce the variables $\bar{A}_x$ and $\bar{P}^\eta$ as in \secref{sec:further change of variables}; the treatment of this further change of variables is not explicitly carried out in \cite{Bojowald:2004af,Bojowald:2004ag,Bojowald:2005cb,Campiglia:2007pr,Gambini:2008dy}. The comparison of notations is given in \tabref{tab:notations}. Note that the spatial coordinates are denoted as $(x,\theta,\phi)$ in this paper, while they are denoted as $(x,\vartheta,\varphi)$ in \cite{Bojowald:2004af,Bojowald:2004ag,Bojowald:2005cb} and $(x,\theta,\varphi)$  in \cite{Campiglia:2007pr,Gambini:2008dy}; in the table, we denote them all as $(x,\theta,\phi)$. Also, the ditto mark \verb+"+ denotes that the case in the right column is identical to that in the middle column.

\begin{table}
\begin{tabular}{c|c|c|c}
in this paper & in \cite{Bojowald:2005cb} (and \cite{Bojowald:2004af,Bojowald:2004ag}) & in \cite{Campiglia:2007pr} (and \cite{Gambini:2008dy})\\
$(x,\theta,\phi)$ & $(x,\theta=\vartheta,\phi=\varphi)$ & $(x,\theta,\phi=\varphi)$ \\ \hline \hline
$A_x,A_1,A_2$ & $A_x,A_1,A_2$ & \verb+"+ \\
$E^x,E^1,E^2$ & $E^x,E^1,E^2$ & \verb+"+ \\ \hline
$\alpha,\beta,\eta$ & $\alpha,\beta,\eta$ & \verb+"+ \\ \hline
$A_\rho,E^\rho$ & $A_\phi,E^\phi$ & \verb+"+ \\ \hline
$\tau_1,\tau_2,\tau_3$ & $\Lambda_1,\Lambda_2,\Lambda_3$ & \verb+"+ \\
$\tau_{1'},\tau_{2'},\tau_3$ & $\Lambda_E^\theta,\Lambda_E^\phi,\Lambda_3$ & \verb+"+ \\
\begin{tabular}{c}
$\tau_{1'}:=\tau_1\cos(\alpha\!+\!\beta)+\tau_2\sin(\alpha\!+\!\beta)$\\
$\tau_{2'}:=-\tau_1\sin(\alpha\!+\!\beta)+\tau_2\cos(\alpha\!+\!\beta)$
\end{tabular}
&
\begin{tabular}{c}
$\Lambda_E^\theta := \Lambda_1\cos(\alpha\!+\!\beta)+\Lambda_2\sin(\alpha\!+\!\beta)$ \\
$\Lambda_E^\phi := -\Lambda_1\sin(\alpha\!+\!\beta)+\Lambda_2\cos(\alpha\!+\!\beta)$
\end{tabular}
&
\verb+"+ & \footnote{In \cite{Bojowald:2004af,Bojowald:2004ag,Bojowald:2005cb,Campiglia:2007pr}, $\Lambda_E^\theta$ and $\Lambda_E^\phi$ are mistaken for each other. This typo is corrected here.}\\
undefined & \begin{tabular}{c}
              $\Lambda_\theta^A := \Lambda_1\cos\beta+\Lambda_2\sin\beta$ \\
              $\Lambda_\phi^A := -\Lambda_1\sin\beta+\Lambda_2\cos\beta$
            \end{tabular}
            & \verb+"+ & \footnote{In \cite{Bojowald:2004af,Bojowald:2004ag,Bojowald:2005cb}, $\Lambda_\theta^A$ and $\Lambda_\phi^A$ are mistaken for each other. This typo is corrected in Equation (36) of \cite{Campiglia:2007pr} and here.} \\ \hline
$e_x^3,e_\theta^{1'},e_\phi^{2'}$ & $e_x,e_\phi,e_\phi\sin\theta$ & $e_x^3,e_\theta^\theta,e_\phi^\phi$ \\
$-\Gamma_\theta^{2'}=\Gamma_\phi^{1'}/\sin\theta=-\frac{E^{x\prime}}{2E^\rho}$ & $\Gamma_\phi:=-\frac{e'_\phi}{e_x}=-\frac{E^{x\prime}}{2E^\phi}$ & $\Gamma_\phi:=-\frac{(e_\theta^\theta)'}{e_x^3}=-\frac{E^{x\prime}}{2E^\phi}$ & \footnote{Equation (29) of \cite{Campiglia:2007pr} reads as $\Gamma_\phi:=-(e_\phi^\phi)'/e_x^3$, which mistakes $e_\theta^\theta$ for $e_\phi^\phi$ as a typo.} \\ \hline
\begin{tabular}{c}
$P^\rho,P^\beta$ defined in \footref{foot:irrelevant variables}; \\
not used elsewhere
\end{tabular}
&
\begin{tabular}{c}
$P^\phi:=2E^\phi\cos\alpha$ \\
$P^\beta:=2A_\phi E^\phi\sin\alpha$
\end{tabular}
& \verb+"+ \\ \hline
$\bar{A}_\rho:=2A_\rho\cos\alpha$ & $2\gamma K_\phi=2A_\phi\cos\alpha$ & $\bar{A}_\phi:=2A_\phi\cos\alpha=2\gamma K_\phi$ \\
$P^\eta:=2A_\rho E^\rho\sin\alpha$ & $P^\eta=P^\beta:=2A_\phi E^\phi\sin\alpha$ & \verb+"+ \\ \hline
$\bar{A}_x:=A_x+(\alpha+\beta)'=\gamma K_x$ & $\gamma K_x=A_x+(\alpha+\beta)'$ & \verb+"+ \\
$\bar{P}^\eta:=P^\eta+E^{x\prime}$ & undefined & undefined
\end{tabular}
\caption{Comparison of notations for polar type variables.}\label{tab:notations}
\end{table}

\section{Poisson bracket between Hamiltonian constraints}\label{app:Poisson bracket}

\subsection{Classical case}

At firs glance, it seems dauntingly tedious to derive \eqnref{pb f}, as the Hamiltonian constraints given in \eqnref{C 2} is rather complicated. A moment of reflection, however, tells that, essentially, only the Poisson brackets between Ashtekar connections and derivatives of conjugate triad variables will contribute. This is because anything like $\int dxdy f(x,y)\{\bar{A}_x(x),E^x(y)\}$ without derivatives inside the Poisson bracket is to be canceled out by the counter term $\int dxdy f(y,x)\{E^x(x),\bar{A}_x(y)\}$. Thus, for example, we can compute the term
\begin{eqnarray}\label{example term}
&&\left\{-\frac{\bar{A}_\rho^2E^\rho}{4\gamma^2\sqrt{\abs{E^x}}}\,(x), -\frac{\sqrt{\abs{E^x}}\,E^{x\prime}E^{\rho\prime}}{(E^\rho)^2}\,(y)\right\}
\sim \frac{E^\rho(x)}{4\gamma^2\sqrt{\abs{E^x(x)}}} \frac{\sqrt{\abs{E^x(y)}}\,E^{x\prime}(y)}{(E^\rho(y))^2} \,\left\{\bar{A}_\rho(x)^2,E^{\rho\prime}(y)\right\}\nonumber\\
&=&2G\gamma\frac{\bar{A}_\rho(x)E^\rho(x)}{2\gamma^2\sqrt{\abs{E^x(x)}}} \frac{\sqrt{\abs{E^x(y)}}\,E^{x\prime}(y)}{(E^\rho(y))^2} \,\partial_y\delta(x-y)
\end{eqnarray}
without worrying $\left\{\bar{A}_\rho(x),1/(E^\rho(y))^2\right\}\neq0$.
This largely simplifies the calculation and it turns out
\begin{subequations}\label{CN CM}
\begin{eqnarray}
\left\{\mathcal{C}[N],\mathcal{C}[M]\right\}
&=&\int dx dy N(x)M(y) \left\{C(x),C(y)\right\}\nonumber\\
\label{CC a}
&=&\int dx \frac{\sgn(E^x)}{2G\gamma\, (E^\rho)^2}
\Big\{\left(\bar{A}_xE^xE^{x'}-\bar{A}_\rho E^xE^{\rho'}
+\bar{A}_\rho E^\rho E^{x'}\right)(MN'-NM')\nonumber\\
&&\qquad \qquad \qquad \qquad +\bar{A}_\rho E^\rho E^x(MN''-NM'')\Big\}\\
\label{CC b}
&=& \frac{1}{2G\gamma} \int dx\, (NM'-MN') \frac{(E^x)^2}{\abs{E^x}(E^\rho)^2} \left(\bar{A}_\rho'E^\rho-\bar{A}_xE^{x'}\right),
\end{eqnarray}
\end{subequations}
which leads to \eqnref{pb f}.
In the above, we have applied integration by parts on the term associated with $\bar{A}_\rho E^\rho E^{x'}$ in \eqnref{CC a} to obtain \eqnref{CC b}.

\subsection{$n\rightarrow\infty$ case}

The same arguments used to derive \eqnref{CN CM} also apply to the case of $\left\{\mathcal{C}^{(\infty)}[N],\mathcal{C}^{(\infty)}[M]\right\}$. That is, only the Poisson brackets between holonomized connections and derivatives of conjugate triad variables will contribute. Consequently, the only relations that matter are \eqnref{pb infty a} and \eqnref{pb infty b}. The factors $f_\sqcap(\mubar_x\bar{A}_x)$ and $f_\sqcap(\mubar_\rho\bar{A}_\rho)$ on the right hand sides of \eqnref{pb infty a} and \eqnref{pb infty b} will give rise to extra factors of the same kinds. For example, \eqnref{example term} is modified accordingly as
\begin{eqnarray}
&&\left\{-\frac{(\bar{A}_\rho^{(\infty)})^2E^\rho}{4\gamma^2\sqrt{\abs{E^x}}}\,(x), -\frac{\sqrt{\abs{E^x}}\,E^{x\prime}E^{\rho\prime}}{(E^\rho)^2}\,(y)\right\}\nonumber\\
&\sim& \frac{E^\rho(x)}{4\gamma^2\sqrt{\abs{E^x(x)}}} \frac{\sqrt{\abs{E^x(y)}}\,E^{x\prime}(y)}{(E^\rho(y))^2} \,\left\{(\bar{A}_\rho^{(\infty)}(x))^2,E^{\rho\prime}(y)\right\}\nonumber\\
&=&2G\gamma f_\sqcap(\mubar_\rho\bar{A}_\rho)\frac{\bar{A}_\rho^{(\infty)}(x)E^\rho(x)}{2\gamma^2\sqrt{\abs{E^x(x)}}} \frac{\sqrt{\abs{E^x(y)}}\,E^{x\prime}(y)}{(E^\rho(y))^2} \,\partial_y\delta(x-y).
\end{eqnarray}
Taking care of all factors of $f_\sqcap(\mubar_x\bar{A}_x)$ and $f_\sqcap(\mubar_\rho\bar{A}_\rho)$, we obtained the counterpart of \eqnref{CN CM} for the case of $n\rightarrow\infty$:
\begin{eqnarray}\label{CN CM infty}
&&\left\{\mathcal{C}^{(\infty)}[N],\mathcal{C}^{(\infty)}[M]\right\}\nonumber\\
&=&\int dx \frac{\sgn(E^x)}{4G\gamma\, (E^\rho)^2}
\Big\{\left[2f_\sqcap(\mubar_x\bar{A}_x)\bar{A}_x^{(\infty)}E^xE^{x'} -2f_\sqcap(\mubar_\rho\bar{A}_\rho)\bar{A}_\rho^{(\infty)} E^xE^{\rho'}\right.\nonumber\\
&&\qquad \qquad \qquad \qquad \left. +\left(f_\sqcap(\mubar_x\bar{A}_x)+f_\sqcap(\mubar_\rho\bar{A}_\rho)\right)\bar{A}_\rho^{(\infty)} E^\rho E^{x'}\right](MN'-NM')\nonumber\\
&&\qquad \qquad \qquad \qquad +2f_\sqcap(\mubar_x\bar{A}_x)\bar{A}_\rho^{(\infty)} E^\rho E^x(MN''-NM'')\Big\}.
\end{eqnarray}
Before and after the bounce, \eqnref{CN CM infty} is largely simplified as
\begin{eqnarray}\label{CN CM infty 2}
\left\{\mathcal{C}^{(\infty)}[N],\mathcal{C}^{(\infty)}[M]\right\}
&=&
\pm\frac{1}{2G\gamma} \int dx\, (NM'-MN') \frac{(E^x)^2}{\abs{E^x}(E^\rho)^2} \left(\bar{A}_\rho^{(\infty)\,'}E^\rho-\bar{A}_x^{(\infty)}E^{x'}\right),\\
&& \pm:\text{ before/after bounce}. \nonumber
\end{eqnarray}
By \eqnref{D across bounce}, we then have
\begin{equation}\label{CN CM infty 3}
\left\{\mathcal{C}^{(\infty)}[N],\mathcal{C}^{(\infty)}[M]\right\}
\doteq
\frac{1}{2G\gamma} \int dx\, (NM'-MN') \frac{(E^x)^2}{\abs{E^x}(E^\rho)^2} \left(D(x)-\eta'\bar{P}^\eta\right),
\end{equation}
which leads to \eqnref{pb f infty}. The symbol $\doteq$ is used in \eqnref{CN CM infty 3} to indicate that the identity is valid only before and after the bounce but breaks down briefly during the transition of the bounce, because $f_\sqcap(\mubar_x\bar{A}_x)$ and $f_\sqcap(\mubar_\rho\bar{A}_\rho)$ flip signs at close but slightly different epochs during the transition and thus the simplification in \eqnref{CN CM infty 2} cannot apply.

\section{Hamilton's equations of heuristic effective dynamics}
At the level of heuristic effective dynamics with the $n$th order holonomy corrections, the corresponding Hamilton's equations is given by \eqnref{Hamilton eqs} with $C$ replaced by $C^{(n)}$ in \eqnref{C nth}. While \eqnref{Hamilton e} and \eqnref{Hamilton f} remain unchanged, \eqnref{Hamilton a}--\eqnref{Hamilton d} are modified accordingly and take the forms as
\begin{subequations}\label{Hamilton eqs nth order}
\begin{eqnarray}
\dot{\bar{A}}_x&=&\eqnref{Hamilton a} \Big|_{\bar{A}_x\rightarrow\bar{A}_x^{(n)},\bar{A}_\rho\rightarrow\bar{A}_\rho^{(n)}}\\
&&-N\left(\sgn(E_x)\frac{\sqrt{\abs{E^x}}\,\bar{A}_\rho^{(n)}}{\gamma}\right) \frac{\partial \bar{A}_x^{(n)}}{\partial E^x}
-N\left(\frac{\bar{A}_\rho^{(n)}E^\rho}{2\gamma\sqrt{\abs{E^x}}} +\sgn(E_x)\frac{\sqrt{\abs{E^x}}\,\bar{A}_x^{(n)}}{\gamma}\right) \frac{\partial \bar{A}_\rho^{(n)}}{\partial E^x},\nonumber\\
\dot{\bar{A}}_\rho&=&\eqnref{Hamilton b} \Big|_{\bar{A}_x\rightarrow\bar{A}_x^{(n)},\bar{A}_\rho\rightarrow\bar{A}_\rho^{(n)}}\\
&&-N\left(\sgn(E_x)\frac{\sqrt{\abs{E^x}}\,\bar{A}_\rho^{(n)}}{\gamma}\right) \frac{\partial \bar{A}_x^{(n)}}{\partial E^\rho}
-N\left(\frac{\bar{A}_\rho^{(n)}E^\rho}{2\gamma\sqrt{\abs{E^x}}} +\sgn(E_x)\frac{\sqrt{\abs{E^x}}\,\bar{A}_x^{(n)}}{\gamma}\right) \frac{\partial \bar{A}_\rho^{(n)}}{\partial E^\rho},\nonumber\\
\dot{E}^x &=& N\left(\sgn(E_x)\frac{\sqrt{\abs{E^x}}\,\bar{A}_\rho^{(n)}}{\gamma}\right)\frac{\partial \bar{A}_x^{(n)}}{\partial \bar{A}_x} +N^xE^{x\prime},\\
\dot{E}^\rho &=& N\left(\sgn(E_x)\frac{\sqrt{\abs{E^x}}\,\bar{A}_x^{(n)}}{\gamma} +\frac{\bar{A}_\rho^{(n)} E^\rho}{2\gamma\sqrt{\abs{E^x}}}\right)\frac{\partial \bar{A}_\rho^{(n)}}{\partial \bar{A}_\rho} +\left(N^xE^\rho\right)',
\end{eqnarray}
\end{subequations}
For the case of $n=\infty$, the Hamilton's equations above are largely simplified: derivatives of $\bar{A}_x^{(\infty)}$ and $\bar{A}_\rho^{(\infty)}$ with respect to canonical variables are given by \eqnref{derivatives of A infty}. Note that prior to the bounce, by \eqnref{triangle wave 2} and \eqnref{square wave}, the Hamilton's equations in \eqnref{Hamilton eqs nth order} for $n=\infty$ are identical to the classical counterparts in \eqnref{Hamilton eqs}.

With the prescription of \eqnref{mu bar modified case 2} in particular, \eqnref{dA dE} yields
\begin{subequations}
\begin{eqnarray}
\frac{\partial\bar{A}_x^{(\infty)}}{\partial E^x} &=&
\left\{
  \begin{array}{lr}
  0, & \text{before } \mubar_x\bar{A}_x \text{ reaches } \pm\pi/2,\\
  \mp\frac{\pi(4l^2-1)}{4\sqrt{\Delta}}\left(\frac{\mathcal{L}^2}{E^x}\right)^l \frac{E^\rho}{(E^x)^{3/2}}, & \text{after } \mubar_x\bar{A}_x, \text{ reaches } \pm\pi/2,
  \end{array}
\right.\\
\frac{\partial\bar{A}_x^{(\infty)}}{\partial E^\rho} &=&
\left\{
  \begin{array}{lr}
  0, & \text{before } \mubar_x\bar{A}_x \text{ reaches } \pm\pi/2,\\
  \pm\frac{\pi(2l-1)}{2\sqrt{\Delta}}\left(\frac{\mathcal{L}^2}{E^x}\right)^l \frac{1}{(E^x)^{1/2}}, & \text{after } \mubar_x\bar{A}_x, \text{ reaches } \pm\pi/2,
  \end{array}
\right.\\
\frac{\partial\bar{A}_\rho^{(\infty)}}{\partial E^x} &=&
\left\{
  \begin{array}{lr}
  0, & \text{before } \mubar_\rho\bar{A}_\rho \text{ reaches } \mp\pi/2,\\
  \pm\frac{\pi(2l-1)}{2\sqrt{\Delta}}\left(\frac{\mathcal{L}^2}{E^x}\right)^l \frac{1}{(E^x)^{1/2}}, & \text{after } \mubar_\rho\bar{A}_\rho, \text{ reaches } \mp\pi/2,
  \end{array}
\right.\\
\frac{\partial\bar{A}_\rho^{(\infty)}}{\partial E^\rho} &=& 0,
\end{eqnarray}
\end{subequations}
where we assume $E^x>0$ for simplicity as $E^x$ does not flip signs throughout evolution.

\end{document}